\documentclass[aps,prb,10pt,showpacs,floatfix,twocolumn,superscriptaddress,longbibliography,nofootinbib]{revtex4-2}

\usepackage{braket}

\usepackage{color}
\usepackage{bm}
\usepackage{hyperref}
\usepackage{verbatim}
\usepackage{soul}
\usepackage{placeins}
\usepackage{amssymb}
\usepackage{amsfonts}
\usepackage{amsmath}

\usepackage{mc}
\usepackage{soul}

\usepackage[caption=false]{subfig}
\usepackage{glossaries}
\usepackage{sidecap}
\usepackage{orcidlink}

\usepackage{xr-hyper}           
\sidecaptionvpos{figure}{c} 
\usepackage{hyperref}
\hypersetup{
    colorlinks=true,
    linkcolor=blue,
    filecolor=magenta,      
    urlcolor=red,
    citecolor=blue,
}
\usepackage{tcolorbox}
\tcbuselibrary{breakable}
\usepackage{xspace}
\usepackage{lipsum}

\newcommand{\Osmate}{Ca\ensuremath{_{3}}LiOsO\ensuremath{_{6}}\xspace}
\newcommand{\Hematite}{Fe\ensuremath{_{2}}O\ensuremath{_{3}}\xspace}
\newcommand{\NiPS}{NiPS\ensuremath{_{3}}\xspace}
\newcommand{\NiCl}{NiCl\ensuremath{_{2}}\xspace}

\newcommand{\be}{\begin{equation}}
\newcommand{\ee}{\end{equation}}
\newcommand{\bea}{\begin{eqnarray}}
\newcommand{\eea}{\end{eqnarray}}

\newacronym{RIXS}{RIXS}{resonant inelastic X-ray scattering}
\newacronym{XAS}{XAS}{X-ray absorption spectroscopy}
\newacronym{INS}{INS}{inelastic neutron scattering}
\newacronym{IXS}{IXS}{inelastic X-ray scattering}
\newacronym{ARPES}{ARPES}{angle-resolved photo-emission}
\newacronym{EELS}{EELS}{electron energy loss spectroscopy} 
\newacronym{STS}{STS}{scanning tunneling spectroscopy}
\newacronym{SOC}{SOC}{spin-orbit coupling}
\newacronym{QSL}{QSL}{quantum spin liquid}
\newacronym{XFEL}{XFEL}{X-ray free electron laser}
\newacronym{ED}{ED}{exact diagonalization}
\newacronym{EPC}{EPC}{electron-phonon coupling}
\newacronym{CDW}{CDW}{charge-density wave}
\newacronym{SM}{SM}{Supplemental Material}
\newacronym{SDW}{SDW}{spin-density wave}
\newacronym{SC}{SC}{superconducting}
\newacronym{KH}{KH}{Kramers-Heisenberg}
\newacronym{QMC}{QMC}{quantum Monte Carlo}
\newacronym{UCL}{UCL}{ultra-short core-hole lifetime}
\newacronym{DMRG}{DMRG}{density matrix renormalization group}
\newacronym{DMFT}{DMFT}{dynamical mean-field theory}
\newacronym{AFM}{AFM}{antiferromagnetic}
\newacronym{DCA}{DCA}{dynamical cluster approximation}
\newacronym{BCS}{BCS}{Bardeen, Cooper, and Schrieffer}
\newacronym{BZ}{BZ}{Brillouin zone}
\newacronym{1D}{1D}{one-dimensional}
\newacronym{2D}{2D}{two-dimensional}
\newacronym{3D}{3D}{three-dimensional}
\newacronym{QFI}{QFI}{quantum Fisher information}
\newacronym{trRIXS}{trRIXS}{time-resolved RIXS}
\newacronym{LCLS}{LCLS}{Linac Coherent Light Source}
\newacronym{HWHM}{HWHM}{half-width at half-maximum}
\newacronym{FWHM}{FWHM}{full-width at half-maximum}
\newacronym{vdW}{vdW}{van der Waals}
\newacronym{GPR}{GPR}{Gaussian process regression}
\newacronym{UCB}{UCB}{Upper Confidence Bound}
\newacronym{NSLS-II}{NSLS-II}{National Synchrotron Light Source II}
\newacronym{SIX}{SIX}{Soft Inelastic X-Ray}
\newacronym{SBI}{SBI}{simulation-based inference}
\newacronym{BOLFI}{BOLFI}{Bayesian Optimization for Likelihood-Free Inference}

\begin{document}

\title{Hamiltonian parameter inference from resonant inelastic x-ray scattering with active learning}

\author{Marton K. Lajer,\orcidlink{0000-0002-1168-8598}}\email[]{mlajer@bnl.gov}
\affiliation{Condensed Matter Physics and Materials Science Department, Brookhaven National Laboratory, Upton, New York 11973, USA\looseness=-1}

\author{Xin Dai\,\orcidlink{0000-0002-3235-1038}}
\affiliation{Computing and Data Sciences Directorate, Brookhaven National Laboratory, Upton, New York 11973, USA}

\author{Kipton Barros\,\orcidlink{0000-0002-1333-5972}}
\affiliation{Theoretical Division and CNLS, Los Alamos National Laboratory, Los Alamos, New Mexico 87545, USA}

\author{Matthew~R.~Carbone\,\orcidlink{0000-0002-5181-9513}}
\affiliation{Computing and Data Sciences Directorate, Brookhaven National Laboratory, Upton, New York 11973, USA}

\author{S. Johnston\,\orcidlink{0000-0002-2343-0113}}\email[]{sjohn145@utk.edu}
\affiliation{Department of Physics and Astronomy, The University of Tennessee, Knoxville, Tennessee 37996, USA}
\affiliation{Institute for Advanced Materials and Manufacturing, University of Tennessee, Knoxville, Tennessee 37996, USA\looseness=-1}

\author{M. P. M. Dean\,\orcidlink{/0000-0001-5139-3543}}\email[]{mdean@bnl.gov}
\affiliation{Condensed Matter Physics and Materials Science Department, Brookhaven National Laboratory, Upton, New York 11973, USA\looseness=-1}
\affiliation{Department of Physics and Astronomy, The University of Tennessee, Knoxville, Tennessee 37996, USA}

\date{\today}

\begin{abstract}
Identifying model Hamiltonians is a vital step toward creating predictive models of materials. Here, we combine Bayesian optimization with the \textsc{EDRIXS} numerical package to infer Hamiltonian parameters from resonant inelastic X-ray scattering (RIXS) spectra within the single atom approximation. To evaluate the efficacy of our method, we test it on experimental RIXS spectra of \NiPS, \NiCl, \Osmate, and \Hematite, and demonstrate that it can reproduce results obtained from hand-fitted parameters to a precision similar to expert human analysis while providing a more systematic mapping of parameter space. Our work provides a key first step toward solving the inverse scattering problem to extract effective multi-orbital models from information-dense RIXS measurements, which can be applied to a host of quantum materials. We also propose atomic model parameter sets for two materials, \Osmate and \Hematite, that were previously missing from the literature.
\end{abstract}

\maketitle

\section{Introduction}
Quantum materials are at the forefront of condensed matter research, due to their rich physics and potential as key components of future technologies~\cite{Basov2017towards, Tokura2017emergent, Giustino2020quantum, Mitrano2024exploring}. These systems are often governed by quantum fluctuations and strong many-body interactions, which are not well described by conventional single-particle theories or {\it ab initio} methods. As such, a central paradigm for progress in quantum materials research has been to identify minimal effective models~\cite{Anderson1997concepts} that capture a material's low-energy properties while remaining tractable. Leveraging such models is also a powerful way to understand and predict new material properties and identify broader organizing principles.  

Reliably identifying a valid minimal effective model for a given system can be extremely challenging. A paradigmatic example of this is the open question of whether the single-band Hubbard model is the correct low-energy model for high-$T_\mathrm{c}$ cuprates~\cite{Gull2013superconductivity, Jiang2019superconductivity, Qin2020absense, Chen2021anomalously, padma2025beyond, scheie2025cooper}. Other examples include magnetic \gls*{vdW} materials like NiPS$_3$ and NiI$_2$, which host novel many-body excitons that depend crucially on the magnetic state of the lattice \cite{Kang2020coherent, He2024MagneticallyPH, Occhialini2024}. Despite extensive studies, the electronic character, mobility, and magnetic interactions of these excitons are still being debated~\cite{Kang2020coherent, He2024MagneticallyPH, hamad2024singletpolarontheorylowenergy}, in part because their low-energy Hamiltonian has yet to be conclusively identified. 

Theoretically, one can attempt to derive an effective low-energy model from a high-energy Hamiltonian by integrating out various degrees of freedom. However, strongly correlated systems often exhibit a near-degeneracy of low-energy states that are easily affected by perturbing interactions. This situation can easily bias results if small-but-relevant interactions are not included in the downfolding process. Alternatively, one can attempt to derive a low-energy model using perturbation theory; however, in some cases, the resulting effective couplings can become non-analytic functions of the high-energy model parameters~\cite{Sharma2023machine}. 

An alternative line of attack is to take a data-driven approach in the so-called ``inverse scattering problem,'' where one attempts to extract the correct effective low-energy Hamiltonian from spectroscopic measurements. This approach has been bolstered by recent advances in instrumentation for spectroscopic techniques such as photoemission and inelastic neutron, electron, and photon scattering spectroscopies, as well as the development of new many-body algorithms. The former allows researchers to collect a large amount of high-quality data for a material in a short time, while the latter provides new capabilities for predicting the spectra for a proposed Hamiltonian (the ``direct scattering problem''). In real experiments, the inverse scattering problem is particularly relevant as model parameters are not known a priori, and obtaining them from first-principles calculations is often a challenging, if not impossible, task. Today, many practitioners of inverse scattering will hand-tune the parameters of the model to match experimental data. Fitting data by hand is laborious, requires considerable expertise, and can be error prone since humans can only visually inspect a rather modest number of candidate solutions.

The past few decades witnessed rapid growth of available computational resources, accompanied by the swift development of sophisticated algorithms to solve the direct scattering problem. At the same time, advances in machine learning have had a tremendous impact on the materials science community~\cite{Carrasquilla2020machine, Bedolla2021machine, 
Johnston2022perspective}. In particular, new insights from machine learning approaches~\cite{Carleo2019machine, Sharma2023machine, Carrasquilla2020machine, Karniadakis2021physics, Chen2021machine, Johnston2022perspective, carbone2019classification, carbone2020machine, torrisi2020random, miles2021machine, sturm2021predicting, kwon2023harnessing, kwon2024spectroscopy, ghose2023uncertainty, rankine2020deep} have made it possible to attack the inverse scattering problem from a purely data-driven framework. \Gls*{INS} provides a suitable testing ground for developing this approach, as its cross-section is well understood, and highly efficient direct solvers are available for a broad class of materials~\cite{Evans2014atomistic, Toth2015linear, Nocera2016spectral, dahlbom2025sunnyjljuliapackagespin}.

Since \gls*{INS} is very sensitive to magnetic excitations, the above-mentioned experimental efforts have so far primarily focused on extracting low-energy spin Hamiltonians for magnetic insulators \cite{Samarakoon2020machine, Samarakoon2021machine, Samarakoon2022integration}. Recent works established machine learning methods in the context of X-ray spectroscopy as well. In \gls*{XAS}, artificial neural networks have been utilized for parameter inference \cite{Luder2021} and surrogate models for the direct solver \cite{Luder2025ML}, while adversarial Bayesian optimization was used for active sampling \cite{zhang2023autonomous}. Bayesian optimization was also used for experimental acquisition in X-ray absorption near edge structure spectroscopy (XANES) \cite{Cakir2024, Du2025}. Our work extends these efforts to \gls*{RIXS} data, which allows us to access a much broader range of materials and Hamiltonians. \gls*{RIXS} is a photon-in photon-out spectroscopic method in which the energy, momentum, and polarization of an incoming x-ray photon are transferred to a material's intrinsic momentum-resolved spin, charge, orbital, and lattice excitations~\cite{Mitrano2024exploring, deGroot2024, Kotani2001, vandenBrink2007, Haverkort2010, Ament2011resonant, Chen2019, Vorwerk2022}. Due to its resonant nature, \gls*{RIXS} can study small samples, including monolayers, while still maintaining bulk sensitivity. \gls*{RIXS} is also suitable for monitoring ultra-fast responses of material in pump-probe experiments \cite{Cao2019ultrafast, Mitrano2020probing}. 

\gls*{RIXS} experiments have access to a large energy window spanning from $\approx 0.02$--$10$~eV and thus provide access to both high- and low-energy sectors of a given material. This aspect makes it an ideal tool for the inverse scattering problem as it can be used to extract effective models for different energy scales. Moreover, methods for computing \gls*{RIXS} spectra for correlated systems have achieved a degree of standardization and become widely available with packages like \textsc{EDRIXS}~\cite{Wang2019EDRIXS}, Quanty~\cite{Zimmermann:ve5082} and CleaRIXS~\cite{Roychoudhury2022}. Given the maturity of the experimental and numerical methods, the time is right to begin developing systematic approaches for the \gls*{RIXS} inverse scattering problem. Here, we undertake such an effort by employing Bayesian optimization with \textsc{EDRIXS}, an exact diagonalization-based open-source \gls*{RIXS} solver~\cite{Wang2019EDRIXS}, to predict model Hamiltonian parameters from experimentally measured spectra. We focus on the single ion model, which can provide quantitative descriptions of $dd$ excitations in transition metal complexes. We demonstrate its power on experimental spectra of \NiPS~\cite{He2024MagneticallyPH}, \NiCl~\cite{Occhialini2024}, \Hematite~\cite{JieminLi2023Fe2O3}, and \Osmate~\cite{Taylor2016SpinOrbitCC}. The resulting parameters are then used to annotate and predict the properties of experimentally observed peaks, including their dependence on temperature and polarization. This work provides the foundation for a fully automated solution to the \gls*{RIXS} inverse scattering problem.

The paper is organized as follows: Sec.~\ref{sec:RIXStheory} describes the atomic approximation used to model the \gls*{RIXS} spectra. Sec.~\ref{sec:Gaussian} introduces the active learning techniques used to solve the inverse scattering problem. Sec.~\ref{sec:implementation} describes how our methods combine the techniques above to infer Hamiltonians from \gls*{RIXS} data. Sec.~\ref{ResultsSection} presents results for several compounds that have been studied previously in the literature. Finally, Sec.~\ref{ConclusionSection} provides some concluding remarks and discussion.

\section{Methods}\label{sec:Theory}

\subsection{RIXS Modeling}\label{sec:RIXStheory}

We compute \gls*{RIXS} using the \textsc{EDRIXS} package~\cite{Wang2019EDRIXS, EDRIXS}. This software implements the \gls*{KH} equation, which is the result of treating the photon-matter interaction using second-order perturbation theory.\footnote{For a detailed derivation of this expression from the light-matter interaction, we refer the reader to the review by Ament {\it et al}.~\cite{Ament2011resonant}.} Denoting the momentum, energy, and polarization of the incoming (outgoing) X-rays as $\hbar \bm{k}$, $\hbar\omega_{\bm{k}}$, and $\hat{\epsilon}$ ($\hbar \bm{k}^\prime$, $\hbar\omega_{\bm{k}^\prime}$, and $\hat{\epsilon}^\prime$), respectively, the intensity for \gls*{RIXS} can be written as
\begin{multline}\label{eq:KH1}
I_{\epsilon\epsilon'}(\hbar\omega_{\bm{k}}, \hbar\omega_{\bm{k}^\prime}, \boldsymbol{k}, \boldsymbol{k}'; T) \propto \frac{1}{\mathcal{Z}(T)}\sum_{i}e^{- E_{i}/(k_\mathrm{B}T)} \\
\times \sum_f |M_{fi}|^2 \delta(E_f + \hbar\omega_{\bm{k}^\prime} - E_i - \hbar\omega_{\bm{k}} ).
\end{multline}
Here, $M_{fi}$ represents the matrix element from the system's initial state $i$ with energy $E_i$ to its final state $f$ with energy $E_f$ via an intermediate state with a core hole, and $\mathcal{Z}(T) = \sum_i e^{- E_i/(k_\mathrm{B}T)}$ is the partition function at temperature $T$. $k_\mathrm{B}$ is the Boltzmann constant.

The examples considered in this work correspond to $L$-edge \gls*{RIXS}, which involves a $d$-electron valence state and a $2p$ core hole. We wish to compute the so-called $dd$ excitations, where the final states entail a reconfiguration of electrons within the $3d$ manifold, and which manifest as peaks at photon energy loss $E_\text{loss} = \hbar (\omega_{\bm{k}} - \omega_{\bm{k}^\prime})$. The matrix element for \gls*{RIXS} within the \gls*{KH} formalism is given by 
\begin{equation}\label{eq:KH_ME}
M_{fi} = \sum_n \frac{\bra{f} {\cal D}^\dagger_{\bm{k}^\prime\hat{\epsilon}^\prime}\ket{n}\bra{n} {\cal D}^{\phantom\dagger}_{\bm{k}\hat{\epsilon}}\ket{i}}{E_n - E_i - \hbar\omega_{\bm{k}}+\mathrm{i}\Gamma_n/2},
\end{equation}
where $\Gamma_n/2$ is the inverse core-hole lifetime in units of energy and $E_n$ is the energy of the intermediate state. $\cal{D}^{\phantom\dagger}_{{\bf k}\hat{\epsilon}}$ is the operator describing the absorption of a photon, promoting a core electron from a $2p$ state into the $d$ valence states. Similarly, $\cal{D}^\dagger_{{\bf k}^\prime\hat{\epsilon}^\prime}$ describes the photon emission process via a $d$ to $2p$ transition. $\ket{i}$ and $\ket{f}$ are the eigenstates of the initial state Hamiltonian $\hat{H}_i$, whereas $\ket{n}$ represents the eigenstates of the intermediate state Hamiltonian $\hat{H}_n$ with a core hole. Due to the attractive potential caused by the core hole, the cross-section is dominated by local transitions, so we take the common approximation of treating the process in the atomic limit, where transitions occur within effective local $3d$ orbitals~\cite{Mitrano2024exploring}. 

Adopting the second quantization formalism, $\hat{H}_i$ and $\hat{H}_n$ take the general form
\begin{equation}
\hat{H} = \sum_{\alpha\beta}t_{\alpha\beta}\hat{f}_{\alpha}^{\dagger}\hat{f}^{\phantom\dagger}_{\beta}+\sum_{\alpha\beta\gamma\delta}U_{\alpha\beta\gamma\delta}\hat{f}_{\alpha}^{\dagger}\hat{f}_{\beta}^{\dagger}\hat{f}^{\phantom\dagger}_{\delta}\hat{f}^{\phantom\dagger}_{\gamma}, \label{eq:secondQ}
\end{equation}
where the indices run through atomic valence and core orbitals and $\hat{f}_{\alpha}^{\dagger}$
creates an electron in spin-orbital $\alpha$. The first term of Eq.~\eqref{eq:secondQ} includes the crystal field as well as spin-orbit coupling and, if appropriate, an applied magnetic field. Here we consider the cubic case in which $t_{\alpha\beta}$ is diagonal in the basis of \emph{real} spherical harmonics: it has eigenvalues $6D_q$ in the $e_g$ orbitals ($d_{x^2-y^2}$ and $d_{z^2}$) and $-4D_q$ for the $t_{2g}$ orbitals ($d_{xy}$, $d_{xz}$ and $d_{yz}$). Valence band spin-orbit coupling for the $d$ states in the initial (intermediate) state is parameterized by $\zeta_{i}$ ($\zeta_{n}$). The core hole spin-orbit coupling is parameterized by $\zeta_{c}$.

The second term of Eq.~\eqref{eq:secondQ} accounts for the Coulomb interactions with the $d$ shell and between the $p$ and $d$ shells and is parameterized by Slater integrals~\cite{Groot2008core}. The relevant terms for the initial Hamiltonian are $F^2_{dd}$ and $F^4_{dd}$. In the intermediate state, these are supplemented by additional parameters describing the interactions between the valence electrons and the core-hole, $F^2_{dp}$, $F^4_{dp}$, $G^1_{dp}$, and $G^3_{dp}$. For an atomic model, as applied here, the parameters $F^0_{dd}$ and $F^{0}_{dp}$ only produce overall shifts in the eigenenergies. In general, the valence Slater integrals can also be modified in the presence of the core hole, but we make the approximation that $F^2_{dd}$ and $F^4_{dd}$ remain the same in the initial and intermediate states. This is reasonable because core-hole effects tend to cause only moderate changes in these values, and because the parameters in the intermediate state affect only the resonant profile of the excitations and not the energies of the final states. Since the resonant profile is already broadened by core-hole lifetime effects, these parameters tend to have only a secondary influence on the quality of the agreement between theory and experiment. As will be seen later, this assumption is validated by the good level of agreement obtained between theory and simulation.

Since the X-ray wavelength tends to be larger than the extent of the atomic orbitals, we treat $\mathcal{D}^{\phantom\dagger}_{{\bf k}\hat{\epsilon}}$ and $\mathcal{D}^\dagger_{{\bf k}^\prime\hat{\epsilon}^\prime}$  within the dipole approximation. 
In the atomic limit this means that $\mathcal{D}^{\phantom\dagger}_{\boldsymbol{k}\hat{\epsilon}} = \sum_{\alpha,\beta}\bra{\phi_\alpha} \hat{\epsilon} \cdot \hat{r} \ket{\phi_\beta} \hat{f}_\alpha^\dagger \hat{f}_\beta^{\phantom{\dagger}}$
,  where $\beta$ indexes a core spin-orbital and $\alpha$ is a valence spin-orbital and $\ket{\phi_{\alpha(\beta)}}$ are the corresponding atomic orbital states.

To account for the finite experimental energy resolution and the finite lifetime of the $dd$ excitations, the delta function in Eq.~\eqref{eq:KH1} is represented by a Lorentzian with \gls*{FWHM} $\gamma$. Both  $\gamma$ and $\Gamma_n$ are considered as constants independent of the state in question. Because the absolute core-hole energy is undefined in this model, we introduce an adjustable energy offset $x_{\text{offset}}$ to align the theoretical spectra with the experimental data.

\begin{figure}[t]
\includegraphics[draft=false,width=\columnwidth]{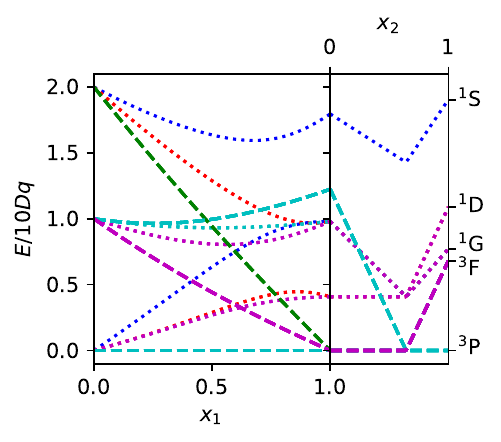}
\caption{Tanabe-Sugano-style plots for $d$-shell electronic (initial) Hamiltonians of the form $x_1 U_\text{coul}^{F^2}+(1-x_1)V_\text{cf} + x_2 U_\text{coul}^{F^4}-(1-x_2)U_\text{coul}^{F^2}$ with occupation numbers 2 or 8. On the left subplot, $x_2=0$ and $x_1$ is varied  ($x_1$ subplot). On the right subplot, $x_1=1$ and $x_2$ is varied. Dotted lines correspond to spin singlets, dashed lines denote triplets. States transforming according to different irreducible representations of the octahedral group are shown in different colors: $A_1$ -- blue, $A_2$ -- green, $E$ -- red, $T_1$ -- teal, $T_2$ -- purple. Ground state energies are subtracted so that occupations $2$ and $8$ yield the same figure. This plot gives an overview of the spectrum of $H_i$ as a function of its three most important parameters $F^2$, $F^4$, and $10Dq$, with other parameters fixed to zero. 
}
\label{fig:TanabeSugano}
\end{figure}

To interpret the \gls*{RIXS} spectra and distinguish different classes of solutions to the inverse scattering problem, it is helpful to be able to label the different $n$-electron eigenvectors of $\hat{H}_i$. In the absence of \gls*{SOC} and crystal field splitting, the eigenvalues of the spin and orbital angular momentum operators are good quantum numbers and form useful labels. This breaks down for the more general case. Still, the eigenvectors can be effectively annotated by the states that transform according to a certain irreducible representations of the octahedral group, as illustrated in Fig.~\ref{fig:TanabeSugano}. When the octahedral symmetry is not exact, we use the weights of the eigenvectors within definite symmetry sectors to approximately characterize the states. Our annotation method is described in Sec.~S1 of the \gls*{SM}~\cite{supplement}.

\subsection{Bayesian optimization via Gaussian process regression}\label{sec:Gaussian}

The inverse scattering problem involves searching a high-dimensional parameter space to identify the model Hamiltonian that best reproduces experimental spectra. Since solving the direct problem (i.e., computing the theoretical \gls*{RIXS} spectrum from a given Hamiltonian) is potentially computationally expensive, it is imperative to minimize the number of function evaluations required during inference. To address this, we employ a Bayesian optimization strategy that builds a surrogate model of the distance function between simulated and experimental spectra, guiding the search for optimal parameters in a sample-efficient manner \cite{Frazier2018tutorial}.

\subsubsection{Distance metrics}
To compare theoretical and experimental spectra, we define a distance function $\chi^2$ operating directly on the \gls*{2D} spectral images. In the following formulae, we denote the entries of the \gls*{2D} arrays containing the experimental (reference) and simulated spectral intensities by $R_{ij}$ and $S_{ij}$, respectively. The summation over indices $i,j,k,l$, etc. spans over each row and column of the corresponding arrays. Several pixel-wise metrics are considered. They include the sum normalized \( L_p \) distance
\bea
  \chi^2_{L_p} = \left(\sum_{ij} \left| r_{ij} - s_{ij} \right|^p\right)^{\frac{1}{p}},
\eea
where 
  \[
    r_{ij}=\frac{R_{ij}}{\sum_{kl} R_{kl}}\quad s_{ij}=\frac{S_{ij}}{\sum_{kl} S_{kl}}; 
  \]
the maximum normalized \( L_p \) distance
\bea
  \chi^2_{L^\prime_p} = \left(\sum_{ij} \left| \frac{R_{ij}}{\max R} - \frac{S_{ij}}{\max S} \right|^p\right)^{\frac{1}{p}}, 
\eea
and the numerical gradient
\bea
\chi^2_\mathrm{g}&=&\sum_{ij}\left|\sqrt{\left(\frac{r_{ij}-r_{i,j-1}}{\Delta \omega_\text{in}}\right)^2+\left(\frac{r_{ij}-r_{i-1,j}}{\Delta E_{\text{loss}}}\right)^2}\right.\cr\cr
&&-(r\leftrightarrow s)\Biggr|, 
\eea
which emphasizes features like sharp peaks or other abrupt changes in the spectrum. Here, $\Delta\omega_\mathrm{in}$ and $\Delta E_\mathrm{loss}$ are the energy spacings along the $\omega_\mathrm{in}$ and $E_\mathrm{loss}$ axes, respectively. 

\subsubsection{Gaussian process regression}

Our approach is based on iteratively constructing and refining a probabilistic model of the distance landscape, defined over the space of Hamiltonian parameters. The central idea is to treat the evaluation of the \gls*{RIXS} spectrum as a black-box function that is expensive to query but smooth and continuous over physically relevant regions.

To model this function, we use \gls*{GPR}, a flexible, non-parametric regression method that defines a distribution over functions \cite{Wang2020GPRIntuitiveTutorial, Rasmussen_Williams_2006}. In our case, the input to the \gls*{GPR} is a vector of Hamiltonian parameters $\boldsymbol{z}$, such as crystal field splitting $10D_q$, Slater integrals, etc. The output is the value of a distance function $\chi^2$, which quantifies the discrepancy between the \gls*{RIXS} spectrum generated by those parameters and the experimental spectrum. Formally, \gls*{GPR} models the function \( \chi^2: \mathbb{R}^n \rightarrow \mathbb{R} \), where $n$ is the number of free parameters being optimized.

A Gaussian process is fully specified by a mean function $f_{\text{prior}}(\boldsymbol{z})$ and a kernel function $k(\boldsymbol{z}_1,\boldsymbol{z}_2)$. Given a set of previously evaluated parameter points and their corresponding distance values, a Gaussian process provides a posterior distribution over functions that can be used to predict the distance $f_{\text{reconst}}(\boldsymbol{z})$ at unseen parameter configurations
\begin{align}
f_{\text{reconst}}(\boldsymbol{z})&=f_{\text{prior}}(\boldsymbol{z})
\cr
&+\sum_{ij}k(\boldsymbol{z},\boldsymbol{z}_i)\left(\boldsymbol{K}^{-1}\right)_{ij}\left[f(\boldsymbol{z}_j)-f_{\text{prior}}(\boldsymbol{z}_j)\right]\cr,
\end{align}
along with a confidence interval 
\begin{align}
\delta f_{\text{reconst}}(\boldsymbol{z})&=k(\boldsymbol{z},\boldsymbol{z})-\sum_{ij}k(\boldsymbol{z},\boldsymbol{z}_i)\left(\boldsymbol{K}^{-1}\right)_{ij}k(\boldsymbol{z}_j,\boldsymbol{z}), 
\end{align}
where the covariance matrix is
 \be
 \boldsymbol{K}=
 \left(
 \begin{matrix}
 k(\boldsymbol{z}_1,\boldsymbol{z}_1) & k(\boldsymbol{z}_1,\boldsymbol{z}_2) & \dots & k(\boldsymbol{z}_1,\boldsymbol{z}_n) \\
 k(\boldsymbol{z}_2,\boldsymbol{z}_1) & \ddots & & \\
 \vdots & & & \vdots \\
 k(\boldsymbol{z}_n,\boldsymbol{z}_1) &\dots& & k(\boldsymbol{z}_n,\boldsymbol{z}_n)
 \end{matrix}
 \right).
 \ee
Throughout this work, we use the Matérn kernel with $\nu=\frac{5}{2}$
\bea
k(\boldsymbol{z}_1,\boldsymbol{z}_2)&=&\left(1+\frac{\sqrt{5}}{l}d(\boldsymbol{z}_1,\boldsymbol{z}_2)+\frac{5}{3l}d(\boldsymbol{z}_1,\boldsymbol{z}_2)^2\right)\cr
&&\quad\times\exp\left(-\frac{\sqrt{5}}{l}d(\boldsymbol{z}_1,\boldsymbol{z}_2)\right), 
\eea
where the length-scale parameter $l$ is optimized internally through the regression process.
The Bayesian prior $f_\text{prior}(\boldsymbol{z})$ is set to zero in the following.

\begin{figure*}[t]
\includegraphics[width=\textwidth]{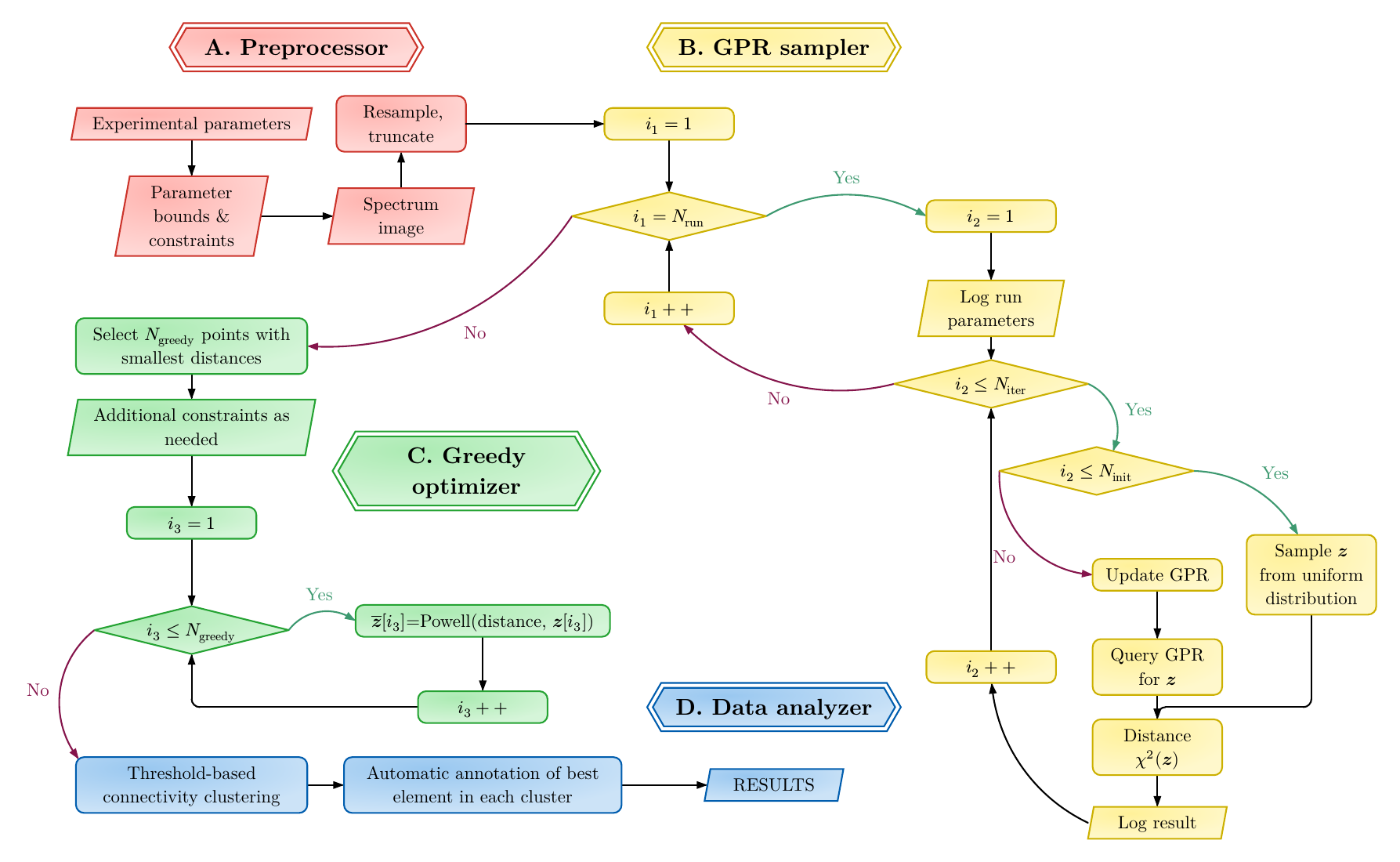}
\caption{Flowchart of the Bayesian optimization algorithm used to solve the inverse scattering problem for \gls*{RIXS}. A. The preprocessor imports and processes a spectrum image from a \gls*{RIXS} experiment. B. The \gls*{GPR} sampler builds a set of $N_\text{run}$ models of the distance function via an active learning protocol. Each run consists of $N_\text{iter}$ \gls*{GPR} iterations, followed by $N_\text{init}$ initial evaluations at random points. C. Parameters obtained from the $N_\text{greedy}$ smallest queried distances are refined by a subsequent greedy optimizer. D. Finally, the results are presented by the data analyzer.} \label{FigFlowchart}
\end{figure*}

\subsubsection{Acquisition function}
 The acquisition function is a central component in active learning, which determines the next sampling point in the parameter space based on the surrogate model constructed by \gls*{GPR}. This function quantifies the trade-off between exploration of regions with large uncertainties and exploitation of regions predicted to yield low distances, i.e., good  descriptions of the experimental spectra.
In our approach, we employ the \gls*{UCB} acquisition function, which balances the mean prediction of the \gls*{GPR} and its associated uncertainty \cite{Kaufmann2012Bayesian}. It is defined as:
\begin{equation}\label{acq_func}
\mathrm{UCB}(\boldsymbol{z}) = f_{\mathrm{reconst}}(\boldsymbol{z}) + \kappa\, \delta f_{\mathrm{reconst}}(\boldsymbol{z}),
\end{equation}
where \( f_{\text{reconst}}(\boldsymbol{z}) \) is the Gaussian process's predicted distance at parameter configuration \(\boldsymbol{z}\), \( \delta f_{\text{reconst}}(\boldsymbol{z}) \) is the model's uncertainty estimate, and \(\kappa\) is a hyperparameter controlling the exploration-exploitation balance. A larger \(\kappa\) encourages exploration by favoring points with high uncertainty, while a lower value promotes exploitation by prioritizing low predicted distances.
The \gls*{UCB} acquisition function enables efficient navigation of the high-dimensional Hamiltonian parameter space by guiding the sampling toward informative regions. This aspect is crucial for minimizing the number of \gls*{RIXS} simulations during the optimization process.

\section{Implementation of the inverse scattering problem}\label{sec:implementation}
Figure~\ref{FigFlowchart} presents a flowchart of the method. We employ a two-stage optimization strategy that combines global exploration with local refinement to efficiently minimize the distance between simulated and experimental \gls*{RIXS} spectra in a high-dimensional parameter space.

In the first stage, the experimental data is preprocessed. In the second stage, we use \gls*{GPR} to construct a surrogate model of the distance function over the Hamiltonian parameter space. The \gls*{GPR} is initialized with a small number of randomly sampled points, and subsequent evaluations are guided by the acquisition function Eq.~\eqref{acq_func}. To improve efficiency and stability, we restrict optimization at this stage to a subset of parameters that have the strongest influence on the \gls*{RIXS} spectra, while holding others such as energy loss broadening $\gamma$ and most spin-orbit couplings fixed. This design choice helps reduce dimensionality and avoids uninformative regions of parameter space. Further details can be found in Sec.~\ref{gpr_sampler}.

In the third stage, we apply a local, derivative-free optimizer (in our case, Powell’s method via \texttt{py-BOBYQA}) to refine the best candidates found by \gls*{GPR} \cite{Cartis2019improving}. This step not only improves the precision of previously optimized parameters but also reintroduces variations for any parameters fixed in the second stage, enabling a full-space optimization. This refinement enhances convergence and consistently leads to lower distance metrics.

Finally, in the fourth stage, the results are processed, peak features are classified according to their symmetry, and the results are output. 

In the following sections, we provide a detailed description of each stage of the process.

\subsection{Preprocessing}
We take the experimental spectrum in the form of a two-dimensional
array of intensities interpolated onto a regular grid. The array indices correspond to different values of the incident energy $\omega_\text{in}$ and energy loss $E_\text{loss}$. The intensities are positive, but given only up to an overall multiplicative constant. The experimental spectrum typically exhibits many peaks, some of which are better described by the single ion model than others. Since we focus on $dd$-excitations, the first step is to crop the spectrum in the energy loss direction, omitting the effect of the elastic  peak, low-energy collective excitations, and excitations to other high-energy orbitals (e.g., charge transfer excitations). Restricting the spectrum to the relevant energy range of the model used is crucial and significantly impacts the reliability of the results. 

\subsection{GPR sampler}\label{gpr_sampler}
Once preprocessing is complete, we provide the processed spectrum to the \gls*{GPR} sampler, which constructs a model of a distance function $\chi^2$ as a function of the model parameters, $\boldsymbol{z}$. Evaluating the distance function involves running the direct scattering problem, that is using \textsc{EDRIXS} to compute the \gls*{RIXS} spectra. 

We have found that it is beneficial to use the sum-normalized $L_1$ distance $\chi^2_{L_1}$ when the \gls*{RIXS} spectrum is not too crowded, e.g., away from a half-filled initial valence shell. This distance function normalizes with respect to the volume under the spectrum, so it gives considerable weight to fainter side peaks that the max-norm measure $\chi^2_{L^\prime_1}$ is more prone to missing. In contrast, when the initial configuration is closer to half-filling, the parameter space consists of a plethora of configurations with intensities widely spread in energy space. This structure leads to a rich and shallow landscape for the sum norm, which makes optimization difficult. Under these circumstances, we have found that the max norm, which gives extra emphasis to the region around the brightest peak, is more robust. We have also found that the gradient norm generally underperformed the sum and max norms. The behaviors of each norm around the final parameter set for each material are reported in Fig.~\ref{NiPS3Distances} of App.~\ref{FitDetails} and Sec.~S2 of the \gls*{SM}~\cite{supplement}.

To build the Gaussian approximation of $\chi^2$, we initially evaluate the distance function at a small number $N_\text{init}=10$ of random points in parameter space. The Gaussian process provides an estimate $\chi^2$ as well as its uncertainty between the calculated points. We then feed this information to the acquisition function [see Eq.~\eqref{acq_func}] and maximize this function to yield the next queried point (at which $\chi^2$ is to be evaluated). Once the value of $\chi^2$ is known at the new point, the \gls*{GPR} model is updated, and the acquisition is maximized again to obtain the next point. The process is iterated for $N_\text{iter}=1000$ iterations. This active learning approach is intended to aid the solution of the global minimization of the distance function with a minimal number of expensive spectral simulations. To improve the results and get a sense of the robustness of the optimization, we repeat the buildup of the Gaussian model from different initial points for $N_\text{run}$ times in total. This yields a total number $N_\text{run}\times\left(N_\text{init}+N_\text{iter}\right)$
evaluations of $\chi^2$ (i.e., calls to the \textsc{EDRIXS} solver). 

Since the number of a priori floating parameters is rather large, we chose a subset of parameters with respect to which $\chi^2$ is optimized by \gls*{GPR}.  This selection is mainly dictated by the sensitivity of the distance function against varying the corresponding parameter. Such dependencies can be predicted theoretically and confirmed numerically, as detailed in Fig.~\ref{NiPS3Distances} and Figs.~S3, S5, and S8 of the \gls*{SM}~\cite{supplement}. Another concern pertaining to the broadening parameters is that floating them often allows the optimization to enter uninteresting valleys of $\chi^2$ where the broadening gets very large and the distinguishing features of the \gls*{RIXS} spectra are washed out. To avoid this, we fix the value of the core-hole broadening $\Gamma/2$ at this stage, as well as (with the exception of \Osmate) all spin-orbit couplings to their atomic values. We also set the final state broadening $\gamma$ empirically to match the typical width of the $dd$ excitations. \Gls*{GPR} then focuses on fitting $10Dq$, the Slater parameters, the offset $x_{\text{offset}}$, and in the case of \Osmate, the initial and intermediate state valence spin-orbit couplings with the temporary constraint $\zeta_{v,i}=\zeta_{v,n}$.

The results of different runs are combined and ranked by distance function value before being passed to a subsequent greedy optimizer. In doing this, one tests for the presence of multiple local minima in the distance function and the extent to which these have similar fit quality. This stage further restricts the parameter space to the most interesting candidate regions and excludes any obviously unphysical local minima identified by \gls*{GPR}. For example, one can exclude small-distance regions where $F^{2}_{dd}\ll F^{4}_{dd}$. 

Although we found that \gls*{GPR} itself is efficient in locating promising regions in parameter space, we found it advantageous to add a further optimization step to refine further the parameters obtained.

\subsection{Greedy optimizer}
Since the \gls*{GPR} sampling balances exploration and optimization in a complex parameter space, it is not necessarily best suited for fine optimization when close to a minimum in the distance landscape. In principle, one could fine-tune the fitting process by adjusting the hyperparameter $\kappa$ controlling the balance between exploration and exploitation. However, we have found it more efficient to instead include a step in which we collect the ``best'' $N_{\text{greedy}}\sim10-30$ evaluated points with the smallest distance function values from the candidate parameter regions. These points are used as initial points to start a greedy optimization of the distance function. The greedy optimizer uses a refined version of Powell's method, provided by the package \texttt{py-BOBYQA} \cite{Cartis2019improving}. At this stage, the optimizer extends the set of fitted parameters to now include the spin-orbit couplings and other parameters that were omitted from \gls*{GPR}. The optimization started from a particular initial point typically converges in a few thousand function evaluations. 

Besides improving on the parameter estimates, the greedy optimization provides valuable information on the performance of the \gls*{GPR} minimization and the robustness of the method in general. The greedy optimizer is used to obtain basic uncertainty intervals for the local solutions.  In principle, it might be possible to use techniques like simulation-based inference to obtain formal distribution functions for the parameters. However, this would not be very practical given the desire to minimize the number of \gls*{RIXS} simulations, and it would not be particularly useful given that the true uncertainties on the parameters have a significant contribution from the approximations used in the atomic model for \gls*{RIXS}.

\subsection{Data analyzer}
The greedy-optimized set of points is grouped based on a threshold-based connectivity clustering approach. (To illustrate the kinds of differences that are typically found in these clustered solutions, Fig.~\ref{FigFe2O3Alternatives} of App.~\ref{FitDetails} plots the spectra obtained from the best fits of three closely packed clusters obtained for \Hematite.) 
Our analysis pipeline presents the user with a list of physically distinct solutions ordered by quality of fit. The robustness of the method is benchmarked by asking the question, how many of the runs provide points that are relaxed into the peak corresponding to the accepted fit. We use the $N_{\text{greedy}}$ best points of the \gls*{GPR} sampler as a proxy to estimate the robustness of the fitting procedure.

Finally, the eigenstates of the initial Hamiltonian for the best fit are annotated by eigenvalues of a set of approximately conserved quantities and discrete symmetry labels. We developed our own code to perform this annotation automatically, which is detailed in Section S1 in the \gls*{SM}~\cite{supplement}.

\section{Results for experimental spectra} \label{ResultsSection}
\subsection{NiPS\texorpdfstring{$_3$}{3}}

\NiPS is a layered van der Waals crystal in which the active Ni $d^8$ ions are coordinated by six S atoms in approximately cubic symmetry \cite{Wildes2022NiPS,Wildes2015NiPS}. It has attracted significant attention for its optoelectronic properties, including a novel magnetic exciton \cite{Kang2020coherent, He2024MagneticallyPH} and its two-dimensional magnetism \cite{Kim2019supression}.  We consider this material because it represents a challenging case in which the atomic model can capture some, but not all, of the physics of this material in light of its small charge-transfer energy \cite{Kang2020coherent, He2024MagneticallyPH}. 

We use the experimental \gls*{RIXS} spectrum from Ref.~\cite{He2024MagneticallyPH}. The spectrum was measured at the SIX 2-ID beamline of the \gls*{NSLS-II} with an energy resolution of 31~meV \gls*{FWHM}. The experimental spectrum was recorded at temperature $T=40\:\text{K}$, well below the N\'{e}el ordering temperature of 140~K. The geometry involved an incident x-ray angle of $\theta_\text{in}=23^{\circ}$ and a scattering angle of $2\Theta=150^{\circ}$. Intensities were recorded in a incident energy range $\omega_\text{in}\in[848.5~\text{eV},857.6~\text{eV}]$ around the Ni $L_3$-edge with $\pi$-polarized incident x-rays. The originally measured spectrum was subsequently truncated to the energy loss range $E_\text{loss}\in[0.5~\text{eV},2.0~\text{eV}]$. It was also resampled into an equidistant grid with $40$ points in the $\omega_\text{in}$ direction and $151$ points in the $E_\text{loss}$ direction.

\begin{figure}
\begin{centering}
\includegraphics[width=\columnwidth]{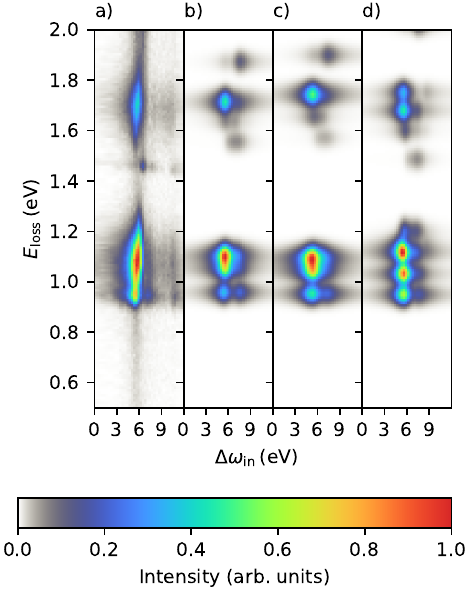}
\par\end{centering}
\caption{Experimental and simulated spectra for \NiPS. (a) experimental data from Ref.~\cite{He2024MagneticallyPH}, (b) hand-fitted (reference), (c) result of GPR,
(d) result after greedy fine-tuning (Powell's method). The corresponding $L_1$ sum distances for the models shown in panels b-d are $\chi^2_{L_1} = 0.692$, $0.694$, and $0.480$, respectively.}
\label{NiPS3_plot}
\end{figure}

\subsubsection{GPR results}
The comparison of experimental spectrum, together with the hand-fit results of Ref.~\cite{He2024MagneticallyPH} (which we refer to as the ``reference'' model hereafter), \gls*{GPR},  and greedy-updated simulated spectra are shown in Fig.~\ref{NiPS3_plot}. Fitted parameters after the greedy refinement are reported along with their reference counterparts in Tab.~\ref{NiPStab}. Reference values included in Tab.~\ref{NiPStab} are taken from Ref.~\cite{He2024MagneticallyPH}. The table includes two different uncertainty intervals for each fitted parameter. Intervals 1.25 and 1.5 correspond to the intervals within which the distance function increases by 1.25 and 1.5, respectively, from their minimum values (see also Fig.~\ref{NiPS3Distances}).
The distance function $\chi^2_{L_1}$ was used in the optimization.

The results obtained from the \gls*{GPR} and after greedy fine-tuning are in excellent agreement with the reference spectrum, and of compatible (if not superior) to the hand-fitted spectrum. 
The classification of the excitations is also similar for both the \gls*{GPR} and greedy fine-tuned solutions. 
The ground state is a triplet $\:^{3}A_{1}$ state with completely filled $t_{2g}$ subshell and two electrons on the $e_{g}$ manifold. Here the upper index denotes spin multiplicity and $A_{1}$ is the symmetry label with respect to the octahedral group. The lowest lying excitations at around $1-1.2$ eV correspond to $dd$ excitations where one electron transfers between the $t_{2g}$ and $e_{g}$ orbitals. These excitations typically have $^{3}T_{2}$ symmetry, with some mixing of $^{1}E$ and $^{3}T_{1}$.  The higher energy loss peaks are predominantly $^{3}T_{1}$ with either one or two electrons transferred to $e_g$ orbitals. These annotations align well with previous findings in the literature \cite{He2024MagneticallyPH}.

It should be noted that the feature around $1.45$ eV, which has previously been identified as a Hund's exciton \cite{He2024MagneticallyPH}, is not perfectly predicted by the model. This difference, and other small discrepancies between the theory and experiment, arise from the atomic approximation used here. In particular, it has been shown that an Anderson impurity model is required to more accurately capture these excitations due to the rather small energetic difference between the Ni $d$ and S $p$ states in this material \cite{Kim2018charge}.

\begin{table}[t]
\caption{NiPS$_{3}$ parameters (in units of eV). The ``interval 1.25'' and ``interval 1.5'' refer to the range of values for each parameter that keep the distance function within $1.25\times$ or $1.5\times$ its minimum value, respectively, with all other parameters held fixed. The abbreviation ``Pt. est.'' is short for point estimate: the set of parameters corresponding to the smallest achieved
minimum of the distance function.}
\begin{centering}
\begin{ruledtabular}
\begin{tabular}{ccccc}\label{NiPStab}
Parameter & Ref. & Pt. est. & interval 1.25 & interval 1.5\\
\hline 
$F^{2}_{dd}$ & 5.26 & 5.50 & {[}5.09, 6.02{]} & {[}4.88, 6.53{]}\\
$F^{2}_{dp}$ & 6.18 & 6.70 & {[}2.15, 10.93{]} & {[}n/a,13.17{]}\\
$F^{4}_{dd}$ & 3.29 & 4.50 & {[}3.66, 5.56{]} & {[}3.08, 6.49{]}\\
$G^{1}_{dp}$ & 2.89 & 2.74 & {[}-0.43, 4.56{]} & {[}-1.81, 5.86{]}\\
$G^{3}_{dp}$ & 1.65 & 2.38 & N/A & N/A\\
$10Dq$ & 1.07 & 1.061 & {[}1.02, 1.09{]} & {[}1.00, 1.18{]}\\
$\Delta\omega_\mathrm{in}$ & N/A & -1.02 & {[}-1.36, -0.65{]} & {[}-1.53, -0.42{]}\\
$\zeta_{i}$ & 0.083 & 0.123 & {[}0.101, 0.147{]}, & {[}0.089, 0.162{]}\\
$\zeta_{n}$ & 0.102 & 0.0 & N/A & N/A\\
$\zeta_{c}$ & 11.2 & 11.46 & {[}10.78, 12.21{]}, & {[}10.43, 12.66{]}\\
$\Gamma_{n}$ & 0.6 & 0.544 & {[}0.273, n/a{]} & {[}0.189, n/a{]}\\
$\gamma$ & 0.05 & 0.06 & (fixed) & (fixed)
\end{tabular}
\end{ruledtabular}
\par\end{centering}
\end{table}

\subsection{NiCl\texorpdfstring{$_{2}$}{2}}

\NiCl is a prototypical van der Waals antiferromagnet and a classic charge-transfer insulator. It adopts a layered rhombohedral structure (space group $R\bar{3}m$) comprised of edge-sharing $\mathrm{NiCl}_6$ octahedra forming a two-dimensional triangular lattice. Each $\mathrm{Ni}^{2+}$ ($3d^8$) center is coordinated by six $\mathrm{Cl}^-$ ligands in an approximately octahedral environment. 

\NiCl's low-lying excitations include extremely sharp spin-singlet $dd$ multiplets ($\:^1A_{1g}/\:^1E_{g}$) stabilized by intra-atomic Hund’s exchange, which are coupled to the ligand environment and exhibit ligand-tuned energies and lifetimes. These aspects make \NiCl a benchmark for studying Hund’s excitons, their interaction with lattice and magnetic degrees of freedom, and their propagation (dispersion) in \gls*{2D} correlated insulators \cite{ Occhialini2024}.

\begin{figure}[t]
\begin{centering}
\includegraphics[width=\columnwidth]{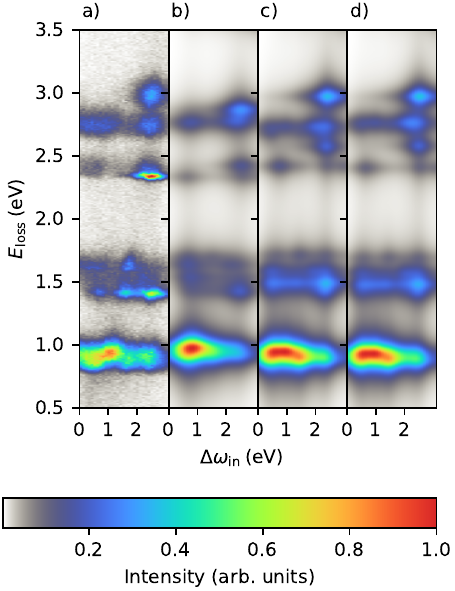}
\par\end{centering}
\caption{Experimental and simulated spectra for \NiCl. (a) experimental data from Ref.~\cite{Occhialini2024}, (b) hand-fitted (reference), (c) result of GPR, (d) result after greedy fine-tuning (Powell's method). The corresponding
L1 sum distances for panels (b)-(d) are $\chi^2_{L_1}=0.496$, $0.406$, and $0.373$, respectively.\label{NiClSpectra}}
\end{figure}

\subsubsection{Experimental setup}
We use the experimental \gls*{RIXS} spectrum from Ref.~\cite{Occhialini2024}. The spectrum was measured at the \gls*{SIX} 2-ID beamline of the \gls*{NSLS-II} with an energy resolution of 31~meV \gls*{FWHM}.
The experiment probed the $\mathrm{Ni}$ $L_3$ edge. The spectrum was recorded at temperature $T=40\:\text{K}$,
with grazing incidence, and the $a^*$ reciprocal-lattice direction lying in the scattering plane. Accordingly, we used scattering angles $\theta_\text{in}=10^{\circ}$, $2\Theta=150^{\circ}$ in the simulation.
 The incident beam was $\sigma$-polarized. Intensities were recorded in the incident energy range of width $3.11$~eV, including $32$ grid points. The examined energy loss range is $E_\text{loss}\in[0.5~\text{eV},3.5~\text{eV}]$, including $311$ grid points in total. 

\subsubsection{GPR results}
We show the experimental spectrum along with the reference, \gls*{GPR}, and greedy refined spectra in Fig.~\ref{NiClSpectra}. Parameter values of the fits are collected in Tab.~\ref{NiClTable}. The reference values in Tab.~\ref{NiClTable} are extracted from Ref.~\cite{Occhialini2024}. Ref. \cite{Occhialini2024} reported a clear double-peak structure at the $L_3$ edge in the incident energy direction, which is well reproduced by the fits. The ground state electronic configuration was characterized as $\:^3A_{2g}$ with $t^6_{2g}$ $e^2_g$ orbital occupation, in complete agreement with our numerics. Ref.~\cite{Occhialini2024} observed two sharp excitonic peaks, which they annotated as spin-singlet excitations with $\:^1A_{1g}$ / $\:^1E_g$ symmetry labels. We observe peaks at similar locations, although it is difficult to comment on their relative widths due to the limitations of the single ion model. The candidate peaks are less pronounced than the experimental spectrum. Our peak assignments made for our fitted spectra approximate, but do not completely overlap  the peak annotations in Ref.~\cite{Occhialini2024}. The peaks at the location of the $^1E_{g}$ exciton at 1.42 eV energy loss are annotated as $^3T_{1}$. In turn, the $^1E_{g}$ peak in the simulation mix with two other states with $^3T_{1}$ symmetry and are positioned upwards in energy by about $0.1$~eV and $0.3$~eV, respectively. The other sharp peak of label $^1A_{1g}$ at $2.43$ eV is present in the computed spectra as well, albeit shifted upwards by $0.05$~eV (\gls*{GPR}) and $0.15$~eV (greedy). The other peaks are annotated consistently with Ref.~\cite{Occhialini2024}, with some variance in the $e_g$ occupation numbers.

\begin{table}
\caption{NiCl$_{2}$ parameters (in units of eV). The ``interval 1.25'' and ``interval 1.5'' refer to the range of values for each parameter that keep the distance function within $1.25\times$ or $1.5\times$ its minimum value, respectively, with all other parameters held fixed.}
\begin{centering}
\begin{ruledtabular}
\begin{tabular}{ccccc}\label{NiClTable}
Parameter & Ref. & Pt. est. & interval 1.25 & interval 1.5\\
\hline 
$F^{2}_{dd}$ & 7.34 (7.8) & 7.91 & {[}7.30, 8.40{]} & {[}6.76, 8.91{]}\\
$F^{2}_{dp}$ & 4.63 & 3.65 & {[}n/a, 7.22{]} & N/A\\
$F^{4}_{dd}$ & 4.56 (4.85) & 5.21 & {[}3.82, 7.56{]} & {[}2.33, 8.90{]}\\
$G^{1}_{dp}$ & 3.93 & 4.16 & {[}3.04, 5.69{]} & {[}2.43, 9.78{]}\\ 
$G^{3}_{dp}$ & 2.24 & 2.41 & {[}n/a, 7.31{]} & N/A\\
$10Dq$ & 0.95 & 0.905 & {[}0.87, 0.94{]}, & {[}0.86, 0.96{]}\\
$\Delta\omega_\mathrm{in}$ & Pt. est. & -4.54 & {[}-4.88, -4.12{]}, & {[}-5.11,-3.77{]}\\
$\zeta_{i}$ & 0.083 & 0.0797 & {[}n/a, 0.132{]} & {[}n/a,0.16{]}\\
$\zeta_{n}$ & 0.102 & 0.125 & {[}n/a, 0.616{]} & N/A\\
$\zeta_{c}$ & 11.507 & 9.774 & {[}9.08, 10.61{]} & {[}8.64, 11.29{]},\\
$\Gamma_{n}$ & 0.5 & 0.4125 & fixed & fixed\\
\end{tabular}
\end{ruledtabular}
\par\end{centering}

\end{table}

\subsection{\texorpdfstring{Fe$_2$O$_3$}{Fe2O3} thin films}
We now consider \gls*{RIXS} measurements on $\alpha-\mathrm{Fe_2 O_3}$ (hematite) thin films~\cite{JieminLi2023Fe2O3}. This material is a well-known insulating antiferromagnet that hosts 
$\mathrm{Fe}^{3+}$ ($d=5$) ($S = 5/2$) ions and supports long‐distance magnon transport without charge flow -- an attractive platform for low‐power spintronic devices. In the corundum structure each $\mathrm{Fe}^{3+}$ is coordinated by six $\mathrm{O}^{2-}$ ligands, forming nearly ideal $\mathrm{FeO}_6$ octahedra \cite{Caliebe19981s2p}. This test case is of particular interest because atomic model fits to Fe $L$-data for this material currently do not exist in the literature. 

\subsubsection{Experimental setup}
We use the experimental $\mathrm{Fe}$ $L_3$ edge spectrum reported in Ref.~\cite{JieminLi2023Fe2O3} for this test case. The measurement took place at the \gls*{SIX} 2-ID beamline of the \gls*{NSLS-II} with an energy resolution of 23~meV \gls*{FWHM}. The experimental spectrum was recorded at $T=100\:\text{K}$ using linearly ($\pi$) polarized light and scattering angles $\theta_\text{in}=20^{\circ}$, $2\Theta=150^{\circ}$, and an azimuth angle $\varphi=0^{\circ}$. Intensities were recorded in the incident energy range of width $6.394$~eV about the Fe $L_3$ resonance, including $30$ grid points. The energy loss range is $E_\text{loss}\in[0.5~\text{eV},2.2~\text{eV}]$ measured on $151$ grid points in total. 

\begin{figure}
\begin{centering}
\includegraphics[width=\columnwidth]{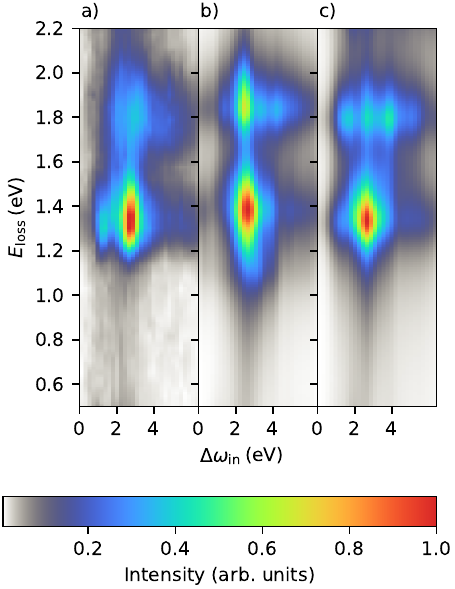}
\par\end{centering}
\caption{Experimental and simulated spectra for \Hematite. (a) experimental data from Ref.~\cite{JieminLi2023Fe2O3}, (b) result of GPR, (c) result after greedy fine-tuning (Powell's method). $L_1$ max distances for panels (b)-(c) are $\chi^2_{L^\prime_1}=158.8$ and $116.8$, respectively.}
\label{Fe2O3_plot}
\end{figure}

\begin{table}
\caption{Fe$_{2}$O$_{3}$ parameters (in units of eV). The ``interval 1.25'' and ``interval 1.5'' refer to the range of values for each parameter that keep the distance function within $1.25\times$ or $1.5\times$ its minimum value, respectively, with all other parameters held fixed.}
\begin{centering}
\begin{ruledtabular}
\begin{tabular}{cccc}\label{Fe2O3_table}
Parameter & Pt. est. & interval 1.25 & interval 1.5\\
\hline 
$F^{2}_{dd}$ & 5.78 & {[}5.60, 5.99{]} & {[}5.50, 6.10{]}\\
$F^{2}_{dp}$ & 0.6 & {[}0, 1.67{]} & {[}0, 2.11{]}\\
$F^{4}_{dd}$ & 3.80 & {[}3.78, 4.03{]} & {[}3.71, 4.10{]}\\
$G^{1}_{dp}$ & 4.65 & {[}3.97, 6.07{]} & {[}3.75, 6.55{]}\\
$G^{3}_{dp}$ & 4.20  & {[}2.85, 5.42{]} & {[}2.24, n/a{]}\\
$10Dq$ &  1.26 & {[}1.21, 1.31{]} & {[}1.18, 1.33{]}\\ 
$\Delta\omega_\mathrm{in}$ & -4.63 & {[}-4.83, -4.29{]} & {[}-4.97, -4.11{]}\\
$\zeta_{i}$ & 0 & {[}0,0.0533{]} & {[}0,0.069{]}\\
$\zeta_{n}$ & 0 &  N/A & N/A\\
$\zeta_{c}$ & 8.44  & {[}7.99, 9.07{]} & {[}7.79, 9.42{]}\\
$\Gamma_{n}$ & 0.34 & {[}0.21, 0.47{]} & {[}0.16, 0.54{]}\\
\end{tabular}
\end{ruledtabular}
\par\end{centering}
\end{table}
\subsubsection{GPR results}
Figure~\ref{Fe2O3_plot} shows the experimental spectrum along with its simulated counterparts. Both the \gls*{GPR} and greedy fine-tuned fits reproduce many aspects of the experimental spectra. Parameter values corresponding to the greedy fit are reported in Tab.~\ref{Fe2O3_table}. 
Ref.~\cite{JieminLi2023Fe2O3} attributed the two main branches of excitations in the $E_\mathrm{loss}\sim1-2$ eV range to $dd$ excitations. They correspond to $\:^6A_{1g}\rightarrow \:^4T_{1g}$ (at 1.4~eV) and $\:^6A_{1g} \rightarrow \:^4T_{2g}$ (at 1.9~eV) transitions, flipping both spin ($S = 5/2 \rightarrow 3/2$) and orbital ($e_g \rightarrow t_{2g}$ ), respectively. This conclusion is largely consistent with the fit results, with the important remark that the fitted spectrum consists of multiple overlapping peaks, some of which feature $^2 T_2 (t_2)^5$ content. Overall, our method successfully captures the major features of the spectrum. In the longer term, incorporating further features into the model, such as explicit treatment of ligand orbitals, cubic symmetry breaking, and polaron physics, could represent possible ways to improve the theory-experiment agreement further. 

\subsection{\texorpdfstring{Ca$_3$LiOsO$_6$}{Ca3LiOsO6}}
As our final test case, we consider 
Os $L_3$-edge \gls*{RIXS} measurements on \Osmate. The osmate material hosts $\mathrm{Os}^{5+}$ ($d^3$) ions that are surrounded by six $O^{2-}$ ligands in an approximately octahedral $\mathrm{OsO}_6$ environment. The octahedra themselves are arranged in a slightly distorted hexagonal $R\bar{3}c$ lattice but remaining very close to ideal Oh symmetry.
\Osmate serves as a model $5d^3$ system in which the local $\mathrm{OsO}_6$ octahedra are relatively isolated, allowing direct access to single ion physics. Previous \gls*{RIXS} measurements on this material have revealed a dramatic spin-orbit–induced splitting of the $t_{2g}$ manifold~\cite{Taylor2016SpinOrbitCC}. As with the hematite case, no full $d$-shell atomic models have been reported for this compound.

\subsubsection{Experimental setup}
We use the experimental spectrum reported in Ref.~\cite{Taylor2016SpinOrbitCC} for this test case. The \gls*{RIXS} measurements were performed at the Advanced Photon Source (APS) at Sector 27 using the MERIX instrumentation. The experiment was performed at the $\mathrm{Os}$ $L_3$ edge using a powdered sample. The spectrum was recorded using a temperature of $T=300\:\mathrm{K}$, scattering angles $\theta_\text{in}=45$, $2\Theta=90^{\circ}$, and an azimuth angle of $\varphi=0^{\circ}$. Intensities were recorded in the incident energy range of width $8$ eV about the Os $L_3$ resonance, including $5$ grid points. The energy loss range is $E_\text{loss}\in[0.5~\text{eV},5.0~\text{eV}]$, including $151$ grid points in total. 

\subsubsection{GPR results}
\begin{figure}[t]
\begin{centering}
\includegraphics[width=\columnwidth]{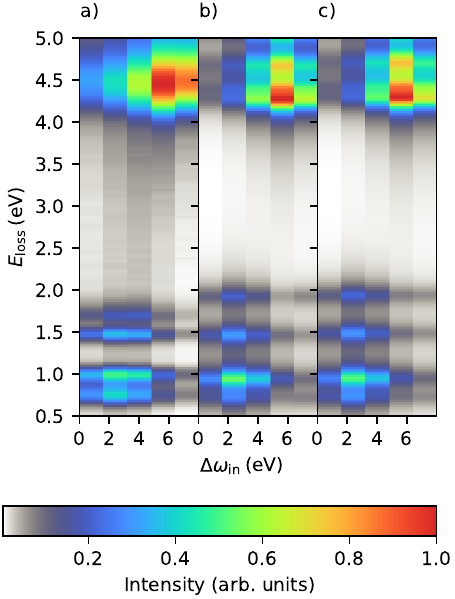}
\par\end{centering}
\caption{Experimental and simulated spectra for \Osmate. (a) experimental data from Ref.~\cite{Taylor2016SpinOrbitCC}, (b) hand-fitted (reference), (c) result of GPR,
result after greedy fine-tuning (Powell's method). $L_1$ sum distances for panels (b)-(c) are $\chi^2_{L_1}=0.313$ and $0.302$, respectively.}
\label{Ca3LiOsO6_plot}
\end{figure}

Figure~\ref{Ca3LiOsO6_plot} shows the experimental \gls*{RIXS} spectra as well as simulated spectra. The overall structure of the data are well reproduced by the fitted models. The corresponding greedy-fitted parameters, along with their confidence intervals are included in Tab.~\ref{Ca3LiOsO6_table}. 
Ref.~\cite{Taylor2016SpinOrbitCC} reported four peaks present at $E_\mathrm{loss} < 2$~eV that resonate at $\omega_\mathrm{in} = 10.874$ keV, and a feature at $E_\mathrm{loss} \approx 4.5$ eV that resonates at $\omega_\mathrm{in} = 10.878$~keV. This behavior indicates that the features below $2$ eV are intra-$t_{2g}$ excitations, whereas the higher energy feature is a transition between the $t_{2g}$ and $e_g$ states. 

Ref.~\cite{Taylor2016SpinOrbitCC} fixes the value of the crystal field to $10Dq = 4.5$~eV; our fitted $10Dq = 4.1$~eV is somewhat lower. 
We assign the dominant ground state character to an $^4A_{2g}$ state, in agreement with 
Ref.~\cite{Taylor2016SpinOrbitCC}, although we find a more significant contribution of $^2T_2$ states mixed in Ref.~\cite{Taylor2016SpinOrbitCC} refers to the first two peaks as hybridized $^2E_g$ and $2T_{1g}$ states. 
The numerics (including simulations we conducted using the parameters in. Ref.~\cite{Taylor2016SpinOrbitCC}), however, predict a more diverse hybridization of $^2E_g$ and $^2T_{1g}$, $^4A_{2g}$ and $^2T_2$. The second peak is a mixture of $^2T_1$ and $^2E$, in agreement with the prior work~\cite{Taylor2016SpinOrbitCC}. The third peak is consistently evaluated to be an almost pure $^2T_2$ state. The fourth peak is predominantly $^2T_2$, with some $^2T_1$ and $^2E$ mixed in. This assignment is also shared between Ref.~\cite{Taylor2016SpinOrbitCC} and our fits. Other heavy-element-based oxide materials have shown that improved model solutions can be obtained from co-fitting $L_3$ and $L_2$ spectra \cite{Frontini2025resonant}. Such an approach would be an interesting extension to the current analysis of \Osmate.

\begin{table}[t]
\caption{Ca$_{3}$LiOsO$_{6}$ parameters (in units of eV). The ``interval 1.25'' and ``interval 1.5'' refer to the range of values for each parameter that keep the distance function within $1.25\times$ or $1.5\times$ its minimum value, respectively, with all other parameters held fixed.}
\begin{centering}
\begin{ruledtabular}
\begin{tabular}{cccc}\label{Ca3LiOsO6_table}
Parameter & Pt. est. & interval 1.25 & interval 1.5\\
\hline 
$F^{2}_{dd}$ & 2.78 & {[}2.17, 3.37{]} & {[}1.53, 3.70{]}\\
$F^{2}_{dp}$ & 0.083 & {[}0,11.61{]} & {[}0,19.17{]}\\
$F^{4}_{dd}$ & 1.37 & {[}0.64, 2.25{]} & {[}0.01,2.62{]}\\
$G^{1}_{dp}$ & 0.84 & {[}n/a,1.92{]} & N/A\\
$G^{3}_{dp}$ & 0.24 & N/A & N/A\\
$10Dq$  & 4.15 & {[}4.02, 4.29{]} & {[}3.91, 4.35{]}\\
$\Delta\omega_{in}$ & -3.34 & {[}-4.41, -2.00{]} & {[}-4.98, -1.38{]}\\
$\zeta_{i}$ &  0.474 & {[}0.393, 0.542{]} & {[}0.313, 0.581{]}\\
$\zeta_{n}$ &  0.551 & {[}0.206, 0.862{]} & N/A\\
$\zeta_{c}$ &  N/A & N/A & N/A\\
$\Gamma_{n}$ & 3.0 (fixed) &  & \\
\end{tabular}
\end{ruledtabular}
\par\end{centering}
\end{table}

\section{Discussion and conclusions} \label{ConclusionSection}
Our results demonstrate that it is possible to automate model Hamiltonian extraction from information-dense \gls*{RIXS} spectra of real materials, dramatically reducing the amount of effort and expertise required compared to traditional hand-tuned fitting.

A common challenge in model Hamiltonian extraction is the non-uniqueness of the inverse problem. Our work shows that multiple distinct parameter configurations can yield visually similar \gls*{RIXS} spectra. This problem is part of why hand-fitting \gls*{RIXS} spectra is challenging since it is difficult for humans to anticipate the number of different local minima of the distance function. Our automated analysis system enables a more comprehensive identification of these local minima. We further outline an automated means to use symmetry labels to more definitively identify distinct solutions.

This research points towards several promising directions for future work. One important issue concerns the choice of the distance function used to quantify similarity between the experimental and simulated spectra. Experimental data are often noisy, and simulated spectra can be affected by systematic model errors, making this comparison nontrivial. In this study, we employed pixel-wise metrics as a straightforward and transparent way to assess visual similarity. While effective in many cases, these metrics are sensitive to local fluctuations and may under represent high-level spectral features. A potentially promising extension of our work would be to incorporate learned perceptual similarity metrics, such as those based on deep neural networks, which have shown remarkable effectiveness in natural image comparison tasks~\cite{zhang2018perceptual}. Whether this approach would improve the quality of the workflow sufficient to justify the additional computational expense is an open question.

An obvious avenue to improve the method is to apply it to models of increasing sophistication such as Anderson impurity models or extended clusters. Such generalizations can provide a better approximation to the experimental spectrum over a wider range of energy loss at the cost of a larger number of floating parameters and a more expensive distance function. Further improvements to the active learning method (such as using deep Gaussian process models or refined distance functions) as well as numerical optimizations of the \gls*{RIXS} simulation together could represent suitable approaches to overcoming these difficulties. Simultaneously incorporating additional information to better constrain the problem is likely to be valuable. This could include first-principles calculations of tight-binding parameters and effective interactions.

Our approach has a close connection to \gls*{SBI} \cite{Diggle1984, Kennedy2001,Cranmer2020}. In \gls*{SBI} one determines a posterior distribution of model parameters, given a set of prior beliefs and a forward simulator mapping parameters to (synthetic) data. Various methods have been proposed to approximate the likelihood from the simulation, including approximate Bayesian computation \cite{Tavare1997, Beaumont2002, Sunnaker2013}, synthetic likelihood \cite{Wood2010}, \Gls*{BOLFI} \cite{Gutmann2016}, and more recent approaches like neural density estimation \cite{Papamakarios2019}.
Our approach is close to \gls*{BOLFI}  in the sense that a distance function between experiment and simulation is defined, which we model with a Gaussian process.  Along the lines of \gls*{BOLFI}, one could define an approximate likelihood function based on the uncertainty of the Gaussian process. The resulting approximate likelihood can be used to sample a posterior distribution of parameters. This approach would be an alternative way of discussing the confidence regions of parameters and formally would provide uncertainty estimates for the parameters. However, the ``probability distributions'' obtained this way would be somewhat artificial, as the major cause of the difference between simulated and experimental spectra is due to systematic omissions in the model (as opposed to e.g., statistical noise), and thus both the Gaussian process model for the uncertainty of the distance and the defined likelihood are somewhat ad-hoc.

Another natural extension is to embed our inference loop into Bayesian optimal experimental design \cite{Lindley1956,Chaloner1995,Huan2013}, where the algorithm selects the next measurement (most directly, the incident energy 
$\omega_{\text{in}}$) to maximally reduce uncertainty in the Hamiltonian parameters. In \gls*{XAS}, Zhang \textit{et al}.~\cite{zhang2023autonomous} implemented an adversarial Bayesian optimization scheme that couples a Hamiltonian-fitting step with a sampling Bayesian optimization step to adaptively choose photon-energy points, achieving accurate models with tens of samples; analogous ideas could be ported to \gls*{RIXS} with suitable low-latency surrogates~\cite{chitturi2023capturing} and streaming objectives. 

In summary, this work establishes that a Bayesian optimization framework can automatically infer multi-orbital Hamiltonian parameters directly from information-dense RIXS spectra with an accuracy that rivals expert hand-fitting. Applied to four representative quantum materials—NiPS$_3$, NiCl$_2$, Fe$_2$O$_3$, and Ca$_3$LiOsO$_6$—the method not only reproduces published parameter sets but also supplies the first quantitative atomic-model parameters for Fe$_2$O$_3$ and Ca$_3$LiOsO$_6$. By rigorously mapping uncertainties, the approach offers a transparent route to evaluating model uniqueness and reliability. 
These results pave the way for extending inverse-problem automation to more elaborate cluster or impurity models and, ultimately, to high-throughput exploration of novel quantum materials with minimal human intervention.

Zenodo repository: \url{https://doi.org/10.5281/zenodo.17353405}

\begin{acknowledgments}
This work was supported by the  U.S.~Department of Energy, Office of Science, Office of Basic Energy Sciences, under Award Number DE-SC0022311. We acknowledge Yilin Wang, Gilberto Fabbris, Gabi Kotliar, Max Rakitin, and Tom Hopkins for their contributions to EDRIXS. We thank Valentina Bisogni, Jiemin Li, and Johnny Pelliciari for sharing data on \Hematite{} \cite{JieminLi2023Fe2O3}, Wei He for sharing data on \NiPS{} \cite{He2024MagneticallyPH}, Connor Occhialini and Riccardo Comin for sharing data on \NiCl{} \cite{Occhialini2024}, and Andy Christianson, and Stuart Calder for sharing data on \Osmate{} \cite{Taylor2016SpinOrbitCC}. MKL gratefully acknowledges Niraj Aryal and Robert Konik for useful discussions.
\end{acknowledgments}

\appendix

\section{Detailed fitting information} \label{FitDetails}
\subsection{\texorpdfstring{NiPS$_{3}$}{NiPS3}}

\begin{figure}[b]  
\begin{centering}
\includegraphics[width=8cm]{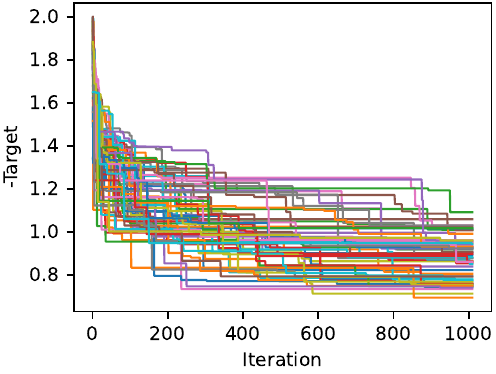}
\par\end{centering}
\caption{\NiPS: Decrease of the sum normalized distance function for $60$ runs of
$1000$ iterations \label{FigNiPSDecrease}}
\end{figure}

\subsubsection{Fitting procedure}
We used the sum-normalized L1 distance $\chi^2_{L1}$ for this analysis. 
Figure~\ref{FigNiPSDecrease} shows the decrease in the best probed distance function as a function of the \gls*{GPR} iterations for $N_\text{run}=60$ independent runs. The variance of the best estimate after a fixed number of iterations is significant, motivating the use of multiple runs.  

\begin{figure}
\begin{centering}
\includegraphics[width=\columnwidth]{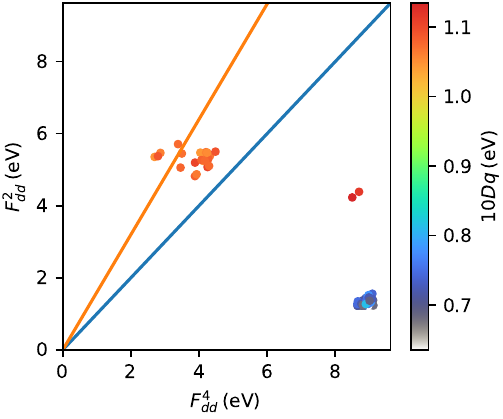}
\includegraphics[width=\columnwidth]{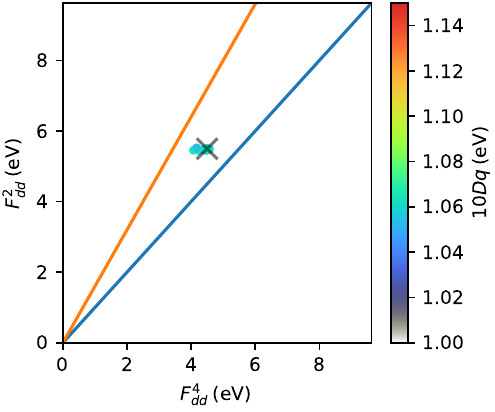}
\par\end{centering}
\caption{Top: distribution of $18$ \gls*{GPR} evaluations with $d\protect\leq1.15\chi^2_{L_1,\text{min},\text{GPR}}$ for \NiPS.
Bottom: results of subsequent greedy optimization starting from the $23$ best \gls*{GPR} points. The chosen point is denoted by a gray $\times$.  On both plots, the lines $F_{dd}^4=F_{dd}^2$ and $F_{dd}^4=0.625\:F_{dd}^2$ are drawn in blue and orange, respectively. The blue line is used to restrict candidate solutions, while the orange is shown as a guideline only. \label{FigNiPSBestPts}}
\end{figure}

The best $73$ points (with distances smaller than $1.15\:\chi^2_{L_1,\text{min},\text{GPR}}$) are shown in the top panel of Fig.~\ref{FigNiPSBestPts}, as a function of $F^2_{dd}$, $F^4_{dd}$ and $10Dq$. These points form three well-separated  clusters, two of which are discarded due to unphysical $F^4_{dd}/F^2_{dd}$ ratios. After throwing out the unphysical point, we are left with $N_\text{greedy}=23$ points, which are further refined with the greedy approach.
The bottom panel of Fig.~\ref{FigNiPSBestPts} shows the result of the greedy optimization starting from the best points of the \gls*{GPR} fit. All points collapse to the same valley of the distance function.

\begin{figure*}
\begin{centering}
\includegraphics[width=\textwidth]{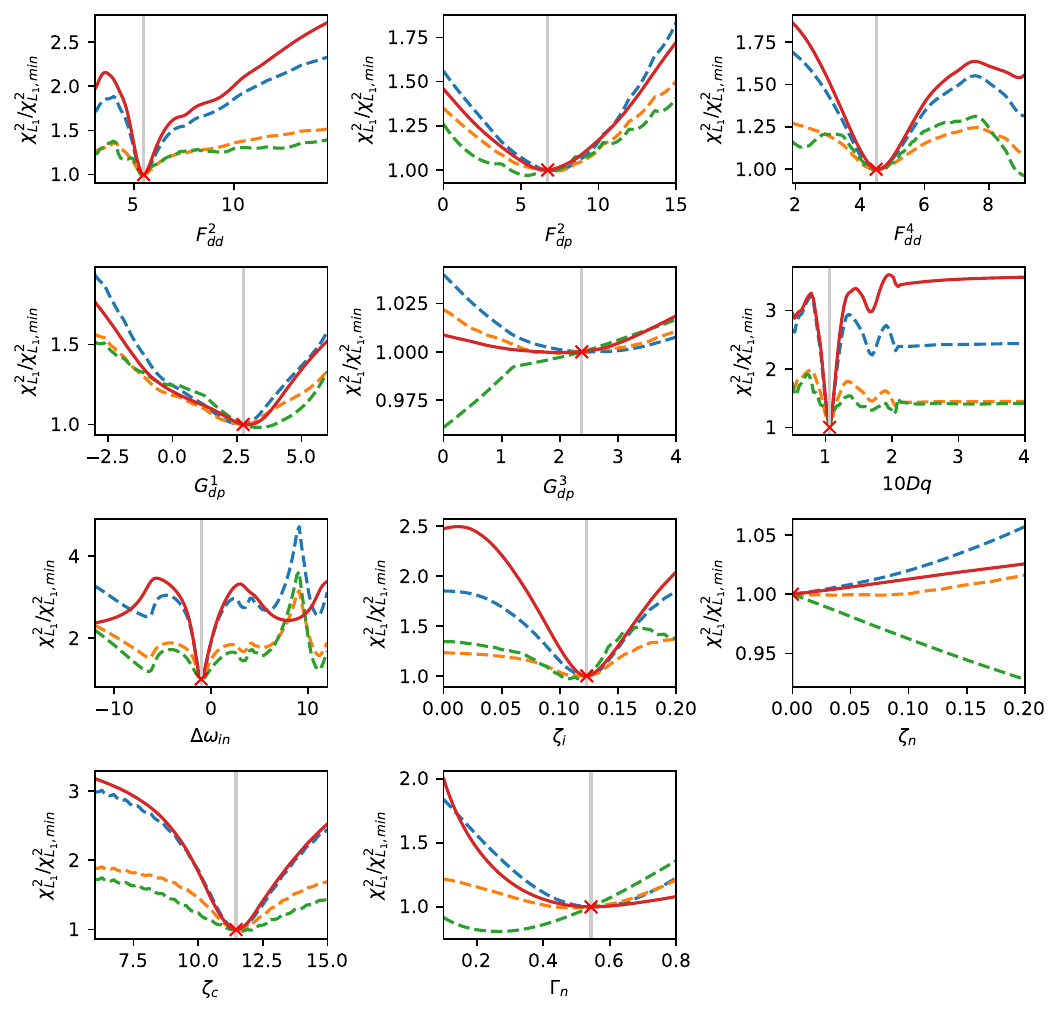}
\par\end{centering}
\caption{\NiPS: The behavior of various distance measures around the fine-tuned minimum.
Solid red: L1 sum normalized, dashed blue: L1 maximum normalized,
dashed orange: L2 sum normalized, dashed green: magnitude of gradient,
maximum normalized. The distance functions are generally sensitive to the initial Slater parameters as well as $10Dq$ and the energy offset $x_\text{offset}$, while they are less sensitive to intermediate state parameters, especially $\zeta_n$. \label{NiPS3Distances}}
\end{figure*}

Figure~\ref{NiPS3Distances} depicts the dependency of the distance functions on each floated parameter, keeping all the rest at their minimum values.

\subsection{\texorpdfstring{NiCl$_{2}$}{NiCl2}}
\subsubsection{Fitting procedure}
We used the sum  normalized L1 distance $\chi^2_{L1}$ for this analysis. 
Figure~S1 of the \gls*{SM}~\cite{supplement} shows the decrease of the best probed distance function as  function of the \gls*{GPR} iterations for $N_\text{run}=60$ independent runs. 
The best $11$ points  (with distances smaller than $1.2\:\chi_{\text{min},\text{GPR}}$) are shown on the top panel of Fig.~S2 (\gls*{SM}~\cite{supplement}), shown as function of $F^2_{dd}$, $F^4_{dd}$ and $10Dq$. 
The bottom panel of Fig.~S2 (\gls*{SM}~\cite{supplement}) exhibits the result of the greedy optimization starting from the outcomes of the \gls*{GPR} fit. Upon the greedy refinement, these points arrange into a single cluster.

Figure~S3 of the \gls*{SM}~\cite{supplement} depicts the dependency of the distance functions on each one floated parameter, keeping all the rest at their minimum values.

\subsection{\texorpdfstring{Fe$_{2}$O$_{3}$}{Fe2O3}}

\begin{figure}
\begin{centering}
\includegraphics[width=\columnwidth]{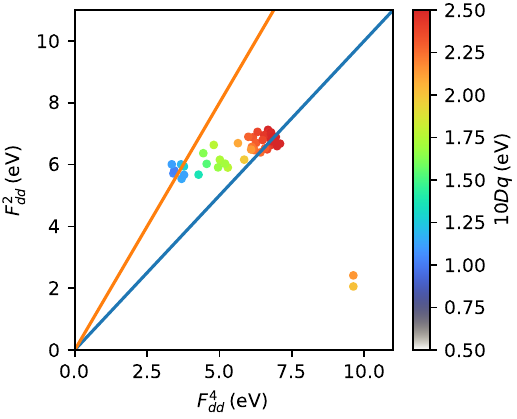}
\includegraphics[width=\columnwidth]{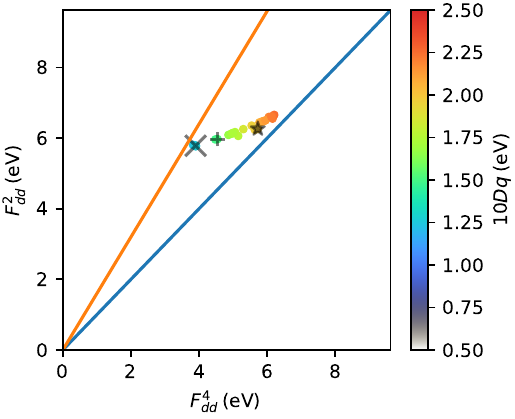}
\par\end{centering}
\caption{Top: distribution of $50$ \gls*{GPR} evaluations with $d\protect\leq1.2\chi^2_{L_1,\text{min},{GPR}}$ and $10Dq < 2.5$ for \Hematite.
Bottom: results of subsequent greedy optimization starting from the
$28$ best \gls*{GPR} points. The overall best point is denoted by a gray X, while alternative solutions are marked with a plus and an asterisk. On both plots, the lines $F_{dd}^4=F_{dd}^2$ and $F_{dd}^4=0.625\:F_{dd}^2$ are drawn in blue and orange, respectively. The blue line is used to restrict candidate solutions, while the orange is shown as a guideline only.\label{FigFe2O3BestPts}}
\end{figure}

\begin{figure}
\begin{centering}
\includegraphics[width=\columnwidth]{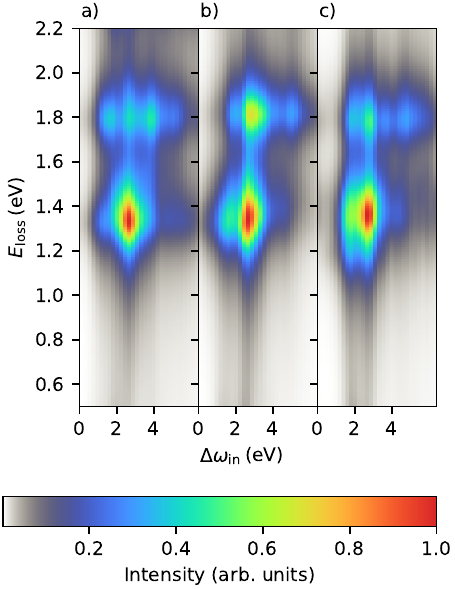}
\par\end{centering}
\caption{\Hematite: Alternative results for the greedy fit.  L1 max distances for panels (a)-(c) are $\chi^2_{L^\prime_1}=116.8$, $129.6$ and $148.7$, respectively. The fits shown in panels a-c correspond to the parameter values indicated by the cross ($\times$), plus ($+$), and star ($\star$) in Fig. \ref{FigFe2O3BestPts}.\label{FigFe2O3Alternatives}}
\end{figure}
\subsubsection{Fitting procedure}
We used the maximum  normalized L1 distance $\chi^2_{L^\prime_1}$ for this analysis. The top panel of Fig.~S4 (see \gls*{SM}~\cite{supplement}) shows the decrease of the best probed distance function as function of the \gls*{GPR} iterations for $N_\text{run}=60$ independent runs.

The best $50$ points (with distances smaller than $1.2\:\chi^2_{L_1^\prime,\text{min},\text{GPR}}$ and $10Dq<2.5$)  are shown on the top panel of Fig.~\ref{FigFe2O3BestPts}.
A subset of the candidate points have an unphysical $F^2_{dd}/F^4_{dd}$ ratio. We thus restrict our analysis to $28$ points with $F^2_{dd}\leq F^4_{dd}$ in the following.
The bottom panel of Fig.~S4 (\gls*{SM}~\cite{supplement}) depicts these remaining candidate points after the restriction.

The bottom panel of Fig.~\ref{FigFe2O3BestPts} exhibits the result of the greedy optimization starting from the outcomes of the \gls*{GPR} fit. 
Three separate but closely packed clusters can be distinguished on this plot. Figure~\ref{FigFe2O3Alternatives} shows alternative fits corresponding to different minima of the greedy optimization.

Figure~S5 in the \gls*{SM}~\cite{supplement} plots the dependency of the distance functions on each one floated parameter, keeping all the rest at their minimum values.

\subsection{\texorpdfstring{Ca$_{3}$LiOsO$_{6}$}{Ca3LiOSO6}}

\subsubsection{Fitting procedure}
We used the sum  normalized L1 distance $\chi^2_{L1}$ for this analysis. 
Figure~S6 in the \gls*{SM}~\cite{supplement}shows the decrease of the best probed distance function as function of the \gls*{GPR} iterations for $N_\text{run}=60$ independent runs. 
The best $24$ points (with distances smaller than $1.06\:\chi^2_{L_1,\text{min},\text{GPR}}$ are shown on the left panel of Fig.~S7 (see \gls*{SM}~\cite{supplement}), shown as function of $F^2_{dd}$, $F^4_{dd}$ and $10Dq$.

The right panel of Fig.~S7 (\gls*{SM}~\cite{supplement}) exhibits the result of the greedy optimization starting from the outcomes of the \gls*{GPR} fit. The refined points organize into a single cluster. 

Figure~S8 (\gls*{SM}~\cite{supplement}) depicts the dependency of the distance functions on each one floated parameter, keeping all the rest at their minimum values.

\textit{Data availability} --- The supporting code for this article is openly available from Github and the Zenodo database \cite{repo}.


\bibliography{refs}

\begin{thebibliography}{1}%
\makeatletter
\providecommand \@ifxundefined [1]{%
 \@ifx{#1\undefined}
}%
\providecommand \@ifnum [1]{%
 \ifnum #1\expandafter \@firstoftwo
 \else \expandafter \@secondoftwo
 \fi
}%
\providecommand \@ifx [1]{%
 \ifx #1\expandafter \@firstoftwo
 \else \expandafter \@secondoftwo
 \fi
}%
\providecommand \natexlab [1]{#1}%
\providecommand \enquote  [1]{``#1''}%
\providecommand \bibnamefont  [1]{#1}%
\providecommand \bibfnamefont [1]{#1}%
\providecommand \citenamefont [1]{#1}%
\providecommand \href@noop [0]{\@secondoftwo}%
\providecommand \href [0]{\begingroup \@sanitize@url \@href}%
\providecommand \@href[1]{\@@startlink{#1}\@@href}%
\providecommand \@@href[1]{\endgroup#1\@@endlink}%
\providecommand \@sanitize@url [0]{\catcode `\\12\catcode `\$12\catcode
  `\&12\catcode `\#12\catcode `\^12\catcode `\_12\catcode `\%12\relax}%
\providecommand \@@startlink[1]{}%
\providecommand \@@endlink[0]{}%
\providecommand \url  [0]{\begingroup\@sanitize@url \@url }%
\providecommand \@url [1]{\endgroup\@href {#1}{\urlprefix }}%
\providecommand \urlprefix  [0]{URL }%
\providecommand \Eprint [0]{\href }%
\providecommand \doibase [0]{https://doi.org/}%
\providecommand \selectlanguage [0]{\@gobble}%
\providecommand \bibinfo  [0]{\@secondoftwo}%
\providecommand \bibfield  [0]{\@secondoftwo}%
\providecommand \translation [1]{[#1]}%
\providecommand \BibitemOpen [0]{}%
\providecommand \bibitemStop [0]{}%
\providecommand \bibitemNoStop [0]{.\EOS\space}%
\providecommand \EOS [0]{\spacefactor3000\relax}%
\providecommand \BibitemShut  [1]{\csname bibitem#1\endcsname}%
\let\auto@bib@innerbib\@empty
\bibitem [{\citenamefont {Dresselhaus}\ \emph {et~al.}(2008)\citenamefont
  {Dresselhaus}, \citenamefont {Dresselhaus},\ and\ \citenamefont
  {Jorio}}]{Dresselhaus2008GroupTheory}%
  \BibitemOpen
  \bibfield  {author} {\bibinfo {author} {\bibfnamefont {M.}~\bibnamefont
  {Dresselhaus}}, \bibinfo {author} {\bibfnamefont {G.}~\bibnamefont
  {Dresselhaus}},\ and\ \bibinfo {author} {\bibfnamefont {A.}~\bibnamefont
  {Jorio}},\ }\href {https://doi.org/10.1007/978-3-540-32899-5} {\emph
  {\bibinfo {title} {Group Theory. Application to the Physics of Condensed
  Matter}}}\ (\bibinfo  {publisher} {Springer},\ \bibinfo {year}
  {2008})\BibitemShut {NoStop}%
\end{thebibliography}%


\begin{thebibliography}{87}%
\makeatletter
\providecommand \@ifxundefined [1]{%
 \@ifx{#1\undefined}
}%
\providecommand \@ifnum [1]{%
 \ifnum #1\expandafter \@firstoftwo
 \else \expandafter \@secondoftwo
 \fi
}%
\providecommand \@ifx [1]{%
 \ifx #1\expandafter \@firstoftwo
 \else \expandafter \@secondoftwo
 \fi
}%
\providecommand \natexlab [1]{#1}%
\providecommand \enquote  [1]{``#1''}%
\providecommand \bibnamefont  [1]{#1}%
\providecommand \bibfnamefont [1]{#1}%
\providecommand \citenamefont [1]{#1}%
\providecommand \href@noop [0]{\@secondoftwo}%
\providecommand \href [0]{\begingroup \@sanitize@url \@href}%
\providecommand \@href[1]{\@@startlink{#1}\@@href}%
\providecommand \@@href[1]{\endgroup#1\@@endlink}%
\providecommand \@sanitize@url [0]{\catcode `\\12\catcode `\$12\catcode
  `\&12\catcode `\#12\catcode `\^12\catcode `\_12\catcode `\%12\relax}%
\providecommand \@@startlink[1]{}%
\providecommand \@@endlink[0]{}%
\providecommand \url  [0]{\begingroup\@sanitize@url \@url }%
\providecommand \@url [1]{\endgroup\@href {#1}{\urlprefix }}%
\providecommand \urlprefix  [0]{URL }%
\providecommand \Eprint [0]{\href }%
\providecommand \doibase [0]{https://doi.org/}%
\providecommand \selectlanguage [0]{\@gobble}%
\providecommand \bibinfo  [0]{\@secondoftwo}%
\providecommand \bibfield  [0]{\@secondoftwo}%
\providecommand \translation [1]{[#1]}%
\providecommand \BibitemOpen [0]{}%
\providecommand \bibitemStop [0]{}%
\providecommand \bibitemNoStop [0]{.\EOS\space}%
\providecommand \EOS [0]{\spacefactor3000\relax}%
\providecommand \BibitemShut  [1]{\csname bibitem#1\endcsname}%
\let\auto@bib@innerbib\@empty
\bibitem [{\citenamefont {Basov}\ \emph {et~al.}(2017)\citenamefont {Basov},
  \citenamefont {Averitt},\ and\ \citenamefont {Hsieh}}]{Basov2017towards}%
  \BibitemOpen
  \bibfield  {author} {\bibinfo {author} {\bibfnamefont {D.}~\bibnamefont
  {Basov}}, \bibinfo {author} {\bibfnamefont {R.}~\bibnamefont {Averitt}},\
  and\ \bibinfo {author} {\bibfnamefont {D.}~\bibnamefont {Hsieh}},\ }\bibfield
   {title} {\bibinfo {title} {Towards properties on demand in quantum
  materials},\ }\href {https://doi.org/10.1038/nmat5017} {\bibfield  {journal}
  {\bibinfo  {journal} {Nat. Mater.}\ }\textbf {\bibinfo {volume} {16}},\
  \bibinfo {pages} {1077} (\bibinfo {year} {2017})}\BibitemShut {NoStop}%
\bibitem [{\citenamefont {Tokura}\ \emph {et~al.}(2017)\citenamefont {Tokura},
  \citenamefont {Kawasaki},\ and\ \citenamefont
  {Nagaosa}}]{Tokura2017emergent}%
  \BibitemOpen
  \bibfield  {author} {\bibinfo {author} {\bibfnamefont {Y.}~\bibnamefont
  {Tokura}}, \bibinfo {author} {\bibfnamefont {M.}~\bibnamefont {Kawasaki}},\
  and\ \bibinfo {author} {\bibfnamefont {N.}~\bibnamefont {Nagaosa}},\
  }\bibfield  {title} {\bibinfo {title} {Emergent functions of quantum
  materials},\ }\href {https://doi.org/10.1038/nphys4274} {\bibfield  {journal}
  {\bibinfo  {journal} {Nat. Phys.}\ }\textbf {\bibinfo {volume} {13}},\
  \bibinfo {pages} {1056} (\bibinfo {year} {2017})}\BibitemShut {NoStop}%
\bibitem [{\citenamefont {Giustino}\ \emph {et~al.}(2021)\citenamefont
  {Giustino}, \citenamefont {Lee}, \citenamefont {Trier}, \citenamefont
  {Bibes}, \citenamefont {Winter}, \citenamefont {Valent{\'\i}}, \citenamefont
  {Son}, \citenamefont {Taillefer}, \citenamefont {Heil}, \citenamefont
  {Figueroa}, \citenamefont {Pla{\c c}ais}, \citenamefont {Wu}, \citenamefont
  {Yazyev}, \citenamefont {Bakkers}, \citenamefont {Nyg{\aa}rd}, \citenamefont
  {Forn-D{\'\i}az}, \citenamefont {Franceschi}, \citenamefont {McIver},
  \citenamefont {Torres}, \citenamefont {Low}, \citenamefont {Kumar},
  \citenamefont {Galceran}, \citenamefont {Valenzuela}, \citenamefont
  {Costache}, \citenamefont {Manchon}, \citenamefont {Kim}, \citenamefont
  {Schleder}, \citenamefont {Fazzio},\ and\ \citenamefont
  {Roche}}]{Giustino2020quantum}%
  \BibitemOpen
  \bibfield  {author} {\bibinfo {author} {\bibfnamefont {F.}~\bibnamefont
  {Giustino}}, \bibinfo {author} {\bibfnamefont {J.~H.}\ \bibnamefont {Lee}},
  \bibinfo {author} {\bibfnamefont {F.}~\bibnamefont {Trier}}, \bibinfo
  {author} {\bibfnamefont {M.}~\bibnamefont {Bibes}}, \bibinfo {author}
  {\bibfnamefont {S.~M.}\ \bibnamefont {Winter}}, \bibinfo {author}
  {\bibfnamefont {R.}~\bibnamefont {Valent{\'\i}}}, \bibinfo {author}
  {\bibfnamefont {Y.-W.}\ \bibnamefont {Son}}, \bibinfo {author} {\bibfnamefont
  {L.}~\bibnamefont {Taillefer}}, \bibinfo {author} {\bibfnamefont
  {C.}~\bibnamefont {Heil}}, \bibinfo {author} {\bibfnamefont {A.~I.}\
  \bibnamefont {Figueroa}}, \bibinfo {author} {\bibfnamefont {B.}~\bibnamefont
  {Pla{\c c}ais}}, \bibinfo {author} {\bibfnamefont {Q.}~\bibnamefont {Wu}},
  \bibinfo {author} {\bibfnamefont {O.~V.}\ \bibnamefont {Yazyev}}, \bibinfo
  {author} {\bibfnamefont {E.~P. A.~M.}\ \bibnamefont {Bakkers}}, \bibinfo
  {author} {\bibfnamefont {J.}~\bibnamefont {Nyg{\aa}rd}}, \bibinfo {author}
  {\bibfnamefont {P.}~\bibnamefont {Forn-D{\'\i}az}}, \bibinfo {author}
  {\bibfnamefont {S.~D.}\ \bibnamefont {Franceschi}}, \bibinfo {author}
  {\bibfnamefont {J.~W.}\ \bibnamefont {McIver}}, \bibinfo {author}
  {\bibfnamefont {L.~E. F.~F.}\ \bibnamefont {Torres}}, \bibinfo {author}
  {\bibfnamefont {T.}~\bibnamefont {Low}}, \bibinfo {author} {\bibfnamefont
  {A.}~\bibnamefont {Kumar}}, \bibinfo {author} {\bibfnamefont
  {R.}~\bibnamefont {Galceran}}, \bibinfo {author} {\bibfnamefont {S.~O.}\
  \bibnamefont {Valenzuela}}, \bibinfo {author} {\bibfnamefont {M.~V.}\
  \bibnamefont {Costache}}, \bibinfo {author} {\bibfnamefont {A.}~\bibnamefont
  {Manchon}}, \bibinfo {author} {\bibfnamefont {E.-A.}\ \bibnamefont {Kim}},
  \bibinfo {author} {\bibfnamefont {G.~R.}\ \bibnamefont {Schleder}}, \bibinfo
  {author} {\bibfnamefont {A.}~\bibnamefont {Fazzio}},\ and\ \bibinfo {author}
  {\bibfnamefont {S.}~\bibnamefont {Roche}},\ }\bibfield  {title} {\bibinfo
  {title} {The 2021 quantum materials roadmap},\ }\href
  {https://doi.org/10.1088/2515-7639/abb74e} {\bibfield  {journal} {\bibinfo
  {journal} {J. Phys. Mater.}\ }\textbf {\bibinfo {volume} {3}},\ \bibinfo
  {pages} {042006} (\bibinfo {year} {2021})}\BibitemShut {NoStop}%
\bibitem [{\citenamefont {Mitrano}\ \emph {et~al.}(2024)\citenamefont
  {Mitrano}, \citenamefont {Johnston}, \citenamefont {Kim},\ and\ \citenamefont
  {Dean}}]{Mitrano2024exploring}%
  \BibitemOpen
  \bibfield  {author} {\bibinfo {author} {\bibfnamefont {M.}~\bibnamefont
  {Mitrano}}, \bibinfo {author} {\bibfnamefont {S.}~\bibnamefont {Johnston}},
  \bibinfo {author} {\bibfnamefont {Y.-J.}\ \bibnamefont {Kim}},\ and\ \bibinfo
  {author} {\bibfnamefont {M.~P.~M.}\ \bibnamefont {Dean}},\ }\bibfield
  {title} {\bibinfo {title} {Exploring quantum materials with resonant
  inelastic x-ray scattering},\ }\href
  {https://doi.org/10.1103/PhysRevX.14.040501} {\bibfield  {journal} {\bibinfo
  {journal} {Phys. Rev. X}\ }\textbf {\bibinfo {volume} {14}},\ \bibinfo
  {pages} {040501} (\bibinfo {year} {2024})}\BibitemShut {NoStop}%
\bibitem [{\citenamefont {Anderson}(1997)}]{Anderson1997concepts}%
  \BibitemOpen
  \bibfield  {author} {\bibinfo {author} {\bibfnamefont {P.~W.}\ \bibnamefont
  {Anderson}},\ }\href {https://doi.org/10.1142/3528} {\emph {\bibinfo {title}
  {Concepts in Solids}}}\ (\bibinfo  {publisher} {World Scientific},\ \bibinfo
  {year} {1997})\BibitemShut {NoStop}%
\bibitem [{\citenamefont {Gull}\ \emph {et~al.}(2013)\citenamefont {Gull},
  \citenamefont {Parcollet},\ and\ \citenamefont
  {Millis}}]{Gull2013superconductivity}%
  \BibitemOpen
  \bibfield  {author} {\bibinfo {author} {\bibfnamefont {E.}~\bibnamefont
  {Gull}}, \bibinfo {author} {\bibfnamefont {O.}~\bibnamefont {Parcollet}},\
  and\ \bibinfo {author} {\bibfnamefont {A.~J.}\ \bibnamefont {Millis}},\
  }\bibfield  {title} {\bibinfo {title} {Superconductivity and the pseudogap in
  the two-dimensional {H}ubbard model},\ }\href
  {https://doi.org/10.1103/PhysRevLett.110.216405} {\bibfield  {journal}
  {\bibinfo  {journal} {Phys. Rev. Lett.}\ }\textbf {\bibinfo {volume} {110}},\
  \bibinfo {pages} {216405} (\bibinfo {year} {2013})}\BibitemShut {NoStop}%
\bibitem [{\citenamefont {Jiang}\ and\ \citenamefont
  {Devereaux}(2019)}]{Jiang2019superconductivity}%
  \BibitemOpen
  \bibfield  {author} {\bibinfo {author} {\bibfnamefont {H.-C.}\ \bibnamefont
  {Jiang}}\ and\ \bibinfo {author} {\bibfnamefont {T.~P.}\ \bibnamefont
  {Devereaux}},\ }\bibfield  {title} {\bibinfo {title} {Superconductivity in
  the doped {H}ubbard model and its interplay with next-nearest hopping
  $t^\prime$},\ }\href {https://doi.org/10.1126/science.aal5304} {\bibfield
  {journal} {\bibinfo  {journal} {Science}\ }\textbf {\bibinfo {volume}
  {365}},\ \bibinfo {pages} {1424} (\bibinfo {year} {2019})}\BibitemShut
  {NoStop}%
\bibitem [{\citenamefont {Qin}\ \emph {et~al.}(2020)\citenamefont {Qin},
  \citenamefont {Chung}, \citenamefont {Shi}, \citenamefont {Vitali},
  \citenamefont {Hubig}, \citenamefont {Schollw\"ock}, \citenamefont {White},\
  and\ \citenamefont {Zhang}}]{Qin2020absense}%
  \BibitemOpen
  \bibfield  {author} {\bibinfo {author} {\bibfnamefont {M.}~\bibnamefont
  {Qin}}, \bibinfo {author} {\bibfnamefont {C.-M.}\ \bibnamefont {Chung}},
  \bibinfo {author} {\bibfnamefont {H.}~\bibnamefont {Shi}}, \bibinfo {author}
  {\bibfnamefont {E.}~\bibnamefont {Vitali}}, \bibinfo {author} {\bibfnamefont
  {C.}~\bibnamefont {Hubig}}, \bibinfo {author} {\bibfnamefont
  {U.}~\bibnamefont {Schollw\"ock}}, \bibinfo {author} {\bibfnamefont {S.~R.}\
  \bibnamefont {White}},\ and\ \bibinfo {author} {\bibfnamefont
  {S.}~\bibnamefont {Zhang}} (\bibinfo {collaboration} {Simons Collaboration on
  the Many-Electron Problem}),\ }\bibfield  {title} {\bibinfo {title} {Absence
  of superconductivity in the pure two-dimensional {H}ubbard model},\ }\href
  {https://doi.org/10.1103/PhysRevX.10.031016} {\bibfield  {journal} {\bibinfo
  {journal} {Phys. Rev. X}\ }\textbf {\bibinfo {volume} {10}},\ \bibinfo
  {pages} {031016} (\bibinfo {year} {2020})}\BibitemShut {NoStop}%
\bibitem [{\citenamefont {Chen}\ \emph
  {et~al.}(2021{\natexlab{a}})\citenamefont {Chen}, \citenamefont {Wang},
  \citenamefont {Rebec}, \citenamefont {Jia}, \citenamefont {Hashimoto},
  \citenamefont {Lu}, \citenamefont {Moritz}, \citenamefont {Moore},
  \citenamefont {Devereaux},\ and\ \citenamefont {Shen}}]{Chen2021anomalously}%
  \BibitemOpen
  \bibfield  {author} {\bibinfo {author} {\bibfnamefont {Z.}~\bibnamefont
  {Chen}}, \bibinfo {author} {\bibfnamefont {Y.}~\bibnamefont {Wang}}, \bibinfo
  {author} {\bibfnamefont {S.~N.}\ \bibnamefont {Rebec}}, \bibinfo {author}
  {\bibfnamefont {T.}~\bibnamefont {Jia}}, \bibinfo {author} {\bibfnamefont
  {M.}~\bibnamefont {Hashimoto}}, \bibinfo {author} {\bibfnamefont
  {D.}~\bibnamefont {Lu}}, \bibinfo {author} {\bibfnamefont {B.}~\bibnamefont
  {Moritz}}, \bibinfo {author} {\bibfnamefont {R.~G.}\ \bibnamefont {Moore}},
  \bibinfo {author} {\bibfnamefont {T.~P.}\ \bibnamefont {Devereaux}},\ and\
  \bibinfo {author} {\bibfnamefont {Z.-X.}\ \bibnamefont {Shen}},\ }\bibfield
  {title} {\bibinfo {title} {Anomalously strong near-neighbor attraction in
  doped {1D} cuprate chains},\ }\href {https://doi.org/10.1126/science.abf5174}
  {\bibfield  {journal} {\bibinfo  {journal} {Science}\ }\textbf {\bibinfo
  {volume} {373}},\ \bibinfo {pages} {1235} (\bibinfo {year}
  {2021}{\natexlab{a}})}\BibitemShut {NoStop}%
\bibitem [{\citenamefont {Padma}\ \emph {et~al.}(2025)\citenamefont {Padma},
  \citenamefont {Thomas}, \citenamefont {TenHuisen}, \citenamefont {He},
  \citenamefont {Guan}, \citenamefont {Li}, \citenamefont {Lee}, \citenamefont
  {Wang}, \citenamefont {Lee}, \citenamefont {Mao}, \citenamefont {Jang},
  \citenamefont {Bisogni}, \citenamefont {Pelliciari}, \citenamefont {Dean},
  \citenamefont {Johnston},\ and\ \citenamefont {Mitrano}}]{padma2025beyond}%
  \BibitemOpen
  \bibfield  {author} {\bibinfo {author} {\bibfnamefont {H.}~\bibnamefont
  {Padma}}, \bibinfo {author} {\bibfnamefont {J.}~\bibnamefont {Thomas}},
  \bibinfo {author} {\bibfnamefont {S.~F.~R.}\ \bibnamefont {TenHuisen}},
  \bibinfo {author} {\bibfnamefont {W.}~\bibnamefont {He}}, \bibinfo {author}
  {\bibfnamefont {Z.}~\bibnamefont {Guan}}, \bibinfo {author} {\bibfnamefont
  {J.}~\bibnamefont {Li}}, \bibinfo {author} {\bibfnamefont {B.}~\bibnamefont
  {Lee}}, \bibinfo {author} {\bibfnamefont {Y.}~\bibnamefont {Wang}}, \bibinfo
  {author} {\bibfnamefont {S.~H.}\ \bibnamefont {Lee}}, \bibinfo {author}
  {\bibfnamefont {Z.}~\bibnamefont {Mao}}, \bibinfo {author} {\bibfnamefont
  {H.}~\bibnamefont {Jang}}, \bibinfo {author} {\bibfnamefont {V.}~\bibnamefont
  {Bisogni}}, \bibinfo {author} {\bibfnamefont {J.}~\bibnamefont {Pelliciari}},
  \bibinfo {author} {\bibfnamefont {M.~P.~M.}\ \bibnamefont {Dean}}, \bibinfo
  {author} {\bibfnamefont {S.}~\bibnamefont {Johnston}},\ and\ \bibinfo
  {author} {\bibfnamefont {M.}~\bibnamefont {Mitrano}},\ }\bibfield  {title}
  {\bibinfo {title} {Beyond-{H}ubbard pairing in a cuprate ladder},\ }\href
  {https://doi.org/10.1103/PhysRevX.15.021049} {\bibfield  {journal} {\bibinfo
  {journal} {Phys. Rev. X}\ }\textbf {\bibinfo {volume} {15}},\ \bibinfo
  {pages} {021049} (\bibinfo {year} {2025})}\BibitemShut {NoStop}%
\bibitem [{\citenamefont {Scheie}\ \emph {et~al.}(2025)\citenamefont {Scheie},
  \citenamefont {Laurell}, \citenamefont {Thomas}, \citenamefont {Sharma},
  \citenamefont {Kolesnikov}, \citenamefont {Granroth}, \citenamefont {Zhang},
  \citenamefont {Lake}, \citenamefont {Jr.}, \citenamefont {Bewley},
  \citenamefont {Eccleston}, \citenamefont {Akimitsu}, \citenamefont {Dagotto},
  \citenamefont {Batista}, \citenamefont {Alvarez}, \citenamefont {Johnston},\
  and\ \citenamefont {Tennant}}]{scheie2025cooper}%
  \BibitemOpen
  \bibfield  {author} {\bibinfo {author} {\bibfnamefont {A.}~\bibnamefont
  {Scheie}}, \bibinfo {author} {\bibfnamefont {P.}~\bibnamefont {Laurell}},
  \bibinfo {author} {\bibfnamefont {J.}~\bibnamefont {Thomas}}, \bibinfo
  {author} {\bibfnamefont {V.}~\bibnamefont {Sharma}}, \bibinfo {author}
  {\bibfnamefont {A.~I.}\ \bibnamefont {Kolesnikov}}, \bibinfo {author}
  {\bibfnamefont {G.~E.}\ \bibnamefont {Granroth}}, \bibinfo {author}
  {\bibfnamefont {Q.}~\bibnamefont {Zhang}}, \bibinfo {author} {\bibfnamefont
  {B.}~\bibnamefont {Lake}}, \bibinfo {author} {\bibfnamefont {M.~M.}\
  \bibnamefont {Jr.}}, \bibinfo {author} {\bibfnamefont {R.~I.}\ \bibnamefont
  {Bewley}}, \bibinfo {author} {\bibfnamefont {R.~S.}\ \bibnamefont
  {Eccleston}}, \bibinfo {author} {\bibfnamefont {J.}~\bibnamefont {Akimitsu}},
  \bibinfo {author} {\bibfnamefont {E.}~\bibnamefont {Dagotto}}, \bibinfo
  {author} {\bibfnamefont {C.~D.}\ \bibnamefont {Batista}}, \bibinfo {author}
  {\bibfnamefont {G.}~\bibnamefont {Alvarez}}, \bibinfo {author} {\bibfnamefont
  {S.}~\bibnamefont {Johnston}},\ and\ \bibinfo {author} {\bibfnamefont
  {D.~A.}\ \bibnamefont {Tennant}},\ }\bibfield  {title} {\bibinfo {title}
  {Cooper-pair localization in the magnetic dynamics of a cuprate ladder},\
  }\href {https://arxiv.org/abs/2501.10296} {\bibfield  {journal} {\bibinfo
  {journal} {arXiv:2501.10296}\ } (\bibinfo {year} {2025})}\BibitemShut
  {NoStop}%
\bibitem [{\citenamefont {Kang}\ \emph {et~al.}(2020)\citenamefont {Kang},
  \citenamefont {Kim}, \citenamefont {Kim}, \citenamefont {Kim}, \citenamefont
  {Sim}, \citenamefont {Lee}, \citenamefont {Lee}, \citenamefont {Park},
  \citenamefont {Yun}, \citenamefont {Kim}, \citenamefont {Nag}, \citenamefont
  {Walters}, \citenamefont {Garcia-Fernandez}, \citenamefont {Li},
  \citenamefont {Chapon}, \citenamefont {Zhou}, \citenamefont {Son},
  \citenamefont {Kim}, \citenamefont {Cheong},\ and\ \citenamefont
  {Park}}]{Kang2020coherent}%
  \BibitemOpen
  \bibfield  {author} {\bibinfo {author} {\bibfnamefont {S.}~\bibnamefont
  {Kang}}, \bibinfo {author} {\bibfnamefont {K.}~\bibnamefont {Kim}}, \bibinfo
  {author} {\bibfnamefont {B.~H.}\ \bibnamefont {Kim}}, \bibinfo {author}
  {\bibfnamefont {J.}~\bibnamefont {Kim}}, \bibinfo {author} {\bibfnamefont
  {K.~I.}\ \bibnamefont {Sim}}, \bibinfo {author} {\bibfnamefont {J.-U.}\
  \bibnamefont {Lee}}, \bibinfo {author} {\bibfnamefont {S.}~\bibnamefont
  {Lee}}, \bibinfo {author} {\bibfnamefont {K.}~\bibnamefont {Park}}, \bibinfo
  {author} {\bibfnamefont {S.}~\bibnamefont {Yun}}, \bibinfo {author}
  {\bibfnamefont {T.}~\bibnamefont {Kim}}, \bibinfo {author} {\bibfnamefont
  {A.}~\bibnamefont {Nag}}, \bibinfo {author} {\bibfnamefont {A.}~\bibnamefont
  {Walters}}, \bibinfo {author} {\bibfnamefont {M.}~\bibnamefont
  {Garcia-Fernandez}}, \bibinfo {author} {\bibfnamefont {J.}~\bibnamefont
  {Li}}, \bibinfo {author} {\bibfnamefont {L.}~\bibnamefont {Chapon}}, \bibinfo
  {author} {\bibfnamefont {K.-J.}\ \bibnamefont {Zhou}}, \bibinfo {author}
  {\bibfnamefont {Y.-W.}\ \bibnamefont {Son}}, \bibinfo {author} {\bibfnamefont
  {J.~H.}\ \bibnamefont {Kim}}, \bibinfo {author} {\bibfnamefont
  {H.}~\bibnamefont {Cheong}},\ and\ \bibinfo {author} {\bibfnamefont {J.-G.}\
  \bibnamefont {Park}},\ }\bibfield  {title} {\bibinfo {title} {Coherent
  many-body exciton in van der {Waals} antiferromagnet {NiPS$_3$}},\ }\href
  {https://doi.org/10.1038/s41586-020-2520-5} {\bibfield  {journal} {\bibinfo
  {journal} {Nature}\ }\textbf {\bibinfo {volume} {583}},\ \bibinfo {pages}
  {785} (\bibinfo {year} {2020})}\BibitemShut {NoStop}%
\bibitem [{\citenamefont {He}\ \emph {et~al.}(2024)\citenamefont {He},
  \citenamefont {Shen}, \citenamefont {Wohlfeld}, \citenamefont {Sears},
  \citenamefont {Li}, \citenamefont {Pelliciari}, \citenamefont {Walicki},
  \citenamefont {Johnston}, \citenamefont {Baldini}, \citenamefont {Bisogni},
  \citenamefont {Mitrano},\ and\ \citenamefont {Dean}}]{He2024MagneticallyPH}%
  \BibitemOpen
  \bibfield  {author} {\bibinfo {author} {\bibfnamefont {W.}~\bibnamefont
  {He}}, \bibinfo {author} {\bibfnamefont {Y.}~\bibnamefont {Shen}}, \bibinfo
  {author} {\bibfnamefont {K.}~\bibnamefont {Wohlfeld}}, \bibinfo {author}
  {\bibfnamefont {J.}~\bibnamefont {Sears}}, \bibinfo {author} {\bibfnamefont
  {J.}~\bibnamefont {Li}}, \bibinfo {author} {\bibfnamefont {J.}~\bibnamefont
  {Pelliciari}}, \bibinfo {author} {\bibfnamefont {M.}~\bibnamefont {Walicki}},
  \bibinfo {author} {\bibfnamefont {S.}~\bibnamefont {Johnston}}, \bibinfo
  {author} {\bibfnamefont {E.}~\bibnamefont {Baldini}}, \bibinfo {author}
  {\bibfnamefont {V.}~\bibnamefont {Bisogni}}, \bibinfo {author} {\bibfnamefont
  {M.}~\bibnamefont {Mitrano}},\ and\ \bibinfo {author} {\bibfnamefont
  {M.~P.~M.}\ \bibnamefont {Dean}},\ }\bibfield  {title} {\bibinfo {title}
  {Magnetically propagating {Hund’s} exciton in van der {Waals}
  antiferromagnet {NiPS$_3$}},\ }\href
  {https://api.semanticscholar.org/CorpusID:269188094} {\bibfield  {journal}
  {\bibinfo  {journal} {Nature Communications}\ }\textbf {\bibinfo {volume}
  {15}} (\bibinfo {year} {2024})}\BibitemShut {NoStop}%
\bibitem [{\citenamefont {Occhialini}\ \emph {et~al.}(2024)\citenamefont
  {Occhialini}, \citenamefont {Tseng}, \citenamefont {Elnaggar}, \citenamefont
  {Song}, \citenamefont {Blei}, \citenamefont {Tongay}, \citenamefont
  {Bisogni}, \citenamefont {de~Groot}, \citenamefont {Pelliciari},\ and\
  \citenamefont {Comin}}]{Occhialini2024}%
  \BibitemOpen
  \bibfield  {author} {\bibinfo {author} {\bibfnamefont {C.~A.}\ \bibnamefont
  {Occhialini}}, \bibinfo {author} {\bibfnamefont {Y.}~\bibnamefont {Tseng}},
  \bibinfo {author} {\bibfnamefont {H.}~\bibnamefont {Elnaggar}}, \bibinfo
  {author} {\bibfnamefont {Q.}~\bibnamefont {Song}}, \bibinfo {author}
  {\bibfnamefont {M.}~\bibnamefont {Blei}}, \bibinfo {author} {\bibfnamefont
  {S.~A.}\ \bibnamefont {Tongay}}, \bibinfo {author} {\bibfnamefont
  {V.}~\bibnamefont {Bisogni}}, \bibinfo {author} {\bibfnamefont {F.~M.~F.}\
  \bibnamefont {de~Groot}}, \bibinfo {author} {\bibfnamefont {J.}~\bibnamefont
  {Pelliciari}},\ and\ \bibinfo {author} {\bibfnamefont {R.}~\bibnamefont
  {Comin}},\ }\bibfield  {title} {\bibinfo {title} {Nature of excitons and
  their ligand-mediated delocalization in nickel dihalide charge-transfer
  insulators},\ }\href {https://doi.org/10.1103/PhysRevX.14.031007} {\bibfield
  {journal} {\bibinfo  {journal} {Phys. Rev. X}\ }\textbf {\bibinfo {volume}
  {14}},\ \bibinfo {pages} {031007} (\bibinfo {year} {2024})}\BibitemShut
  {NoStop}%
\bibitem [{\citenamefont {Hamad}\ \emph {et~al.}(2024)\citenamefont {Hamad},
  \citenamefont {Helman}, \citenamefont {Manuel}, \citenamefont {Feiguin},\
  and\ \citenamefont {Aligia}}]{hamad2024singletpolarontheorylowenergy}%
  \BibitemOpen
  \bibfield  {author} {\bibinfo {author} {\bibfnamefont {I.~J.}\ \bibnamefont
  {Hamad}}, \bibinfo {author} {\bibfnamefont {C.~S.}\ \bibnamefont {Helman}},
  \bibinfo {author} {\bibfnamefont {L.~O.}\ \bibnamefont {Manuel}}, \bibinfo
  {author} {\bibfnamefont {A.~E.}\ \bibnamefont {Feiguin}},\ and\ \bibinfo
  {author} {\bibfnamefont {A.~A.}\ \bibnamefont {Aligia}},\ }\bibfield  {title}
  {\bibinfo {title} {Singlet polaron theory of low-energy optical excitations
  in {${\mathrm{NiPS}}_{3}$}},\ }\href
  {https://doi.org/10.1103/PhysRevLett.133.146502} {\bibfield  {journal}
  {\bibinfo  {journal} {Phys. Rev. Lett.}\ }\textbf {\bibinfo {volume} {133}},\
  \bibinfo {pages} {146502} (\bibinfo {year} {2024})}\BibitemShut {NoStop}%
\bibitem [{\citenamefont {Sharma}\ \emph {et~al.}(2023)\citenamefont {Sharma},
  \citenamefont {Wang},\ and\ \citenamefont {Batista}}]{Sharma2023machine}%
  \BibitemOpen
  \bibfield  {author} {\bibinfo {author} {\bibfnamefont {V.}~\bibnamefont
  {Sharma}}, \bibinfo {author} {\bibfnamefont {Z.}~\bibnamefont {Wang}},\ and\
  \bibinfo {author} {\bibfnamefont {C.~D.}\ \bibnamefont {Batista}},\
  }\bibfield  {title} {\bibinfo {title} {Machine learning assisted derivation
  of minimal low-energy models for metallic magnets},\ }\href
  {https://doi.org/10.1038/s41524-023-01137-x} {\bibfield  {journal} {\bibinfo
  {journal} {npj Computational Materials}\ }\textbf {\bibinfo {volume} {9}},\
  \bibinfo {pages} {192} (\bibinfo {year} {2023})}\BibitemShut {NoStop}%
\bibitem [{\citenamefont {Carrasquilla}(2020)}]{Carrasquilla2020machine}%
  \BibitemOpen
  \bibfield  {author} {\bibinfo {author} {\bibfnamefont {J.}~\bibnamefont
  {Carrasquilla}},\ }\bibfield  {title} {\bibinfo {title} {Machine learning for
  quantum matter},\ }\href {https://doi.org/10.1080/23746149.2020.1797528}
  {\bibfield  {journal} {\bibinfo  {journal} {Advances in Physics: X}\ }\textbf
  {\bibinfo {volume} {5}},\ \bibinfo {pages} {1797528} (\bibinfo {year}
  {2020})}\BibitemShut {NoStop}%
\bibitem [{\citenamefont {Bedolla}\ \emph {et~al.}(2020)\citenamefont
  {Bedolla}, \citenamefont {Padierna},\ and\ \citenamefont
  {Casta{\~n}eda-Priego}}]{Bedolla2021machine}%
  \BibitemOpen
  \bibfield  {author} {\bibinfo {author} {\bibfnamefont {E.}~\bibnamefont
  {Bedolla}}, \bibinfo {author} {\bibfnamefont {L.~C.}\ \bibnamefont
  {Padierna}},\ and\ \bibinfo {author} {\bibfnamefont {R.}~\bibnamefont
  {Casta{\~n}eda-Priego}},\ }\bibfield  {title} {\bibinfo {title} {Machine
  learning for condensed matter physics},\ }\href
  {https://doi.org/10.1088/1361-648X/abb895} {\bibfield  {journal} {\bibinfo
  {journal} {Journal of Physics: Condensed Matter}\ }\textbf {\bibinfo {volume}
  {33}},\ \bibinfo {pages} {053001} (\bibinfo {year} {2020})}\BibitemShut
  {NoStop}%
\bibitem [{\citenamefont {Johnston}\ \emph {et~al.}(2022)\citenamefont
  {Johnston}, \citenamefont {Khatami},\ and\ \citenamefont
  {Scalettar}}]{Johnston2022perspective}%
  \BibitemOpen
  \bibfield  {author} {\bibinfo {author} {\bibfnamefont {S.}~\bibnamefont
  {Johnston}}, \bibinfo {author} {\bibfnamefont {E.}~\bibnamefont {Khatami}},\
  and\ \bibinfo {author} {\bibfnamefont {R.}~\bibnamefont {Scalettar}},\
  }\bibfield  {title} {\bibinfo {title} {A perspective on machine learning and
  data science for strongly correlated electron problems},\ }\href
  {https://doi.org/https://doi.org/10.1016/j.cartre.2022.100231} {\bibfield
  {journal} {\bibinfo  {journal} {Carbon Trends}\ }\textbf {\bibinfo {volume}
  {9}},\ \bibinfo {pages} {100231} (\bibinfo {year} {2022})}\BibitemShut
  {NoStop}%
\bibitem [{\citenamefont {Carleo}\ \emph {et~al.}(2019)\citenamefont {Carleo},
  \citenamefont {Cirac}, \citenamefont {Cranmer}, \citenamefont {Daudet},
  \citenamefont {Schuld}, \citenamefont {Tishby}, \citenamefont
  {Vogt-Maranto},\ and\ \citenamefont {Zdeborov\'a}}]{Carleo2019machine}%
  \BibitemOpen
  \bibfield  {author} {\bibinfo {author} {\bibfnamefont {G.}~\bibnamefont
  {Carleo}}, \bibinfo {author} {\bibfnamefont {I.}~\bibnamefont {Cirac}},
  \bibinfo {author} {\bibfnamefont {K.}~\bibnamefont {Cranmer}}, \bibinfo
  {author} {\bibfnamefont {L.}~\bibnamefont {Daudet}}, \bibinfo {author}
  {\bibfnamefont {M.}~\bibnamefont {Schuld}}, \bibinfo {author} {\bibfnamefont
  {N.}~\bibnamefont {Tishby}}, \bibinfo {author} {\bibfnamefont
  {L.}~\bibnamefont {Vogt-Maranto}},\ and\ \bibinfo {author} {\bibfnamefont
  {L.}~\bibnamefont {Zdeborov\'a}},\ }\bibfield  {title} {\bibinfo {title}
  {Machine learning and the physical sciences},\ }\href
  {https://doi.org/10.1103/RevModPhys.91.045002} {\bibfield  {journal}
  {\bibinfo  {journal} {Rev. Mod. Phys.}\ }\textbf {\bibinfo {volume} {91}},\
  \bibinfo {pages} {045002} (\bibinfo {year} {2019})}\BibitemShut {NoStop}%
\bibitem [{\citenamefont {Karniadakis}\ \emph {et~al.}(2021)\citenamefont
  {Karniadakis}, \citenamefont {Kevrekidis}, \citenamefont {Lu}, \citenamefont
  {Perdikaris}, \citenamefont {Wang},\ and\ \citenamefont
  {Yang}}]{Karniadakis2021physics}%
  \BibitemOpen
  \bibfield  {author} {\bibinfo {author} {\bibfnamefont {G.~E.}\ \bibnamefont
  {Karniadakis}}, \bibinfo {author} {\bibfnamefont {I.~G.}\ \bibnamefont
  {Kevrekidis}}, \bibinfo {author} {\bibfnamefont {L.}~\bibnamefont {Lu}},
  \bibinfo {author} {\bibfnamefont {P.}~\bibnamefont {Perdikaris}}, \bibinfo
  {author} {\bibfnamefont {S.}~\bibnamefont {Wang}},\ and\ \bibinfo {author}
  {\bibfnamefont {L.}~\bibnamefont {Yang}},\ }\bibfield  {title} {\bibinfo
  {title} {Physics-informed machine learning},\ }\href
  {https://doi.org/10.1038/s42254-021-00314-5} {\bibfield  {journal} {\bibinfo
  {journal} {Nature Reviews Physics}\ }\textbf {\bibinfo {volume} {3}},\
  \bibinfo {pages} {422} (\bibinfo {year} {2021})}\BibitemShut {NoStop}%
\bibitem [{\citenamefont {Chen}\ \emph
  {et~al.}(2021{\natexlab{b}})\citenamefont {Chen}, \citenamefont {Andrejevic},
  \citenamefont {Drucker}, \citenamefont {Nguyen}, \citenamefont {Xian},
  \citenamefont {Smidt}, \citenamefont {Wang}, \citenamefont {Ernstorfer},
  \citenamefont {Tennant}, \citenamefont {Chan},\ and\ \citenamefont
  {Li}}]{Chen2021machine}%
  \BibitemOpen
  \bibfield  {author} {\bibinfo {author} {\bibfnamefont {Z.}~\bibnamefont
  {Chen}}, \bibinfo {author} {\bibfnamefont {N.}~\bibnamefont {Andrejevic}},
  \bibinfo {author} {\bibfnamefont {N.~C.}\ \bibnamefont {Drucker}}, \bibinfo
  {author} {\bibfnamefont {T.}~\bibnamefont {Nguyen}}, \bibinfo {author}
  {\bibfnamefont {R.~P.}\ \bibnamefont {Xian}}, \bibinfo {author}
  {\bibfnamefont {T.}~\bibnamefont {Smidt}}, \bibinfo {author} {\bibfnamefont
  {Y.}~\bibnamefont {Wang}}, \bibinfo {author} {\bibfnamefont {R.}~\bibnamefont
  {Ernstorfer}}, \bibinfo {author} {\bibfnamefont {D.~A.}\ \bibnamefont
  {Tennant}}, \bibinfo {author} {\bibfnamefont {M.}~\bibnamefont {Chan}},\ and\
  \bibinfo {author} {\bibfnamefont {M.}~\bibnamefont {Li}},\ }\bibfield
  {title} {\bibinfo {title} {Machine learning on neutron and x-ray scattering
  and spectroscopies},\ }\href {https://doi.org/10.1063/5.0049111} {\bibfield
  {journal} {\bibinfo  {journal} {Chemical Physics Reviews}\ }\textbf {\bibinfo
  {volume} {2}},\ \bibinfo {pages} {031301} (\bibinfo {year}
  {2021}{\natexlab{b}})}\BibitemShut {NoStop}%
\bibitem [{\citenamefont {Carbone}\ \emph {et~al.}(2019)\citenamefont
  {Carbone}, \citenamefont {Yoo}, \citenamefont {Topsakal},\ and\ \citenamefont
  {Lu}}]{carbone2019classification}%
  \BibitemOpen
  \bibfield  {author} {\bibinfo {author} {\bibfnamefont {M.~R.}\ \bibnamefont
  {Carbone}}, \bibinfo {author} {\bibfnamefont {S.}~\bibnamefont {Yoo}},
  \bibinfo {author} {\bibfnamefont {M.}~\bibnamefont {Topsakal}},\ and\
  \bibinfo {author} {\bibfnamefont {D.}~\bibnamefont {Lu}},\ }\bibfield
  {title} {\bibinfo {title} {Classification of local chemical environments from
  x-ray absorption spectra using supervised machine learning},\ }\href
  {https://doi.org/10.1103/PhysRevMaterials.3.033604} {\bibfield  {journal}
  {\bibinfo  {journal} {Phys. Rev. Mater.}\ }\textbf {\bibinfo {volume} {3}},\
  \bibinfo {pages} {033604} (\bibinfo {year} {2019})}\BibitemShut {NoStop}%
\bibitem [{\citenamefont {Carbone}\ \emph {et~al.}(2020)\citenamefont
  {Carbone}, \citenamefont {Topsakal}, \citenamefont {Lu},\ and\ \citenamefont
  {Yoo}}]{carbone2020machine}%
  \BibitemOpen
  \bibfield  {author} {\bibinfo {author} {\bibfnamefont {M.~R.}\ \bibnamefont
  {Carbone}}, \bibinfo {author} {\bibfnamefont {M.}~\bibnamefont {Topsakal}},
  \bibinfo {author} {\bibfnamefont {D.}~\bibnamefont {Lu}},\ and\ \bibinfo
  {author} {\bibfnamefont {S.}~\bibnamefont {Yoo}},\ }\bibfield  {title}
  {\bibinfo {title} {Machine-learning x-ray absorption spectra to quantitative
  accuracy},\ }\href {https://doi.org/10.1103/PhysRevLett.124.156401}
  {\bibfield  {journal} {\bibinfo  {journal} {Phys. Rev. Lett.}\ }\textbf
  {\bibinfo {volume} {124}},\ \bibinfo {pages} {156401} (\bibinfo {year}
  {2020})}\BibitemShut {NoStop}%
\bibitem [{\citenamefont {Torrisi}\ \emph {et~al.}(2020)\citenamefont
  {Torrisi}, \citenamefont {Carbone}, \citenamefont {Rohr}, \citenamefont
  {Montoya}, \citenamefont {Ha}, \citenamefont {Yano}, \citenamefont {Suram},\
  and\ \citenamefont {Hung}}]{torrisi2020random}%
  \BibitemOpen
  \bibfield  {author} {\bibinfo {author} {\bibfnamefont {S.~B.}\ \bibnamefont
  {Torrisi}}, \bibinfo {author} {\bibfnamefont {M.~R.}\ \bibnamefont
  {Carbone}}, \bibinfo {author} {\bibfnamefont {B.~A.}\ \bibnamefont {Rohr}},
  \bibinfo {author} {\bibfnamefont {J.~H.}\ \bibnamefont {Montoya}}, \bibinfo
  {author} {\bibfnamefont {Y.}~\bibnamefont {Ha}}, \bibinfo {author}
  {\bibfnamefont {J.}~\bibnamefont {Yano}}, \bibinfo {author} {\bibfnamefont
  {S.~K.}\ \bibnamefont {Suram}},\ and\ \bibinfo {author} {\bibfnamefont
  {L.}~\bibnamefont {Hung}},\ }\bibfield  {title} {\bibinfo {title} {Random
  forest machine learning models for interpretable x-ray absorption near-edge
  structure spectrum-property relationships},\ }\href
  {https://www.nature.com/articles/s41524-020-00376-6} {\bibfield  {journal}
  {\bibinfo  {journal} {npj Comput. Mater.}\ }\textbf {\bibinfo {volume} {6}},\
  \bibinfo {pages} {1} (\bibinfo {year} {2020})}\BibitemShut {NoStop}%
\bibitem [{\citenamefont {Miles}\ \emph {et~al.}(2021)\citenamefont {Miles},
  \citenamefont {Carbone}, \citenamefont {Sturm}, \citenamefont {Lu},
  \citenamefont {Weichselbaum}, \citenamefont {Barros},\ and\ \citenamefont
  {Konik}}]{miles2021machine}%
  \BibitemOpen
  \bibfield  {author} {\bibinfo {author} {\bibfnamefont {C.}~\bibnamefont
  {Miles}}, \bibinfo {author} {\bibfnamefont {M.~R.}\ \bibnamefont {Carbone}},
  \bibinfo {author} {\bibfnamefont {E.~J.}\ \bibnamefont {Sturm}}, \bibinfo
  {author} {\bibfnamefont {D.}~\bibnamefont {Lu}}, \bibinfo {author}
  {\bibfnamefont {A.}~\bibnamefont {Weichselbaum}}, \bibinfo {author}
  {\bibfnamefont {K.}~\bibnamefont {Barros}},\ and\ \bibinfo {author}
  {\bibfnamefont {R.~M.}\ \bibnamefont {Konik}},\ }\bibfield  {title} {\bibinfo
  {title} {Machine learning of {Kondo} physics using variational autoencoders
  and symbolic regression},\ }\href
  {https://doi.org/10.1103/PhysRevB.104.235111} {\bibfield  {journal} {\bibinfo
   {journal} {Phys. Rev. B}\ }\textbf {\bibinfo {volume} {104}},\ \bibinfo
  {pages} {235111} (\bibinfo {year} {2021})}\BibitemShut {NoStop}%
\bibitem [{\citenamefont {Sturm}\ \emph {et~al.}(2021)\citenamefont {Sturm},
  \citenamefont {Carbone}, \citenamefont {Lu}, \citenamefont {Weichselbaum},\
  and\ \citenamefont {Konik}}]{sturm2021predicting}%
  \BibitemOpen
  \bibfield  {author} {\bibinfo {author} {\bibfnamefont {E.~J.}\ \bibnamefont
  {Sturm}}, \bibinfo {author} {\bibfnamefont {M.~R.}\ \bibnamefont {Carbone}},
  \bibinfo {author} {\bibfnamefont {D.}~\bibnamefont {Lu}}, \bibinfo {author}
  {\bibfnamefont {A.}~\bibnamefont {Weichselbaum}},\ and\ \bibinfo {author}
  {\bibfnamefont {R.~M.}\ \bibnamefont {Konik}},\ }\bibfield  {title} {\bibinfo
  {title} {Predicting impurity spectral functions using machine learning},\
  }\href {https://doi.org/10.1103/PhysRevB.103.245118} {\bibfield  {journal}
  {\bibinfo  {journal} {Phys. Rev. B}\ }\textbf {\bibinfo {volume} {103}},\
  \bibinfo {pages} {245118} (\bibinfo {year} {2021})}\BibitemShut {NoStop}%
\bibitem [{\citenamefont {Kwon}\ \emph {et~al.}(2023)\citenamefont {Kwon},
  \citenamefont {Sun}, \citenamefont {Hsu}, \citenamefont {Jeong},
  \citenamefont {Aydin}, \citenamefont {Sharma}, \citenamefont {Meng},
  \citenamefont {Carbone}, \citenamefont {Chen}, \citenamefont {Lu} \emph
  {et~al.}}]{kwon2023harnessing}%
  \BibitemOpen
  \bibfield  {author} {\bibinfo {author} {\bibfnamefont {H.}~\bibnamefont
  {Kwon}}, \bibinfo {author} {\bibfnamefont {W.}~\bibnamefont {Sun}}, \bibinfo
  {author} {\bibfnamefont {T.}~\bibnamefont {Hsu}}, \bibinfo {author}
  {\bibfnamefont {W.}~\bibnamefont {Jeong}}, \bibinfo {author} {\bibfnamefont
  {F.}~\bibnamefont {Aydin}}, \bibinfo {author} {\bibfnamefont
  {S.}~\bibnamefont {Sharma}}, \bibinfo {author} {\bibfnamefont
  {F.}~\bibnamefont {Meng}}, \bibinfo {author} {\bibfnamefont {M.~R.}\
  \bibnamefont {Carbone}}, \bibinfo {author} {\bibfnamefont {X.}~\bibnamefont
  {Chen}}, \bibinfo {author} {\bibfnamefont {D.}~\bibnamefont {Lu}}, \emph
  {et~al.},\ }\bibfield  {title} {\bibinfo {title} {Harnessing neural networks
  for elucidating x-ray absorption structure--spectrum relationships in
  amorphous carbon},\ }\href
  {https://pubs.acs.org/doi/10.1021/acs.jpcc.3c02029} {\bibfield  {journal}
  {\bibinfo  {journal} {The Journal of Physical Chemistry C}\ }\textbf
  {\bibinfo {volume} {127}},\ \bibinfo {pages} {16473} (\bibinfo {year}
  {2023})}\BibitemShut {NoStop}%
\bibitem [{\citenamefont {Kwon}\ \emph {et~al.}(2024)\citenamefont {Kwon},
  \citenamefont {Hsu}, \citenamefont {Sun}, \citenamefont {Jeong},
  \citenamefont {Aydin}, \citenamefont {Chapman}, \citenamefont {Chen},
  \citenamefont {Lordi}, \citenamefont {Carbone}, \citenamefont {Lu},
  \citenamefont {Zhou},\ and\ \citenamefont {Anh~Pham}}]{kwon2024spectroscopy}%
  \BibitemOpen
  \bibfield  {author} {\bibinfo {author} {\bibfnamefont {H.}~\bibnamefont
  {Kwon}}, \bibinfo {author} {\bibfnamefont {T.}~\bibnamefont {Hsu}}, \bibinfo
  {author} {\bibfnamefont {W.}~\bibnamefont {Sun}}, \bibinfo {author}
  {\bibfnamefont {W.}~\bibnamefont {Jeong}}, \bibinfo {author} {\bibfnamefont
  {F.}~\bibnamefont {Aydin}}, \bibinfo {author} {\bibfnamefont
  {J.}~\bibnamefont {Chapman}}, \bibinfo {author} {\bibfnamefont
  {X.}~\bibnamefont {Chen}}, \bibinfo {author} {\bibfnamefont {V.}~\bibnamefont
  {Lordi}}, \bibinfo {author} {\bibfnamefont {M.~R.}\ \bibnamefont {Carbone}},
  \bibinfo {author} {\bibfnamefont {D.}~\bibnamefont {Lu}}, \bibinfo {author}
  {\bibfnamefont {F.}~\bibnamefont {Zhou}},\ and\ \bibinfo {author}
  {\bibfnamefont {T.}~\bibnamefont {Anh~Pham}},\ }\bibfield  {title} {\bibinfo
  {title} {Spectroscopy-guided discovery of three-dimensional structures of
  disordered materials with diffusion models},\ }\href
  {https://doi.org/10.1088/2632-2153/ad8c10} {\bibfield  {journal} {\bibinfo
  {journal} {Machine Learning: Science and Technology}\ }\textbf {\bibinfo
  {volume} {5}},\ \bibinfo {pages} {045037} (\bibinfo {year}
  {2024})}\BibitemShut {NoStop}%
\bibitem [{\citenamefont {Ghose}\ \emph {et~al.}(2023)\citenamefont {Ghose},
  \citenamefont {Segal}, \citenamefont {Meng}, \citenamefont {Liang},
  \citenamefont {Hybertsen}, \citenamefont {Qu}, \citenamefont {Stavitski},
  \citenamefont {Yoo}, \citenamefont {Lu},\ and\ \citenamefont
  {Carbone}}]{ghose2023uncertainty}%
  \BibitemOpen
  \bibfield  {author} {\bibinfo {author} {\bibfnamefont {A.}~\bibnamefont
  {Ghose}}, \bibinfo {author} {\bibfnamefont {M.}~\bibnamefont {Segal}},
  \bibinfo {author} {\bibfnamefont {F.}~\bibnamefont {Meng}}, \bibinfo {author}
  {\bibfnamefont {Z.}~\bibnamefont {Liang}}, \bibinfo {author} {\bibfnamefont
  {M.~S.}\ \bibnamefont {Hybertsen}}, \bibinfo {author} {\bibfnamefont
  {X.}~\bibnamefont {Qu}}, \bibinfo {author} {\bibfnamefont {E.}~\bibnamefont
  {Stavitski}}, \bibinfo {author} {\bibfnamefont {S.}~\bibnamefont {Yoo}},
  \bibinfo {author} {\bibfnamefont {D.}~\bibnamefont {Lu}},\ and\ \bibinfo
  {author} {\bibfnamefont {M.~R.}\ \bibnamefont {Carbone}},\ }\bibfield
  {title} {\bibinfo {title} {Uncertainty-aware predictions of molecular x-ray
  absorption spectra using neural network ensembles},\ }\href
  {https://doi.org/10.1103/PhysRevResearch.5.013180} {\bibfield  {journal}
  {\bibinfo  {journal} {Phys. Rev. Res.}\ }\textbf {\bibinfo {volume} {5}},\
  \bibinfo {pages} {013180} (\bibinfo {year} {2023})}\BibitemShut {NoStop}%
\bibitem [{\citenamefont {Rankine}\ \emph {et~al.}(2020)\citenamefont
  {Rankine}, \citenamefont {Madkhali},\ and\ \citenamefont
  {Penfold}}]{rankine2020deep}%
  \BibitemOpen
  \bibfield  {author} {\bibinfo {author} {\bibfnamefont {C.~D.}\ \bibnamefont
  {Rankine}}, \bibinfo {author} {\bibfnamefont {M.~M.}\ \bibnamefont
  {Madkhali}},\ and\ \bibinfo {author} {\bibfnamefont {T.~J.}\ \bibnamefont
  {Penfold}},\ }\bibfield  {title} {\bibinfo {title} {A deep neural network for
  the rapid prediction of x-ray absorption spectra},\ }\href
  {https://pubs.acs.org/doi/10.1021/acs.jpca.0c03723} {\bibfield  {journal}
  {\bibinfo  {journal} {The Journal of Physical Chemistry A}\ }\textbf
  {\bibinfo {volume} {124}},\ \bibinfo {pages} {4263} (\bibinfo {year}
  {2020})}\BibitemShut {NoStop}%
\bibitem [{\citenamefont {Evans}\ \emph {et~al.}(2014)\citenamefont {Evans},
  \citenamefont {Fan}, \citenamefont {Chureemart}, \citenamefont {Ostler},
  \citenamefont {Ellis},\ and\ \citenamefont {Chantrell}}]{Evans2014atomistic}%
  \BibitemOpen
  \bibfield  {author} {\bibinfo {author} {\bibfnamefont {R.~F.~L.}\
  \bibnamefont {Evans}}, \bibinfo {author} {\bibfnamefont {W.~J.}\ \bibnamefont
  {Fan}}, \bibinfo {author} {\bibfnamefont {P.}~\bibnamefont {Chureemart}},
  \bibinfo {author} {\bibfnamefont {T.~A.}\ \bibnamefont {Ostler}}, \bibinfo
  {author} {\bibfnamefont {M.~O.~A.}\ \bibnamefont {Ellis}},\ and\ \bibinfo
  {author} {\bibfnamefont {R.~W.}\ \bibnamefont {Chantrell}},\ }\bibfield
  {title} {\bibinfo {title} {Atomistic spin model simulations of magnetic
  nanomaterials},\ }\href {https://doi.org/10.1088/0953-8984/26/10/103202}
  {\bibfield  {journal} {\bibinfo  {journal} {Journal of Physics: Condensed
  Matter}\ }\textbf {\bibinfo {volume} {26}},\ \bibinfo {pages} {103202}
  (\bibinfo {year} {2014})}\BibitemShut {NoStop}%
\bibitem [{\citenamefont {Toth}\ and\ \citenamefont
  {Lake}(2015)}]{Toth2015linear}%
  \BibitemOpen
  \bibfield  {author} {\bibinfo {author} {\bibfnamefont {S.}~\bibnamefont
  {Toth}}\ and\ \bibinfo {author} {\bibfnamefont {B.}~\bibnamefont {Lake}},\
  }\bibfield  {title} {\bibinfo {title} {Linear spin wave theory for single-{Q}
  incommensurate magnetic structures},\ }\href
  {https://doi.org/10.1088/0953-8984/27/16/166002} {\bibfield  {journal}
  {\bibinfo  {journal} {Journal of Physics: Condensed Matter}\ }\textbf
  {\bibinfo {volume} {27}},\ \bibinfo {pages} {166002} (\bibinfo {year}
  {2015})}\BibitemShut {NoStop}%
\bibitem [{\citenamefont {Nocera}\ and\ \citenamefont
  {Alvarez}(2016)}]{Nocera2016spectral}%
  \BibitemOpen
  \bibfield  {author} {\bibinfo {author} {\bibfnamefont {A.}~\bibnamefont
  {Nocera}}\ and\ \bibinfo {author} {\bibfnamefont {G.}~\bibnamefont
  {Alvarez}},\ }\bibfield  {title} {\bibinfo {title} {Spectral functions with
  the density matrix renormalization group: {K}rylov-space approach for
  correction vectors},\ }\href {https://doi.org/10.1103/PhysRevE.94.053308}
  {\bibfield  {journal} {\bibinfo  {journal} {Phys. Rev. E}\ }\textbf {\bibinfo
  {volume} {94}},\ \bibinfo {pages} {053308} (\bibinfo {year}
  {2016})}\BibitemShut {NoStop}%
\bibitem [{\citenamefont {Dahlbom}\ \emph {et~al.}(2025)\citenamefont
  {Dahlbom}, \citenamefont {Zhang}, \citenamefont {Miles}, \citenamefont
  {Quinn}, \citenamefont {Niraula}, \citenamefont {Thipe}, \citenamefont
  {Wilson}, \citenamefont {Matin}, \citenamefont {Mankad}, \citenamefont
  {Hahn}, \citenamefont {Pajerowski}, \citenamefont {Johnston}, \citenamefont
  {Wang}, \citenamefont {Lane}, \citenamefont {Li}, \citenamefont {Bai},
  \citenamefont {Mourigal}, \citenamefont {Batista},\ and\ \citenamefont
  {Barros}}]{dahlbom2025sunnyjljuliapackagespin}%
  \BibitemOpen
  \bibfield  {author} {\bibinfo {author} {\bibfnamefont {D.}~\bibnamefont
  {Dahlbom}}, \bibinfo {author} {\bibfnamefont {H.}~\bibnamefont {Zhang}},
  \bibinfo {author} {\bibfnamefont {C.}~\bibnamefont {Miles}}, \bibinfo
  {author} {\bibfnamefont {S.}~\bibnamefont {Quinn}}, \bibinfo {author}
  {\bibfnamefont {A.}~\bibnamefont {Niraula}}, \bibinfo {author} {\bibfnamefont
  {B.}~\bibnamefont {Thipe}}, \bibinfo {author} {\bibfnamefont
  {M.}~\bibnamefont {Wilson}}, \bibinfo {author} {\bibfnamefont
  {S.}~\bibnamefont {Matin}}, \bibinfo {author} {\bibfnamefont
  {H.}~\bibnamefont {Mankad}}, \bibinfo {author} {\bibfnamefont
  {S.}~\bibnamefont {Hahn}}, \bibinfo {author} {\bibfnamefont {D.}~\bibnamefont
  {Pajerowski}}, \bibinfo {author} {\bibfnamefont {S.}~\bibnamefont
  {Johnston}}, \bibinfo {author} {\bibfnamefont {Z.}~\bibnamefont {Wang}},
  \bibinfo {author} {\bibfnamefont {H.}~\bibnamefont {Lane}}, \bibinfo {author}
  {\bibfnamefont {Y.~W.}\ \bibnamefont {Li}}, \bibinfo {author} {\bibfnamefont
  {X.}~\bibnamefont {Bai}}, \bibinfo {author} {\bibfnamefont {M.}~\bibnamefont
  {Mourigal}}, \bibinfo {author} {\bibfnamefont {C.~D.}\ \bibnamefont
  {Batista}},\ and\ \bibinfo {author} {\bibfnamefont {K.}~\bibnamefont
  {Barros}},\ }\bibfield  {title} {\bibinfo {title} {Sunny.jl: A julia package
  for spin dynamics},\ }\href {https://arxiv.org/abs/2501.13095} {\bibfield
  {journal} {\bibinfo  {journal} {arXiv:2501.13095}\ } (\bibinfo {year}
  {2025})}\BibitemShut {NoStop}%
\bibitem [{\citenamefont {Samarakoon}\ \emph {et~al.}(2020)\citenamefont
  {Samarakoon}, \citenamefont {Barros}, \citenamefont {Li}, \citenamefont
  {Eisenbach}, \citenamefont {Zhang}, \citenamefont {Ye}, \citenamefont
  {Sharma}, \citenamefont {Dun}, \citenamefont {Zhou}, \citenamefont {Grigera},
  \citenamefont {Batista},\ and\ \citenamefont
  {Tennant}}]{Samarakoon2020machine}%
  \BibitemOpen
  \bibfield  {author} {\bibinfo {author} {\bibfnamefont {A.~M.}\ \bibnamefont
  {Samarakoon}}, \bibinfo {author} {\bibfnamefont {K.}~\bibnamefont {Barros}},
  \bibinfo {author} {\bibfnamefont {Y.~W.}\ \bibnamefont {Li}}, \bibinfo
  {author} {\bibfnamefont {M.}~\bibnamefont {Eisenbach}}, \bibinfo {author}
  {\bibfnamefont {Q.}~\bibnamefont {Zhang}}, \bibinfo {author} {\bibfnamefont
  {F.}~\bibnamefont {Ye}}, \bibinfo {author} {\bibfnamefont {V.}~\bibnamefont
  {Sharma}}, \bibinfo {author} {\bibfnamefont {Z.~L.}\ \bibnamefont {Dun}},
  \bibinfo {author} {\bibfnamefont {H.}~\bibnamefont {Zhou}}, \bibinfo {author}
  {\bibfnamefont {S.~A.}\ \bibnamefont {Grigera}}, \bibinfo {author}
  {\bibfnamefont {C.~D.}\ \bibnamefont {Batista}},\ and\ \bibinfo {author}
  {\bibfnamefont {D.~A.}\ \bibnamefont {Tennant}},\ }\bibfield  {title}
  {\bibinfo {title} {Machine-learning-assisted insight into spin ice
  {Dy$_2$Ti$_2$O$_7$}},\ }\href {https://doi.org/10.1038/s41467-020-14660-y}
  {\bibfield  {journal} {\bibinfo  {journal} {Nature Communications}\ }\textbf
  {\bibinfo {volume} {11}},\ \bibinfo {pages} {892} (\bibinfo {year}
  {2020})}\BibitemShut {NoStop}%
\bibitem [{\citenamefont {Samarakoon}\ and\ \citenamefont
  {Tennant}(2021)}]{Samarakoon2021machine}%
  \BibitemOpen
  \bibfield  {author} {\bibinfo {author} {\bibfnamefont {A.~M.}\ \bibnamefont
  {Samarakoon}}\ and\ \bibinfo {author} {\bibfnamefont {D.~A.}\ \bibnamefont
  {Tennant}},\ }\bibfield  {title} {\bibinfo {title} {Machine learning for
  magnetic phase diagrams and inverse scattering problems},\ }\href
  {https://doi.org/10.1088/1361-648X/abe818} {\bibfield  {journal} {\bibinfo
  {journal} {Journal of Physics: Condensed Matter}\ }\textbf {\bibinfo {volume}
  {34}},\ \bibinfo {pages} {044002} (\bibinfo {year} {2021})}\BibitemShut
  {NoStop}%
\bibitem [{\citenamefont {Samarakoon}\ \emph {et~al.}(2022)\citenamefont
  {Samarakoon}, \citenamefont {Tennant}, \citenamefont {Ye}, \citenamefont
  {Zhang},\ and\ \citenamefont {Grigera}}]{Samarakoon2022integration}%
  \BibitemOpen
  \bibfield  {author} {\bibinfo {author} {\bibfnamefont {A.}~\bibnamefont
  {Samarakoon}}, \bibinfo {author} {\bibfnamefont {D.~A.}\ \bibnamefont
  {Tennant}}, \bibinfo {author} {\bibfnamefont {F.}~\bibnamefont {Ye}},
  \bibinfo {author} {\bibfnamefont {Q.}~\bibnamefont {Zhang}},\ and\ \bibinfo
  {author} {\bibfnamefont {S.~A.}\ \bibnamefont {Grigera}},\ }\bibfield
  {title} {\bibinfo {title} {Integration of machine learning with neutron
  scattering for the hamiltonian tuning of spin ice under pressure},\ }\href
  {https://doi.org/10.1038/s43246-022-00306-7} {\bibfield  {journal} {\bibinfo
  {journal} {Communications Materials}\ }\textbf {\bibinfo {volume} {3}},\
  \bibinfo {pages} {84} (\bibinfo {year} {2022})}\BibitemShut {NoStop}%
\bibitem [{\citenamefont {L\"uder}(2021)}]{Luder2021}%
  \BibitemOpen
  \bibfield  {author} {\bibinfo {author} {\bibfnamefont {J.}~\bibnamefont
  {L\"uder}},\ }\bibfield  {title} {\bibinfo {title} {Determining electronic
  properties from {$L$}-edge x-ray absorption spectra of transition metal
  compounds with artificial neural networks},\ }\href
  {https://doi.org/10.1103/PhysRevB.103.045140} {\bibfield  {journal} {\bibinfo
   {journal} {Phys. Rev. B}\ }\textbf {\bibinfo {volume} {103}},\ \bibinfo
  {pages} {045140} (\bibinfo {year} {2021})}\BibitemShut {NoStop}%
\bibitem [{\citenamefont {L\"uder}(2025)}]{Luder2025ML}%
  \BibitemOpen
  \bibfield  {author} {\bibinfo {author} {\bibfnamefont {J.}~\bibnamefont
  {L\"uder}},\ }\bibfield  {title} {\bibinfo {title} {Machine learning approach
  to predict {$L$}-edge x-ray absorption spectra of light transition metal ion
  compounds},\ }\href {https://doi.org/10.1103/PhysRevB.111.085110} {\bibfield
  {journal} {\bibinfo  {journal} {Phys. Rev. B}\ }\textbf {\bibinfo {volume}
  {111}},\ \bibinfo {pages} {085110} (\bibinfo {year} {2025})}\BibitemShut
  {NoStop}%
\bibitem [{\citenamefont {Zhang}\ \emph {et~al.}(2023)\citenamefont {Zhang},
  \citenamefont {Xie}, \citenamefont {Long}, \citenamefont {G{\"u}nzing},
  \citenamefont {Wende}, \citenamefont {Ollefs},\ and\ \citenamefont
  {Zhang}}]{zhang2023autonomous}%
  \BibitemOpen
  \bibfield  {author} {\bibinfo {author} {\bibfnamefont {Y.}~\bibnamefont
  {Zhang}}, \bibinfo {author} {\bibfnamefont {R.}~\bibnamefont {Xie}}, \bibinfo
  {author} {\bibfnamefont {T.}~\bibnamefont {Long}}, \bibinfo {author}
  {\bibfnamefont {D.}~\bibnamefont {G{\"u}nzing}}, \bibinfo {author}
  {\bibfnamefont {H.}~\bibnamefont {Wende}}, \bibinfo {author} {\bibfnamefont
  {K.~J.}\ \bibnamefont {Ollefs}},\ and\ \bibinfo {author} {\bibfnamefont
  {H.}~\bibnamefont {Zhang}},\ }\bibfield  {title} {\bibinfo {title}
  {Autonomous atomic {H}amiltonian construction and active sampling of x-ray
  absorption spectroscopy by adversarial {B}ayesian optimization},\ }\href
  {https://www.nature.com/articles/s41524-023-00994-w} {\bibfield  {journal}
  {\bibinfo  {journal} {npj Computational Materials}\ }\textbf {\bibinfo
  {volume} {9}},\ \bibinfo {pages} {46} (\bibinfo {year} {2023})}\BibitemShut
  {NoStop}%
\bibitem [{\citenamefont {Cakir}\ \emph {et~al.}(2024)\citenamefont {Cakir},
  \citenamefont {Bogoclu}, \citenamefont {Emmerling}, \citenamefont {Streli},
  \citenamefont {Guilherme~Buzanich},\ and\ \citenamefont
  {Radtke}}]{Cakir2024}%
  \BibitemOpen
  \bibfield  {author} {\bibinfo {author} {\bibfnamefont {C.~T.}\ \bibnamefont
  {Cakir}}, \bibinfo {author} {\bibfnamefont {C.}~\bibnamefont {Bogoclu}},
  \bibinfo {author} {\bibfnamefont {F.}~\bibnamefont {Emmerling}}, \bibinfo
  {author} {\bibfnamefont {C.}~\bibnamefont {Streli}}, \bibinfo {author}
  {\bibfnamefont {A.}~\bibnamefont {Guilherme~Buzanich}},\ and\ \bibinfo
  {author} {\bibfnamefont {M.}~\bibnamefont {Radtke}},\ }\bibfield  {title}
  {\bibinfo {title} {Machine learning for efficient grazing-exit x-ray
  absorption near edge structure spectroscopy analysis: {B}ayesian optimization
  approach},\ }\href {https://doi.org/10.1088/2632-2153/ad4253} {\bibfield
  {journal} {\bibinfo  {journal} {Machine Learning: Science and Technology}\
  }\textbf {\bibinfo {volume} {5}},\ \bibinfo {pages} {025037} (\bibinfo {year}
  {2024})}\BibitemShut {NoStop}%
\bibitem [{\citenamefont {Du}\ \emph {et~al.}(2025)\citenamefont {Du},
  \citenamefont {Wolfman}, \citenamefont {Sun}, \citenamefont {Kelly},\ and\
  \citenamefont {Cherukara}}]{Du2025}%
  \BibitemOpen
  \bibfield  {author} {\bibinfo {author} {\bibfnamefont {M.}~\bibnamefont
  {Du}}, \bibinfo {author} {\bibfnamefont {M.}~\bibnamefont {Wolfman}},
  \bibinfo {author} {\bibfnamefont {C.}~\bibnamefont {Sun}}, \bibinfo {author}
  {\bibfnamefont {S.~D.}\ \bibnamefont {Kelly}},\ and\ \bibinfo {author}
  {\bibfnamefont {M.~J.}\ \bibnamefont {Cherukara}},\ }\bibfield  {title}
  {\bibinfo {title} {Demonstration of an {AI}-driven workflow for dynamic x-ray
  spectroscopy},\ }\href {https://arxiv.org/abs/2504.17124} {\bibfield
  {journal} {\bibinfo  {journal} {arXiv:2504.17124}\ } (\bibinfo {year}
  {2025})}\BibitemShut {NoStop}%
\bibitem [{\citenamefont {{de Groot}}\ \emph {et~al.}(2024)\citenamefont {{de
  Groot}}, \citenamefont {Haverkort}, \citenamefont {Elnaggar}, \citenamefont
  {Juhin}, \citenamefont {Zhou},\ and\ \citenamefont {Glatzel}}]{deGroot2024}%
  \BibitemOpen
  \bibfield  {author} {\bibinfo {author} {\bibfnamefont {F.~M.~F.}\
  \bibnamefont {{de Groot}}}, \bibinfo {author} {\bibfnamefont {M.~W.}\
  \bibnamefont {Haverkort}}, \bibinfo {author} {\bibfnamefont {H.}~\bibnamefont
  {Elnaggar}}, \bibinfo {author} {\bibfnamefont {A.}~\bibnamefont {Juhin}},
  \bibinfo {author} {\bibfnamefont {K.-J.}\ \bibnamefont {Zhou}},\ and\
  \bibinfo {author} {\bibfnamefont {P.}~\bibnamefont {Glatzel}},\ }\bibfield
  {title} {\bibinfo {title} {Resonant inelastic x-ray scattering},\ }\href
  {https://doi.org/10.1038/s43586-024-00322-6} {\bibfield  {journal} {\bibinfo
  {journal} {Nature Reviews Methods Primers}\ }\textbf {\bibinfo {volume}
  {4}},\ \bibinfo {pages} {45} (\bibinfo {year} {2024})}\BibitemShut {NoStop}%
\bibitem [{\citenamefont {Kotani}\ and\ \citenamefont
  {Shin}(2001)}]{Kotani2001}%
  \BibitemOpen
  \bibfield  {author} {\bibinfo {author} {\bibfnamefont {A.}~\bibnamefont
  {Kotani}}\ and\ \bibinfo {author} {\bibfnamefont {S.}~\bibnamefont {Shin}},\
  }\bibfield  {title} {\bibinfo {title} {Resonant inelastic x-ray scattering
  spectra for electrons in solids},\ }\href
  {https://doi.org/10.1103/RevModPhys.73.203} {\bibfield  {journal} {\bibinfo
  {journal} {Rev. Mod. Phys.}\ }\textbf {\bibinfo {volume} {73}},\ \bibinfo
  {pages} {203} (\bibinfo {year} {2001})}\BibitemShut {NoStop}%
\bibitem [{\citenamefont {{van den Brink}}(2007)}]{vandenBrink2007}%
  \BibitemOpen
  \bibfield  {author} {\bibinfo {author} {\bibfnamefont {J.}~\bibnamefont {{van
  den Brink}}},\ }\bibfield  {title} {\bibinfo {title} {{The theory of indirect
  resonant inelastic X-ray scattering on magnons}},\ }\href
  {https://doi.org/10.1209/0295-5075/80/47003} {\bibfield  {journal} {\bibinfo
  {journal} {EPL (Europhysics Letters)}\ }\textbf {\bibinfo {volume} {80}},\
  \bibinfo {pages} {47003} (\bibinfo {year} {2007})}\BibitemShut {NoStop}%
\bibitem [{\citenamefont {Haverkort}(2010)}]{Haverkort2010}%
  \BibitemOpen
  \bibfield  {author} {\bibinfo {author} {\bibfnamefont {M.~W.}\ \bibnamefont
  {Haverkort}},\ }\bibfield  {title} {\bibinfo {title} {Theory of resonant
  inelastic x-ray scattering by collective magnetic excitations},\ }\href
  {https://doi.org/10.1103/PhysRevLett.105.167404} {\bibfield  {journal}
  {\bibinfo  {journal} {Phys. Rev. Lett.}\ }\textbf {\bibinfo {volume} {105}},\
  \bibinfo {pages} {167404} (\bibinfo {year} {2010})}\BibitemShut {NoStop}%
\bibitem [{\citenamefont {Ament}\ \emph {et~al.}(2011)\citenamefont {Ament},
  \citenamefont {van Veenendaal}, \citenamefont {Devereaux}, \citenamefont
  {Hill},\ and\ \citenamefont {van~den Brink}}]{Ament2011resonant}%
  \BibitemOpen
  \bibfield  {author} {\bibinfo {author} {\bibfnamefont {L.~J.~P.}\
  \bibnamefont {Ament}}, \bibinfo {author} {\bibfnamefont {M.}~\bibnamefont
  {van Veenendaal}}, \bibinfo {author} {\bibfnamefont {T.~P.}\ \bibnamefont
  {Devereaux}}, \bibinfo {author} {\bibfnamefont {J.~P.}\ \bibnamefont
  {Hill}},\ and\ \bibinfo {author} {\bibfnamefont {J.}~\bibnamefont {van~den
  Brink}},\ }\bibfield  {title} {\bibinfo {title} {Resonant inelastic x-ray
  scattering studies of elementary excitations},\ }\href
  {https://doi.org/10.1103/RevModPhys.83.705} {\bibfield  {journal} {\bibinfo
  {journal} {Rev. Mod. Phys.}\ }\textbf {\bibinfo {volume} {83}},\ \bibinfo
  {pages} {705} (\bibinfo {year} {2011})}\BibitemShut {NoStop}%
\bibitem [{\citenamefont {Chen}\ \emph {et~al.}(2019)\citenamefont {Chen},
  \citenamefont {Wang}, \citenamefont {Jia}, \citenamefont {Moritz},
  \citenamefont {Shvaika}, \citenamefont {Freericks},\ and\ \citenamefont
  {Devereaux}}]{Chen2019}%
  \BibitemOpen
  \bibfield  {author} {\bibinfo {author} {\bibfnamefont {Y.}~\bibnamefont
  {Chen}}, \bibinfo {author} {\bibfnamefont {Y.}~\bibnamefont {Wang}}, \bibinfo
  {author} {\bibfnamefont {C.}~\bibnamefont {Jia}}, \bibinfo {author}
  {\bibfnamefont {B.}~\bibnamefont {Moritz}}, \bibinfo {author} {\bibfnamefont
  {A.~M.}\ \bibnamefont {Shvaika}}, \bibinfo {author} {\bibfnamefont {J.~K.}\
  \bibnamefont {Freericks}},\ and\ \bibinfo {author} {\bibfnamefont {T.~P.}\
  \bibnamefont {Devereaux}},\ }\bibfield  {title} {\bibinfo {title} {Theory for
  time-resolved resonant inelastic x-ray scattering},\ }\href
  {https://doi.org/10.1103/PhysRevB.99.104306} {\bibfield  {journal} {\bibinfo
  {journal} {Phys. Rev. B}\ }\textbf {\bibinfo {volume} {99}},\ \bibinfo
  {pages} {104306} (\bibinfo {year} {2019})}\BibitemShut {NoStop}%
\bibitem [{\citenamefont {Vorwerk}\ \emph {et~al.}(2022)\citenamefont
  {Vorwerk}, \citenamefont {Sottile},\ and\ \citenamefont
  {Draxl}}]{Vorwerk2022}%
  \BibitemOpen
  \bibfield  {author} {\bibinfo {author} {\bibfnamefont {C.}~\bibnamefont
  {Vorwerk}}, \bibinfo {author} {\bibfnamefont {F.}~\bibnamefont {Sottile}},\
  and\ \bibinfo {author} {\bibfnamefont {C.}~\bibnamefont {Draxl}},\ }\bibfield
   {title} {\bibinfo {title} {All-electron many-body approach to resonant
  inelastic x-ray scattering},\ }\href {https://doi.org/10.1039/D2CP00994C}
  {\bibfield  {journal} {\bibinfo  {journal} {Phys. Chem. Chem. Phys.}\
  }\textbf {\bibinfo {volume} {24}},\ \bibinfo {pages} {17439} (\bibinfo {year}
  {2022})}\BibitemShut {NoStop}%
\bibitem [{\citenamefont {Cao}\ \emph {et~al.}(2019)\citenamefont {Cao},
  \citenamefont {Mazzone}, \citenamefont {Meyers}, \citenamefont {Hill},
  \citenamefont {Liu}, \citenamefont {Wall},\ and\ \citenamefont
  {Dean}}]{Cao2019ultrafast}%
  \BibitemOpen
  \bibfield  {author} {\bibinfo {author} {\bibfnamefont {Y.}~\bibnamefont
  {Cao}}, \bibinfo {author} {\bibfnamefont {D.~G.}\ \bibnamefont {Mazzone}},
  \bibinfo {author} {\bibfnamefont {D.}~\bibnamefont {Meyers}}, \bibinfo
  {author} {\bibfnamefont {J.~P.}\ \bibnamefont {Hill}}, \bibinfo {author}
  {\bibfnamefont {X.}~\bibnamefont {Liu}}, \bibinfo {author} {\bibfnamefont
  {S.}~\bibnamefont {Wall}},\ and\ \bibinfo {author} {\bibfnamefont {M.~P.~M.}\
  \bibnamefont {Dean}},\ }\bibfield  {title} {\bibinfo {title} {Ultrafast
  dynamics of spin and orbital correlations in quantum materials: {An} energy-
  and momentum-resolved perspective},\ }\href
  {https://doi.org/10.1098/rsta.2017.0480} {\bibfield  {journal} {\bibinfo
  {journal} {Philos. Trans. R. Soc. A}\ }\textbf {\bibinfo {volume} {377}},\
  \bibinfo {pages} {20170480} (\bibinfo {year} {2019})}\BibitemShut {NoStop}%
\bibitem [{\citenamefont {Mitrano}\ and\ \citenamefont
  {Wang}(2020)}]{Mitrano2020probing}%
  \BibitemOpen
  \bibfield  {author} {\bibinfo {author} {\bibfnamefont {M.}~\bibnamefont
  {Mitrano}}\ and\ \bibinfo {author} {\bibfnamefont {Y.}~\bibnamefont {Wang}},\
  }\bibfield  {title} {\bibinfo {title} {Probing light-driven quantum materials
  with ultrafast resonant inelastic x-ray scattering},\ }\href
  {https://doi.org/10.1038/s42005-020-00447-6} {\bibfield  {journal} {\bibinfo
  {journal} {Comm. Phys.}\ }\textbf {\bibinfo {volume} {3}},\ \bibinfo {pages}
  {184} (\bibinfo {year} {2020})}\BibitemShut {NoStop}%
\bibitem [{\citenamefont {Wang}\ \emph {et~al.}(2019)\citenamefont {Wang},
  \citenamefont {Fabbris}, \citenamefont {Dean},\ and\ \citenamefont
  {Kotliar}}]{Wang2019EDRIXS}%
  \BibitemOpen
  \bibfield  {author} {\bibinfo {author} {\bibfnamefont {Y.}~\bibnamefont
  {Wang}}, \bibinfo {author} {\bibfnamefont {G.}~\bibnamefont {Fabbris}},
  \bibinfo {author} {\bibfnamefont {M.}~\bibnamefont {Dean}},\ and\ \bibinfo
  {author} {\bibfnamefont {G.}~\bibnamefont {Kotliar}},\ }\bibfield  {title}
  {\bibinfo {title} {{EDRIXS}: An open source toolkit for simulating spectra of
  resonant inelastic x-ray scattering},\ }\href
  {https://doi.org/https://doi.org/10.1016/j.cpc.2019.04.018} {\bibfield
  {journal} {\bibinfo  {journal} {Computer Physics Communications}\ }\textbf
  {\bibinfo {volume} {243}},\ \bibinfo {pages} {151} (\bibinfo {year}
  {2019})}\BibitemShut {NoStop}%
\bibitem [{\citenamefont {Zimmermann}\ \emph {et~al.}(2018)\citenamefont
  {Zimmermann}, \citenamefont {Green}, \citenamefont {Haverkort},\ and\
  \citenamefont {de~Groot}}]{Zimmermann:ve5082}%
  \BibitemOpen
  \bibfield  {author} {\bibinfo {author} {\bibfnamefont {P.}~\bibnamefont
  {Zimmermann}}, \bibinfo {author} {\bibfnamefont {R.~J.}\ \bibnamefont
  {Green}}, \bibinfo {author} {\bibfnamefont {M.~W.}\ \bibnamefont
  {Haverkort}},\ and\ \bibinfo {author} {\bibfnamefont {F.~M.~F.}\ \bibnamefont
  {de~Groot}},\ }\bibfield  {title} {\bibinfo {title} {{{\it Quanty4RIXS}: a
  program for crystal field multiplet calculations of RIXS and RIXS{--}MCD
  spectra using {\it Quanty}}},\ }\href
  {https://doi.org/10.1107/S1600577518004058} {\bibfield  {journal} {\bibinfo
  {journal} {Journal of Synchrotron Radiation}\ }\textbf {\bibinfo {volume}
  {25}},\ \bibinfo {pages} {899} (\bibinfo {year} {2018})}\BibitemShut
  {NoStop}%
\bibitem [{\citenamefont {Roychoudhury}\ and\ \citenamefont
  {Prendergast}(2022)}]{Roychoudhury2022}%
  \BibitemOpen
  \bibfield  {author} {\bibinfo {author} {\bibfnamefont {S.}~\bibnamefont
  {Roychoudhury}}\ and\ \bibinfo {author} {\bibfnamefont {D.}~\bibnamefont
  {Prendergast}},\ }\bibfield  {title} {\bibinfo {title} {{CleaRIXS}: A fast
  and accurate first-principles method for simulation and analysis of resonant
  inelastic x-ray scattering},\ }\href
  {https://doi.org/10.1103/PhysRevB.106.115115} {\bibfield  {journal} {\bibinfo
   {journal} {Phys. Rev. B}\ }\textbf {\bibinfo {volume} {106}},\ \bibinfo
  {pages} {115115} (\bibinfo {year} {2022})}\BibitemShut {NoStop}%
\bibitem [{\citenamefont {{Li}}\ \emph {et~al.}(2023)\citenamefont {{Li}},
  \citenamefont {{Gu}}, \citenamefont {{Takahashi}}, \citenamefont {{Higashi}},
  \citenamefont {{Kim}}, \citenamefont {{Cheng}}, \citenamefont {{Yang}},
  \citenamefont {{Kune{\v{s}}}}, \citenamefont {{Pelliciari}}, \citenamefont
  {{Hariki}},\ and\ \citenamefont {{Bisogni}}}]{JieminLi2023Fe2O3}%
  \BibitemOpen
  \bibfield  {author} {\bibinfo {author} {\bibfnamefont {J.}~\bibnamefont
  {{Li}}}, \bibinfo {author} {\bibfnamefont {Y.}~\bibnamefont {{Gu}}}, \bibinfo
  {author} {\bibfnamefont {Y.}~\bibnamefont {{Takahashi}}}, \bibinfo {author}
  {\bibfnamefont {K.}~\bibnamefont {{Higashi}}}, \bibinfo {author}
  {\bibfnamefont {T.}~\bibnamefont {{Kim}}}, \bibinfo {author} {\bibfnamefont
  {Y.}~\bibnamefont {{Cheng}}}, \bibinfo {author} {\bibfnamefont
  {F.}~\bibnamefont {{Yang}}}, \bibinfo {author} {\bibfnamefont
  {J.}~\bibnamefont {{Kune{\v{s}}}}}, \bibinfo {author} {\bibfnamefont
  {J.}~\bibnamefont {{Pelliciari}}}, \bibinfo {author} {\bibfnamefont
  {A.}~\bibnamefont {{Hariki}}},\ and\ \bibinfo {author} {\bibfnamefont
  {V.}~\bibnamefont {{Bisogni}}},\ }\bibfield  {title} {\bibinfo {title}
  {{Single- and Multimagnon Dynamics in Antiferromagnetic {\ensuremath{\alpha}}
  -Fe$_{2}$O$_{3}$ Thin Films}},\ }\href
  {https://doi.org/10.1103/PhysRevX.13.011012} {\bibfield  {journal} {\bibinfo
  {journal} {Physical Review X}\ }\textbf {\bibinfo {volume} {13}},\ \bibinfo
  {eid} {011012} (\bibinfo {year} {2023})}\BibitemShut {NoStop}%
\bibitem [{\citenamefont {Taylor}\ \emph {et~al.}(2016)\citenamefont {Taylor},
  \citenamefont {Calder}, \citenamefont {Morrow}, \citenamefont {Feng},
  \citenamefont {Upton}, \citenamefont {Lumsden}, \citenamefont {Yamaura},
  \citenamefont {Woodward}, \citenamefont {Christianson},\ and\ \citenamefont
  {Christianson}}]{Taylor2016SpinOrbitCC}%
  \BibitemOpen
  \bibfield  {author} {\bibinfo {author} {\bibfnamefont {A.~E.}\ \bibnamefont
  {Taylor}}, \bibinfo {author} {\bibfnamefont {S.}~\bibnamefont {Calder}},
  \bibinfo {author} {\bibfnamefont {R.}~\bibnamefont {Morrow}}, \bibinfo
  {author} {\bibfnamefont {H.~L.}\ \bibnamefont {Feng}}, \bibinfo {author}
  {\bibfnamefont {M.}~\bibnamefont {Upton}}, \bibinfo {author} {\bibfnamefont
  {M.~D.}\ \bibnamefont {Lumsden}}, \bibinfo {author} {\bibfnamefont
  {K.}~\bibnamefont {Yamaura}}, \bibinfo {author} {\bibfnamefont {P.~M.}\
  \bibnamefont {Woodward}}, \bibinfo {author} {\bibfnamefont {A.~D.}\
  \bibnamefont {Christianson}},\ and\ \bibinfo {author} {\bibfnamefont {A.~D.}\
  \bibnamefont {Christianson}},\ }\bibfield  {title} {\bibinfo {title}
  {Spin-orbit coupling controlled {$J=3/2$} electronic ground state in $5d^{3}$
  oxides.},\ }\href {https://api.semanticscholar.org/CorpusID:42304916}
  {\bibfield  {journal} {\bibinfo  {journal} {Physical review letters}\
  }\textbf {\bibinfo {volume} {118 20}},\ \bibinfo {pages} {207202} (\bibinfo
  {year} {2016})}\BibitemShut {NoStop}%
\bibitem [{EDR(2025)}]{EDRIXS}%
  \BibitemOpen
  \href@noop {} {\bibinfo {title} {{EDRIXS} website}},\ \bibinfo {howpublished}
  {\url{https://github.com/NSLS-II/edrixs}} (\bibinfo {year} {2025}),\ \bibinfo
  {note} {accessed: 2025-06-01}\BibitemShut {NoStop}%
\bibitem [{\citenamefont {de~Groot}\ and\ \citenamefont
  {Kotani}(2008)}]{Groot2008core}%
  \BibitemOpen
  \bibfield  {author} {\bibinfo {author} {\bibfnamefont {F.}~\bibnamefont
  {de~Groot}}\ and\ \bibinfo {author} {\bibfnamefont {A.}~\bibnamefont
  {Kotani}},\ }\href@noop {} {\emph {\bibinfo {title} {Core Level Spectroscopy
  of Solids}}}\ (\bibinfo  {publisher} {CRC Press},\ \bibinfo {year}
  {2008})\BibitemShut {NoStop}%
\bibitem [{sup()}]{supplement}%
  \BibitemOpen
  \href@noop {} {}\bibinfo {note} {See the online supplementary materials,
  available at: \url{URL supplied by the publisher}, which also includes
  Ref.~\cite{Dresselhaus2008GroupTheory}.}\BibitemShut {Stop}%
\bibitem [{\citenamefont {Frazier}(2018)}]{Frazier2018tutorial}%
  \BibitemOpen
  \bibfield  {author} {\bibinfo {author} {\bibfnamefont {P.~I.}\ \bibnamefont
  {Frazier}},\ }\bibfield  {title} {\bibinfo {title} {A tutorial on {Bayesian}
  optimization},\ }\href {https://arxiv.org/abs/1807.02811} {\bibfield
  {journal} {\bibinfo  {journal} {arXiv:1807.02811}\ } (\bibinfo {year}
  {2018})}\BibitemShut {NoStop}%
\bibitem [{\citenamefont {{Wang}}(2023)}]{Wang2020GPRIntuitiveTutorial}%
  \BibitemOpen
  \bibfield  {author} {\bibinfo {author} {\bibfnamefont {J.}~\bibnamefont
  {{Wang}}},\ }\bibfield  {title} {\bibinfo {title} {{An Intuitive Tutorial to
  Gaussian Process Regression}},\ }\href
  {https://doi.org/10.1109/MCSE.2023.3342149} {\bibfield  {journal} {\bibinfo
  {journal} {Computing in Science and Engineering}\ }\textbf {\bibinfo {volume}
  {25}},\ \bibinfo {pages} {4} (\bibinfo {year} {2023})}\BibitemShut {NoStop}%
\bibitem [{\citenamefont {Rasmussen}\ and\ \citenamefont
  {Williams}(2006)}]{Rasmussen_Williams_2006}%
  \BibitemOpen
  \bibfield  {author} {\bibinfo {author} {\bibfnamefont {C.~E.}\ \bibnamefont
  {Rasmussen}}\ and\ \bibinfo {author} {\bibfnamefont {C.~K.~I.}\ \bibnamefont
  {Williams}},\ }\href@noop {} {\emph {\bibinfo {title} {Gaussian process for
  machine learning}}}\ (\bibinfo  {publisher} {The MIT Press},\ \bibinfo {year}
  {2006})\BibitemShut {NoStop}%
\bibitem [{\citenamefont {Kaufmann}\ \emph {et~al.}(2012)\citenamefont
  {Kaufmann}, \citenamefont {Cappe},\ and\ \citenamefont
  {Garivier}}]{Kaufmann2012Bayesian}%
  \BibitemOpen
  \bibfield  {author} {\bibinfo {author} {\bibfnamefont {E.}~\bibnamefont
  {Kaufmann}}, \bibinfo {author} {\bibfnamefont {O.}~\bibnamefont {Cappe}},\
  and\ \bibinfo {author} {\bibfnamefont {A.}~\bibnamefont {Garivier}},\
  }\bibfield  {title} {\bibinfo {title} {On {B}ayesian upper confidence bounds
  for bandit problems},\ }in\ \href
  {https://proceedings.mlr.press/v22/kaufmann12.html} {\emph {\bibinfo
  {booktitle} {Proceedings of the Fifteenth International Conference on
  Artificial Intelligence and Statistics}}},\ \bibinfo {series} {Proceedings of
  Machine Learning Research}, Vol.~\bibinfo {volume} {22},\ \bibinfo {editor}
  {edited by\ \bibinfo {editor} {\bibfnamefont {N.~D.}\ \bibnamefont
  {Lawrence}}\ and\ \bibinfo {editor} {\bibfnamefont {M.}~\bibnamefont
  {Girolami}}}\ (\bibinfo  {publisher} {PMLR},\ \bibinfo {address} {La Palma,
  Canary Islands},\ \bibinfo {year} {2012})\ pp.\ \bibinfo {pages}
  {592--600}\BibitemShut {NoStop}%
\bibitem [{\citenamefont {Cartis}\ \emph {et~al.}(2019)\citenamefont {Cartis},
  \citenamefont {Fiala}, \citenamefont {Marteau},\ and\ \citenamefont
  {Roberts}}]{Cartis2019improving}%
  \BibitemOpen
  \bibfield  {author} {\bibinfo {author} {\bibfnamefont {C.}~\bibnamefont
  {Cartis}}, \bibinfo {author} {\bibfnamefont {J.}~\bibnamefont {Fiala}},
  \bibinfo {author} {\bibfnamefont {B.}~\bibnamefont {Marteau}},\ and\ \bibinfo
  {author} {\bibfnamefont {L.}~\bibnamefont {Roberts}},\ }\bibfield  {title}
  {\bibinfo {title} {Improving the flexibility and robustness of model-based
  derivative-free optimization solvers},\ }\bibfield  {journal} {\bibinfo
  {journal} {ACM Trans. Math. Softw.}\ }\textbf {\bibinfo {volume} {45}},\
  \href {https://doi.org/10.1145/3338517} {10.1145/3338517} (\bibinfo {year}
  {2019})\BibitemShut {NoStop}%
\bibitem [{\citenamefont {Wildes}\ \emph {et~al.}(2022)\citenamefont {Wildes},
  \citenamefont {Stewart}, \citenamefont {Le}, \citenamefont {Ewings},
  \citenamefont {Rule}, \citenamefont {Deng},\ and\ \citenamefont
  {Anand}}]{Wildes2022NiPS}%
  \BibitemOpen
  \bibfield  {author} {\bibinfo {author} {\bibfnamefont {A.~R.}\ \bibnamefont
  {Wildes}}, \bibinfo {author} {\bibfnamefont {J.~R.}\ \bibnamefont {Stewart}},
  \bibinfo {author} {\bibfnamefont {M.~D.}\ \bibnamefont {Le}}, \bibinfo
  {author} {\bibfnamefont {R.~A.}\ \bibnamefont {Ewings}}, \bibinfo {author}
  {\bibfnamefont {K.~C.}\ \bibnamefont {Rule}}, \bibinfo {author}
  {\bibfnamefont {G.}~\bibnamefont {Deng}},\ and\ \bibinfo {author}
  {\bibfnamefont {K.}~\bibnamefont {Anand}},\ }\bibfield  {title} {\bibinfo
  {title} {Magnetic dynamics of {${\mathrm{NiPS}}_{3}$}},\ }\href
  {https://doi.org/10.1103/PhysRevB.106.174422} {\bibfield  {journal} {\bibinfo
   {journal} {Phys. Rev. B}\ }\textbf {\bibinfo {volume} {106}},\ \bibinfo
  {pages} {174422} (\bibinfo {year} {2022})}\BibitemShut {NoStop}%
\bibitem [{\citenamefont {Wildes}\ \emph {et~al.}(2015)\citenamefont {Wildes},
  \citenamefont {Simonet}, \citenamefont {Ressouche}, \citenamefont {McIntyre},
  \citenamefont {Avdeev}, \citenamefont {Suard}, \citenamefont {Kimber},
  \citenamefont {Lan\ifmmode~\mbox{\c{c}}\else \c{c}\fi{}on}, \citenamefont
  {Pepe}, \citenamefont {Moubaraki},\ and\ \citenamefont
  {Hicks}}]{Wildes2015NiPS}%
  \BibitemOpen
  \bibfield  {author} {\bibinfo {author} {\bibfnamefont {A.~R.}\ \bibnamefont
  {Wildes}}, \bibinfo {author} {\bibfnamefont {V.}~\bibnamefont {Simonet}},
  \bibinfo {author} {\bibfnamefont {E.}~\bibnamefont {Ressouche}}, \bibinfo
  {author} {\bibfnamefont {G.~J.}\ \bibnamefont {McIntyre}}, \bibinfo {author}
  {\bibfnamefont {M.}~\bibnamefont {Avdeev}}, \bibinfo {author} {\bibfnamefont
  {E.}~\bibnamefont {Suard}}, \bibinfo {author} {\bibfnamefont {S.~A.~J.}\
  \bibnamefont {Kimber}}, \bibinfo {author} {\bibfnamefont {D.}~\bibnamefont
  {Lan\ifmmode~\mbox{\c{c}}\else \c{c}\fi{}on}}, \bibinfo {author}
  {\bibfnamefont {G.}~\bibnamefont {Pepe}}, \bibinfo {author} {\bibfnamefont
  {B.}~\bibnamefont {Moubaraki}},\ and\ \bibinfo {author} {\bibfnamefont
  {T.~J.}\ \bibnamefont {Hicks}},\ }\bibfield  {title} {\bibinfo {title}
  {Magnetic structure of the quasi-two-dimensional antiferromagnet
  {${\text{NiPS}}_{3}$}},\ }\href {https://doi.org/10.1103/PhysRevB.92.224408}
  {\bibfield  {journal} {\bibinfo  {journal} {Phys. Rev. B}\ }\textbf {\bibinfo
  {volume} {92}},\ \bibinfo {pages} {224408} (\bibinfo {year}
  {2015})}\BibitemShut {NoStop}%
\bibitem [{\citenamefont {Kim}\ \emph {et~al.}(2019)\citenamefont {Kim},
  \citenamefont {Lim}, \citenamefont {Lee}, \citenamefont {Lee}, \citenamefont
  {Kim}, \citenamefont {Park}, \citenamefont {Jeon}, \citenamefont {Park},
  \citenamefont {Park},\ and\ \citenamefont {Cheong}}]{Kim2019supression}%
  \BibitemOpen
  \bibfield  {author} {\bibinfo {author} {\bibfnamefont {K.}~\bibnamefont
  {Kim}}, \bibinfo {author} {\bibfnamefont {S.~Y.}\ \bibnamefont {Lim}},
  \bibinfo {author} {\bibfnamefont {J.-U.}\ \bibnamefont {Lee}}, \bibinfo
  {author} {\bibfnamefont {S.}~\bibnamefont {Lee}}, \bibinfo {author}
  {\bibfnamefont {T.~Y.}\ \bibnamefont {Kim}}, \bibinfo {author} {\bibfnamefont
  {K.}~\bibnamefont {Park}}, \bibinfo {author} {\bibfnamefont {G.~S.}\
  \bibnamefont {Jeon}}, \bibinfo {author} {\bibfnamefont {C.-H.}\ \bibnamefont
  {Park}}, \bibinfo {author} {\bibfnamefont {J.-G.}\ \bibnamefont {Park}},\
  and\ \bibinfo {author} {\bibfnamefont {H.}~\bibnamefont {Cheong}},\
  }\bibfield  {title} {\bibinfo {title} {Suppression of magnetic ordering in
  {$XXZ$}-type antiferromagnetic monolayer {NiPS$_3$}},\ }\href
  {https://doi.org/10.1038/s41467-018-08284-6} {\bibfield  {journal} {\bibinfo
  {journal} {Nature Communications}\ }\textbf {\bibinfo {volume} {10}},\
  \bibinfo {pages} {345} (\bibinfo {year} {2019})}\BibitemShut {NoStop}%
\bibitem [{\citenamefont {Kim}\ \emph {et~al.}(2018)\citenamefont {Kim},
  \citenamefont {Kim}, \citenamefont {Sandilands}, \citenamefont {Sinn},
  \citenamefont {Lee}, \citenamefont {Son}, \citenamefont {Lee}, \citenamefont
  {Choi}, \citenamefont {Kim}, \citenamefont {Park}, \citenamefont {Jeon},
  \citenamefont {Kim}, \citenamefont {Park}, \citenamefont {Park},
  \citenamefont {Moon},\ and\ \citenamefont {Noh}}]{Kim2018charge}%
  \BibitemOpen
  \bibfield  {author} {\bibinfo {author} {\bibfnamefont {S.~Y.}\ \bibnamefont
  {Kim}}, \bibinfo {author} {\bibfnamefont {T.~Y.}\ \bibnamefont {Kim}},
  \bibinfo {author} {\bibfnamefont {L.~J.}\ \bibnamefont {Sandilands}},
  \bibinfo {author} {\bibfnamefont {S.}~\bibnamefont {Sinn}}, \bibinfo {author}
  {\bibfnamefont {M.-C.}\ \bibnamefont {Lee}}, \bibinfo {author} {\bibfnamefont
  {J.}~\bibnamefont {Son}}, \bibinfo {author} {\bibfnamefont {S.}~\bibnamefont
  {Lee}}, \bibinfo {author} {\bibfnamefont {K.-Y.}\ \bibnamefont {Choi}},
  \bibinfo {author} {\bibfnamefont {W.}~\bibnamefont {Kim}}, \bibinfo {author}
  {\bibfnamefont {B.-G.}\ \bibnamefont {Park}}, \bibinfo {author}
  {\bibfnamefont {C.}~\bibnamefont {Jeon}}, \bibinfo {author} {\bibfnamefont
  {H.-D.}\ \bibnamefont {Kim}}, \bibinfo {author} {\bibfnamefont {C.-H.}\
  \bibnamefont {Park}}, \bibinfo {author} {\bibfnamefont {J.-G.}\ \bibnamefont
  {Park}}, \bibinfo {author} {\bibfnamefont {S.~J.}\ \bibnamefont {Moon}},\
  and\ \bibinfo {author} {\bibfnamefont {T.~W.}\ \bibnamefont {Noh}},\
  }\bibfield  {title} {\bibinfo {title} {Charge-spin correlation in van der
  {Waals} antiferromagnet {${\mathrm{NiPS}}_{3}$}},\ }\href
  {https://doi.org/10.1103/PhysRevLett.120.136402} {\bibfield  {journal}
  {\bibinfo  {journal} {Phys. Rev. Lett.}\ }\textbf {\bibinfo {volume} {120}},\
  \bibinfo {pages} {136402} (\bibinfo {year} {2018})}\BibitemShut {NoStop}%
\bibitem [{\citenamefont {Caliebe}\ \emph {et~al.}(1998)\citenamefont
  {Caliebe}, \citenamefont {Kao}, \citenamefont {Hastings}, \citenamefont
  {Taguchi}, \citenamefont {Kotani}, \citenamefont {Uozumi},\ and\
  \citenamefont {de~Groot}}]{Caliebe19981s2p}%
  \BibitemOpen
  \bibfield  {author} {\bibinfo {author} {\bibfnamefont {W.~A.}\ \bibnamefont
  {Caliebe}}, \bibinfo {author} {\bibfnamefont {C.-C.}\ \bibnamefont {Kao}},
  \bibinfo {author} {\bibfnamefont {J.~B.}\ \bibnamefont {Hastings}}, \bibinfo
  {author} {\bibfnamefont {M.}~\bibnamefont {Taguchi}}, \bibinfo {author}
  {\bibfnamefont {A.}~\bibnamefont {Kotani}}, \bibinfo {author} {\bibfnamefont
  {T.}~\bibnamefont {Uozumi}},\ and\ \bibinfo {author} {\bibfnamefont
  {F.~M.~F.}\ \bibnamefont {de~Groot}},\ }\bibfield  {title} {\bibinfo {title}
  {$1s2p$ resonant inelastic x-ray scattering in
  {$\ensuremath{\alpha}\ensuremath{-}{\mathrm{Fe}}_{2}{\mathrm{O}}_{3}$}},\
  }\href {https://doi.org/10.1103/PhysRevB.58.13452} {\bibfield  {journal}
  {\bibinfo  {journal} {Phys. Rev. B}\ }\textbf {\bibinfo {volume} {58}},\
  \bibinfo {pages} {13452} (\bibinfo {year} {1998})}\BibitemShut {NoStop}%
\bibitem [{\citenamefont {Frontini}\ \emph {et~al.}(2025)\citenamefont
  {Frontini}, \citenamefont {Heath}, \citenamefont {Yuan}, \citenamefont
  {Thompson}, \citenamefont {Greedan}, \citenamefont {Hauser}, \citenamefont
  {Yang}, \citenamefont {Dean}, \citenamefont {Upton}, \citenamefont {Casa},\
  and\ \citenamefont {Kim}}]{Frontini2025resonant}%
  \BibitemOpen
  \bibfield  {author} {\bibinfo {author} {\bibfnamefont {F.~I.}\ \bibnamefont
  {Frontini}}, \bibinfo {author} {\bibfnamefont {C.~J.~S.}\ \bibnamefont
  {Heath}}, \bibinfo {author} {\bibfnamefont {B.}~\bibnamefont {Yuan}},
  \bibinfo {author} {\bibfnamefont {C.~M.}\ \bibnamefont {Thompson}}, \bibinfo
  {author} {\bibfnamefont {J.}~\bibnamefont {Greedan}}, \bibinfo {author}
  {\bibfnamefont {A.~J.}\ \bibnamefont {Hauser}}, \bibinfo {author}
  {\bibfnamefont {F.~Y.}\ \bibnamefont {Yang}}, \bibinfo {author}
  {\bibfnamefont {M.~P.~M.}\ \bibnamefont {Dean}}, \bibinfo {author}
  {\bibfnamefont {M.~H.}\ \bibnamefont {Upton}}, \bibinfo {author}
  {\bibfnamefont {D.~M.}\ \bibnamefont {Casa}},\ and\ \bibinfo {author}
  {\bibfnamefont {Y.-J.}\ \bibnamefont {Kim}},\ }\bibfield  {title} {\bibinfo
  {title} {Resonant inelastic x-ray scattering investigation of hund's and
  spin-orbit coupling in $5{d}^{2}$ double perovskites},\ }\href
  {https://doi.org/10.1103/k64w-cvsh} {\bibfield  {journal} {\bibinfo
  {journal} {Phys. Rev. B}\ }\textbf {\bibinfo {volume} {112}},\ \bibinfo
  {pages} {165103} (\bibinfo {year} {2025})}\BibitemShut {NoStop}%
\bibitem [{\citenamefont {Zhang}\ \emph {et~al.}(2018)\citenamefont {Zhang},
  \citenamefont {Isola}, \citenamefont {Efros}, \citenamefont {Shechtman},\
  and\ \citenamefont {Wang}}]{zhang2018perceptual}%
  \BibitemOpen
  \bibfield  {author} {\bibinfo {author} {\bibfnamefont {R.}~\bibnamefont
  {Zhang}}, \bibinfo {author} {\bibfnamefont {P.}~\bibnamefont {Isola}},
  \bibinfo {author} {\bibfnamefont {A.~A.}\ \bibnamefont {Efros}}, \bibinfo
  {author} {\bibfnamefont {E.}~\bibnamefont {Shechtman}},\ and\ \bibinfo
  {author} {\bibfnamefont {O.}~\bibnamefont {Wang}},\ }\bibfield  {title}
  {\bibinfo {title} {The unreasonable effectiveness of deep features as a
  perceptual metric},\ }in\ \href
  {https://openaccess.thecvf.com/content_cvpr_2018/html/Zhang_The_Unreasonable_Effectiveness_CVPR_2018_paper.html}
  {\emph {\bibinfo {booktitle} {CVPR}}}\ (\bibinfo {year} {2018})\BibitemShut
  {NoStop}%
\bibitem [{\citenamefont {Diggle}\ and\ \citenamefont
  {Gratton}(1984)}]{Diggle1984}%
  \BibitemOpen
  \bibfield  {author} {\bibinfo {author} {\bibfnamefont {P.~J.}\ \bibnamefont
  {Diggle}}\ and\ \bibinfo {author} {\bibfnamefont {R.~J.}\ \bibnamefont
  {Gratton}},\ }\bibfield  {title} {\bibinfo {title} {Monte {C}arlo methods of
  inference for implicit statistical models},\ }\href
  {http://www.jstor.org/stable/2345504} {\bibfield  {journal} {\bibinfo
  {journal} {Journal of the Royal Statistical Society. Series B
  (Methodological)}\ }\textbf {\bibinfo {volume} {46}},\ \bibinfo {pages} {193}
  (\bibinfo {year} {1984})}\BibitemShut {NoStop}%
\bibitem [{\citenamefont {Kennedy}\ and\ \citenamefont
  {O'Hagan}(2001)}]{Kennedy2001}%
  \BibitemOpen
  \bibfield  {author} {\bibinfo {author} {\bibfnamefont {M.~C.}\ \bibnamefont
  {Kennedy}}\ and\ \bibinfo {author} {\bibfnamefont {A.}~\bibnamefont
  {O'Hagan}},\ }\bibfield  {title} {\bibinfo {title} {Bayesian calibration of
  computer models},\ }\href
  {https://doi.org/https://doi.org/10.1111/1467-9868.00294} {\bibfield
  {journal} {\bibinfo  {journal} {Journal of the Royal Statistical Society:
  Series B (Statistical Methodology)}\ }\textbf {\bibinfo {volume} {63}},\
  \bibinfo {pages} {425} (\bibinfo {year} {2001})}\BibitemShut {NoStop}%
\bibitem [{\citenamefont {Cranmer}\ \emph {et~al.}(2020)\citenamefont
  {Cranmer}, \citenamefont {Brehmer},\ and\ \citenamefont
  {Louppe}}]{Cranmer2020}%
  \BibitemOpen
  \bibfield  {author} {\bibinfo {author} {\bibfnamefont {K.}~\bibnamefont
  {Cranmer}}, \bibinfo {author} {\bibfnamefont {J.}~\bibnamefont {Brehmer}},\
  and\ \bibinfo {author} {\bibfnamefont {G.}~\bibnamefont {Louppe}},\
  }\bibfield  {title} {\bibinfo {title} {The frontier of simulation-based
  inference},\ }\href {https://doi.org/10.1073/pnas.1912789117} {\bibfield
  {journal} {\bibinfo  {journal} {Proceedings of the National Academy of
  Sciences}\ }\textbf {\bibinfo {volume} {117}},\ \bibinfo {pages} {30055}
  (\bibinfo {year} {2020})}\BibitemShut {NoStop}%
\bibitem [{\citenamefont {Tavaré}\ \emph {et~al.}(1997)\citenamefont
  {Tavaré}, \citenamefont {Balding}, \citenamefont {Griffiths},\ and\
  \citenamefont {Donnelly}}]{Tavare1997}%
  \BibitemOpen
  \bibfield  {author} {\bibinfo {author} {\bibfnamefont {S.}~\bibnamefont
  {Tavaré}}, \bibinfo {author} {\bibfnamefont {D.~J.}\ \bibnamefont
  {Balding}}, \bibinfo {author} {\bibfnamefont {R.~C.}\ \bibnamefont
  {Griffiths}},\ and\ \bibinfo {author} {\bibfnamefont {P.}~\bibnamefont
  {Donnelly}},\ }\bibfield  {title} {\bibinfo {title} {Inferring coalescence
  times from dna sequence data},\ }\href
  {https://doi.org/10.1093/genetics/145.2.505} {\bibfield  {journal} {\bibinfo
  {journal} {Genetics}\ }\textbf {\bibinfo {volume} {145}},\ \bibinfo {pages}
  {505} (\bibinfo {year} {1997})}\BibitemShut {NoStop}%
\bibitem [{\citenamefont {Beaumont}\ \emph {et~al.}(2002)\citenamefont
  {Beaumont}, \citenamefont {Zhang},\ and\ \citenamefont
  {Balding}}]{Beaumont2002}%
  \BibitemOpen
  \bibfield  {author} {\bibinfo {author} {\bibfnamefont {M.~A.}\ \bibnamefont
  {Beaumont}}, \bibinfo {author} {\bibfnamefont {W.}~\bibnamefont {Zhang}},\
  and\ \bibinfo {author} {\bibfnamefont {D.~J.}\ \bibnamefont {Balding}},\
  }\bibfield  {title} {\bibinfo {title} {Approximate {B}ayesian computation in
  population genetics},\ }\href {https://doi.org/10.1093/genetics/162.4.2025}
  {\bibfield  {journal} {\bibinfo  {journal} {Genetics}\ }\textbf {\bibinfo
  {volume} {162}},\ \bibinfo {pages} {2025} (\bibinfo {year}
  {2002})}\BibitemShut {NoStop}%
\bibitem [{\citenamefont {Sunn{\aa}ker}\ \emph {et~al.}(2013)\citenamefont
  {Sunn{\aa}ker}, \citenamefont {Busetto}, \citenamefont {Numminen},
  \citenamefont {Corander}, \citenamefont {Foll},\ and\ \citenamefont
  {Dessimoz}}]{Sunnaker2013}%
  \BibitemOpen
  \bibfield  {author} {\bibinfo {author} {\bibfnamefont {M.}~\bibnamefont
  {Sunn{\aa}ker}}, \bibinfo {author} {\bibfnamefont {A.~G.}\ \bibnamefont
  {Busetto}}, \bibinfo {author} {\bibfnamefont {E.}~\bibnamefont {Numminen}},
  \bibinfo {author} {\bibfnamefont {J.}~\bibnamefont {Corander}}, \bibinfo
  {author} {\bibfnamefont {M.}~\bibnamefont {Foll}},\ and\ \bibinfo {author}
  {\bibfnamefont {C.}~\bibnamefont {Dessimoz}},\ }\bibfield  {title} {\bibinfo
  {title} {Approximate {B}ayesian computation},\ }\href
  {https://doi.org/10.1371/journal.pcbi.1002803} {\bibfield  {journal}
  {\bibinfo  {journal} {PLoS Computational Biology}\ }\textbf {\bibinfo
  {volume} {9}},\ \bibinfo {pages} {e1002803} (\bibinfo {year}
  {2013})}\BibitemShut {NoStop}%
\bibitem [{\citenamefont {Wood}(2010)}]{Wood2010}%
  \BibitemOpen
  \bibfield  {author} {\bibinfo {author} {\bibfnamefont {S.~N.}\ \bibnamefont
  {Wood}},\ }\bibfield  {title} {\bibinfo {title} {Statistical inference for
  noisy nonlinear ecological dynamic systems},\ }\href
  {https://doi.org/10.1038/nature09319} {\bibfield  {journal} {\bibinfo
  {journal} {Nature}\ }\textbf {\bibinfo {volume} {466}},\ \bibinfo {pages}
  {1102} (\bibinfo {year} {2010})}\BibitemShut {NoStop}%
\bibitem [{\citenamefont {Gutmann}\ and\ \citenamefont
  {Corander}(2016)}]{Gutmann2016}%
  \BibitemOpen
  \bibfield  {author} {\bibinfo {author} {\bibfnamefont {M.~U.}\ \bibnamefont
  {Gutmann}}\ and\ \bibinfo {author} {\bibfnamefont {J.}~\bibnamefont
  {Corander}},\ }\bibfield  {title} {\bibinfo {title} {Bayesian optimization
  for likelihood-free inference of simulator-based statistical models},\ }\href
  {http://jmlr.org/papers/v17/15-017.html} {\bibfield  {journal} {\bibinfo
  {journal} {Journal of Machine Learning Research}\ }\textbf {\bibinfo {volume}
  {17}},\ \bibinfo {pages} {1} (\bibinfo {year} {2016})}\BibitemShut {NoStop}%
\bibitem [{\citenamefont {Papamakarios}\ \emph {et~al.}(2019)\citenamefont
  {Papamakarios}, \citenamefont {Sterratt},\ and\ \citenamefont
  {Murray}}]{Papamakarios2019}%
  \BibitemOpen
  \bibfield  {author} {\bibinfo {author} {\bibfnamefont {G.}~\bibnamefont
  {Papamakarios}}, \bibinfo {author} {\bibfnamefont {D.}~\bibnamefont
  {Sterratt}},\ and\ \bibinfo {author} {\bibfnamefont {I.}~\bibnamefont
  {Murray}},\ }\bibfield  {title} {\bibinfo {title} {Sequential neural
  likelihood: Fast likelihood-free inference with autoregressive flows},\ }in\
  \href {https://proceedings.mlr.press/v89/papamakarios19a.html} {\emph
  {\bibinfo {booktitle} {Proceedings of the Twenty-Second International
  Conference on Artificial Intelligence and Statistics}}},\ \bibinfo {series}
  {Proceedings of Machine Learning Research}, Vol.~\bibinfo {volume} {89},\
  \bibinfo {editor} {edited by\ \bibinfo {editor} {\bibfnamefont
  {K.}~\bibnamefont {Chaudhuri}}\ and\ \bibinfo {editor} {\bibfnamefont
  {M.}~\bibnamefont {Sugiyama}}}\ (\bibinfo  {publisher} {PMLR},\ \bibinfo
  {year} {2019})\ pp.\ \bibinfo {pages} {837--848}\BibitemShut {NoStop}%
\bibitem [{\citenamefont {Lindley}(1956)}]{Lindley1956}%
  \BibitemOpen
  \bibfield  {author} {\bibinfo {author} {\bibfnamefont {D.~V.}\ \bibnamefont
  {Lindley}},\ }\bibfield  {title} {\bibinfo {title} {{On a Measure of the
  Information Provided by an Experiment}},\ }\href
  {https://doi.org/10.1214/aoms/1177728069} {\bibfield  {journal} {\bibinfo
  {journal} {The Annals of Mathematical Statistics}\ }\textbf {\bibinfo
  {volume} {27}},\ \bibinfo {pages} {986 } (\bibinfo {year}
  {1956})}\BibitemShut {NoStop}%
\bibitem [{\citenamefont {Chaloner}\ and\ \citenamefont
  {Verdinelli}(1995)}]{Chaloner1995}%
  \BibitemOpen
  \bibfield  {author} {\bibinfo {author} {\bibfnamefont {K.}~\bibnamefont
  {Chaloner}}\ and\ \bibinfo {author} {\bibfnamefont {I.}~\bibnamefont
  {Verdinelli}},\ }\bibfield  {title} {\bibinfo {title} {{Bayesian Experimental
  Design: A Review}},\ }\href {https://doi.org/10.1214/ss/1177009939}
  {\bibfield  {journal} {\bibinfo  {journal} {Statistical Science}\ }\textbf
  {\bibinfo {volume} {10}},\ \bibinfo {pages} {273 } (\bibinfo {year}
  {1995})}\BibitemShut {NoStop}%
\bibitem [{\citenamefont {Huan}\ and\ \citenamefont
  {Marzouk}(2013)}]{Huan2013}%
  \BibitemOpen
  \bibfield  {author} {\bibinfo {author} {\bibfnamefont {X.}~\bibnamefont
  {Huan}}\ and\ \bibinfo {author} {\bibfnamefont {Y.~M.}\ \bibnamefont
  {Marzouk}},\ }\bibfield  {title} {\bibinfo {title} {Simulation-based optimal
  {B}ayesian experimental design for nonlinear systems},\ }\href
  {https://doi.org/https://doi.org/10.1016/j.jcp.2012.08.013} {\bibfield
  {journal} {\bibinfo  {journal} {Journal of Computational Physics}\ }\textbf
  {\bibinfo {volume} {232}},\ \bibinfo {pages} {288} (\bibinfo {year}
  {2013})}\BibitemShut {NoStop}%
\bibitem [{\citenamefont {Chitturi}\ \emph {et~al.}(2023)\citenamefont
  {Chitturi}, \citenamefont {Ji}, \citenamefont {Petsch}, \citenamefont {Peng},
  \citenamefont {Chen}, \citenamefont {Plumley}, \citenamefont {Dunne},
  \citenamefont {Mardanya}, \citenamefont {Chowdhury}, \citenamefont {Chen}
  \emph {et~al.}}]{chitturi2023capturing}%
  \BibitemOpen
  \bibfield  {author} {\bibinfo {author} {\bibfnamefont {S.~R.}\ \bibnamefont
  {Chitturi}}, \bibinfo {author} {\bibfnamefont {Z.}~\bibnamefont {Ji}},
  \bibinfo {author} {\bibfnamefont {A.~N.}\ \bibnamefont {Petsch}}, \bibinfo
  {author} {\bibfnamefont {C.}~\bibnamefont {Peng}}, \bibinfo {author}
  {\bibfnamefont {Z.}~\bibnamefont {Chen}}, \bibinfo {author} {\bibfnamefont
  {R.}~\bibnamefont {Plumley}}, \bibinfo {author} {\bibfnamefont
  {M.}~\bibnamefont {Dunne}}, \bibinfo {author} {\bibfnamefont
  {S.}~\bibnamefont {Mardanya}}, \bibinfo {author} {\bibfnamefont
  {S.}~\bibnamefont {Chowdhury}}, \bibinfo {author} {\bibfnamefont
  {H.}~\bibnamefont {Chen}}, \emph {et~al.},\ }\bibfield  {title} {\bibinfo
  {title} {Capturing dynamical correlations using implicit neural
  representations},\ }\href
  {https://www.nature.com/articles/s41467-023-41378-4} {\bibfield  {journal}
  {\bibinfo  {journal} {Nature Communications}\ }\textbf {\bibinfo {volume}
  {14}},\ \bibinfo {pages} {5852} (\bibinfo {year} {2023})}\BibitemShut
  {NoStop}%
\bibitem [{\citenamefont {Lajer}\ \emph {et~al.}(2025)\citenamefont {Lajer},
  \citenamefont {Dai}, \citenamefont {Barros}, \citenamefont {Carbone},
  \citenamefont {Johnston}, ,\ and\ \citenamefont {Dean}}]{repo}%
  \BibitemOpen
  \bibfield  {author} {\bibinfo {author} {\bibfnamefont {M.~K.}\ \bibnamefont
  {Lajer}}, \bibinfo {author} {\bibfnamefont {X.}~\bibnamefont {Dai}}, \bibinfo
  {author} {\bibfnamefont {K.}~\bibnamefont {Barros}}, \bibinfo {author}
  {\bibfnamefont {M.~R.}\ \bibnamefont {Carbone}}, \bibinfo {author}
  {\bibfnamefont {S.}~\bibnamefont {Johnston}}, ,\ and\ \bibinfo {author}
  {\bibfnamefont {M.~P.~M.}\ \bibnamefont {Dean}},\ }\href@noop {} {\bibinfo
  {title} {Respository for: Hamiltonian parameter inference from rixs spectra
  with active learning. {Github} respository:
  \url{https://github.com/mpmdean/lajer2025Hamiltonian}; {Zenodo repository:
  \url{https://doi.org/10.5281/zenodo.17353405}}}} (\bibinfo {year} {2025}),\
  \bibinfo {note} {to be assigned}\BibitemShut {NoStop}%
\bibitem [{\citenamefont {Dresselhaus}\ \emph {et~al.}(2008)\citenamefont
  {Dresselhaus}, \citenamefont {Dresselhaus},\ and\ \citenamefont
  {Jorio}}]{Dresselhaus2008GroupTheory}%
  \BibitemOpen
  \bibfield  {author} {\bibinfo {author} {\bibfnamefont {M.}~\bibnamefont
  {Dresselhaus}}, \bibinfo {author} {\bibfnamefont {G.}~\bibnamefont
  {Dresselhaus}},\ and\ \bibinfo {author} {\bibfnamefont {A.}~\bibnamefont
  {Jorio}},\ }\href {https://doi.org/10.1007/978-3-540-32899-5} {\emph
  {\bibinfo {title} {Group Theory. Application to the Physics of Condensed
  Matter}}}\ (\bibinfo  {publisher} {Springer},\ \bibinfo {year}
  {2008})\BibitemShut {NoStop}%
\end{thebibliography}%
\end{document}


\title{Supplementary Material for ``Hamiltonian parameter inference from resonant inelastic x-ray scattering with active learning''}

\author{Marton K. Lajer,\orcidlink{0000-0002-1168-8598}}\email[]{mlajer@bnl.gov}
\affiliation{Condensed Matter Physics and Materials Science Department, Brookhaven National Laboratory, Upton, New York 11973, USA\looseness=-1}

\author{Xin Dai\,\orcidlink{0000-0002-3235-1038}}
\affiliation{Computing and Data Sciences Directorate, Brookhaven National Laboratory, Upton, New York 11973, USA}

\author{Kipton Barros\,\orcidlink{0000-0002-1333-5972}}
\affiliation{Theoretical Division and CNLS, Los Alamos National Laboratory, Los Alamos, New Mexico 87545, USA}

\author{Matthew~R.~Carbone\,\orcidlink{0000-0002-5181-9513}}
\affiliation{Computing and Data Sciences Directorate, Brookhaven National Laboratory, Upton, New York 11973, USA}

\author{S. Johnston\,\orcidlink{0000-0002-2343-0113}}\email[]{sjohn145@utk.edu}
\affiliation{Department of Physics and Astronomy, The University of Tennessee, Knoxville, Tennessee 37996, USA}
\affiliation{Institute for Advanced Materials and Manufacturing, University of Tennessee, Knoxville, Tennessee 37996, USA\looseness=-1}

\author{M. P. M. Dean\,\orcidlink{/0000-0001-5139-3543}}\email[]{mdean@bnl.gov}
\affiliation{Condensed Matter Physics and Materials Science Department, Brookhaven National Laboratory, Upton, New York 11973, USA\looseness=-1}
\affiliation{Department of Physics and Astronomy, The University of Tennessee, Knoxville, Tennessee 37996, USA}

\date{\today}

\maketitle

\section{Determining symmetry labels}\label{AnnotationAppendix}
\subsection{Quantum numbers}
Let $\mathcal{H}$ be a Hilbert space, and $Q_{i}$, $i\in\left\{ 1,2,\dots n_{q}\right\} $
mutually commuting operators acting in $\mathcal{H}$ such that 
\[
\left[Q_{i},Q_{j}\right]=0,\quad\forall i,j.
\]
The set $\left\{ Q_{i}\right\} $ has a common set of eigenvectors
$\left|\left\{ q_{i}\right\} ;r_{i}\right\rangle $ labeled by the
set of eigenvalues $Q_{j}\left|\left\{ q_{i}\right\} ;r_{i}\right\rangle =q_{j}\left|\left\{ q_{i}\right\} ;r_{i}\right\rangle $.
The numbers $q_{i}$ are called quantum numbers. The index $r_{i}$ runs through the degenerate subspaces with completely coinciding quantum numbers $\left\{ q_{i}\right\} $.

EDRIXS works in the Fock basis where basis states are labeled by the occupation numbers of individual orbitals. However, we are often interested in the eigenbasis of operators like $S_{z}$, $S^{2}$,
$L_{z}$, $L^{2}$, $J_{z}=\left(L_{z}+S_{z}\right)$ or $J^{2}$.
These operators don't all commute with the occupation number operators,
so a basis vector in the occupation basis will not generally have a definite angular momentum eigenvalue.

Let us construct an auxiliary operator $A$ such that
\[
A=\sum_{i=1}^{n_{q}}a_{i}Q_{i}
\]
with $a_{i}$ chosen to be algebraically independent. For example, a convenient choice is to take $a_{i}=\eta\sqrt{p_{i}}$, where $\eta$ is an arbitrary constant and $p_{i}$ is the sequence such that $p_{1}=1$ and $p_{i}$ is the $i-1$ smallest prime number for $i>1$. If the matrix elements of $Q_{i}$ are computed in the occupation basis,
the unitary matrix built from the eigenvectors of $A$ provides the
basis transform to the basis with definite symmetry labels. The set
of $q$ values labeling a degenerate subspace can by obtained from the
eigenvalues of $A$:
\[
A\left|\phi_{i}^{A}\right\rangle =\eta\sum_{i=1}^{n_{q}}\sqrt{p_{i}}q_{i}\left|\phi_{i}^{A}\right\rangle 
\]
using e.g. simple table lookup. Alternatively, one can act the symmetry
operators directly, $Q_{i}\left|\phi_{j}^{A}\right\rangle =q_{i}^{\left(j\right)}\left|\phi_{j}^{A}\right\rangle $.

\subsection{Discrete symmetry labels}

To decompose the degenerate subspaces of $A$ further, we consider
random symmetric matrices $R_{\left\{ q_{i}\right\} }$ with matrix
elements sampled from an uniform distribution. We intend to build matrices 
invariant with respect to a discrete symmetry group $G$. To this
end, we construct the operator
\[
\bar{R}_{\left\{ q_{i}\right\} }=\frac{1}{|G|}\sum_{g\in G}D\left(g\right)R_{\left\{ q_{i}\right\} }D^{\dagger}\left(g\right)
\]
It is easy to see that $\bar{R}_{\left\{ q_{i}\right\} }$ is invariant
with respect to $G$ by conjugating it with an arbitrary
group element $g^{\prime}\in G$:
\begin{align*}
D\left(g^{\prime}\right)\bar{R}_{\left\{ q_{i}\right\} }D^{\dagger}\left(g^{\prime}\right) & =\frac{1}{|G|}\sum_{g\in G}D\left(g^{\prime}\right)D\left(g\right)R_{\left\{ q_{i}\right\} }D^{\dagger}\left(g\right)D^{\dagger}\left(g^{\prime}\right)\\
 & =\frac{1}{|G|}\sum_{g\in G}D\left(g^{\prime}g\right)R_{\left\{ q_{i}\right\} }D^{\dagger}\left(g^{\prime}g\right)\\
 & =\frac{1}{|G|}\sum_{h\in G}D\left(h\right)R_{\left\{ q_{i}\right\} }D^{\dagger}\left(h\right)\\
 & =\bar{R}_{\left\{ q_{i}\right\} }.
\end{align*}
Since $\bar{R}_{\left\{ q_{i}\right\} }$ commutes with all group elements, its eigenvectors are restricted to transform within its degenerate subspaces. 
A priori it could happen that one such subspace contains multiple \glspl*{irrep}. However, as we constructed $R$ with random matrix elements, it is guaranteed
(with probability $\sim1$ up to machine precision) that $\bar{R}_{\left\{ q_{i}\right\} }$
does not have accidental degeneracies in its eigenvalues apart from that strictly required by its group invariance. The unlikely event of an accidental degeneracy is under control (see below) and can be corrected by repeating with another random matrix $R$. Therefore the eigenvectors of $\bar{R}_{\left\{ q_{i}\right\} }$ can be assumed to
transform according to individual \glspl*{irrep} of $G$. In turn, the eigenvectors can be labeled by the particular \gls*{irrep} copy they belong to. 

Given the character table of the group $G$, we construct the square matrix $T$ with matrix elements corresponding to the entries of the character table. We then pick an exemplar group element from each conjugate class and calculate the trace of their representations in the degenerate subspace of $\bar{R}_{\left\{ q_{i}\right\} }$ with eigenvalue $\lambda$. Thus, in each subspace $\lambda$, we get a vector $\mathcal{X}_\lambda$ of characters. The multiplicity of each \gls*{irrep} in the subspace is given by the expression
\be
\mu_\lambda=\mathcal{X}_\lambda T^{-1}. \label{getannotfromT}
\ee
If the degenerate subspace indeed corresponds to an \gls*{irrep}, the vector $\mu$ is a standard basis vector with one element being $1$ and the rest are zeros. We can read the type of the \gls*{irrep} by comparing the position of the one in $\mu$ and the arrangement of the rows of the character table.

Let us consider a Hamiltonian $H$ acting in the Hilbert space $\mathcal{H}$.
We insist that $H$ is not necessarily invariant under the symmetry
operations generated by $Q_{i}$:
\[
\left[H,Q_{i}\right]\neq0.
\]
We also do not require invariance with respect to the finite group $G$. Instead, after we  obtain the eigenvectors of $
H\left|\psi\right\rangle=E_\psi\left|\psi\right\rangle$, we calculate the squares of overlaps between $\left|\psi\right\rangle$ and the completely annotated basis vectors $\left|\phi_{q_i;\lambda;r}\right\rangle$
\be
w_{\psi, q_i;\lambda}=\sum_{r=1}^{\dim D_\lambda}\left|\left\langle \psi | \phi_{q_i;\lambda;r}\right\rangle\right|^2.
\ee

Since the wavefunction is normalized and the annotated vectors span the Hilbert space $\mathcal{H}$,
\be
 \sum_{q_i;\lambda} w_{\psi, q_i;\lambda}=1.
\ee
The weight of any combination of symmetry labels in $\left|\psi\right\rangle$ is easily calculated by summing up the appropriate subset of elementary weights $ w_{\psi, q_i;\lambda}$.

We note that projections of $\left|\psi\right\rangle$ to isotypic subspaces of the group $G$ corresponding to an \gls*{irrep} $\Gamma$ could also be calculated by the projection formula (see e.g. Ref.~\cite{Dresselhaus2008GroupTheory})
\be
\sum_{\substack{
\lambda\in\text{copies} \\ 
\text{of irrep }\Gamma }}w_{\psi, q_i;\lambda} =
\left|\frac{\dim \Gamma}{|G|} \sum_{g \in G} \chi^{\Gamma}(g)^* \left\langle \psi | D_{q_i}(g) | \psi \right\rangle \right|^2, 
\ee
where $D_{q_i}(g)$ is the (generally reducible) representation of $G$ that acts in the degenerate subspace $\{q_i\}$ of the matrix $A$. This approach avoids the construction of the commutant $\bar{R}$, but does not give access to weights in individual \gls*{irrep} copies. Such extra information can sometimes be useful, as we discuss in Section \ref{IrrepCopies}.

This concludes our general strategy of eigenstate annotation. We stress that the commutant method is applicable for any finite group.\footnote{Although tangential to the present discussion, we remark that this method entails a didactic, systematic way to obtain character tables from multiplication tables of finite matrices. One then starts by constructing the regular representation of dimension $|G|$. All inequivalent \glspl*{irrep} are present in the regular representation. Solving for the spectrum of $R'$ provides the invariant subspaces of the group. The characters are then obtained from the traces of group elements projected into the invariant subspaces. This method allows the computation of character tables for groups of order up to a few tens of thousands on a PC, depending on available memory.}

\subsection{The octahedral group}
We focus on the octahedral group Oh, and specifically its orientation-preserving subgroup O. The group Oh is of order $48$, while the group O has $24$ elements. The reason of our focus is that we intend to annotate the states of the initial Hamiltonian, which are linear combinations of $d$-shell excitations. Therefore, all eigenfunctions are invariant under inversion.

The group elements of the octahedral group take the form
\[
g_{\alpha}=e^{\mathrm{i}\alpha_{x}L_{x}+\mathrm{i}\alpha_{y}L_{y}+\mathrm{i}\alpha_{z}L_{z}}.
\]
Elements of the group O are organized into $5$ conjugacy classes as follows.
\begin{enumerate}
\item ($6C_4$ -- order 4) 6 rotations about the cube edges with angles $n\frac{\pi}{2}$,
$n\in\left\{ 1,3\right\} $:
\[
\alpha_{3n-2}=n\frac{\pi}{2}\left(1,0,0\right),\quad\alpha_{3n-1}=n\frac{\pi}{2}\left(0,1,0\right),
\quad
\alpha_{3n}=n\frac{\pi}{2}\left(0,0,1\right).
\]
\item ($3C_2\sim (C_4)^2$ -- order 2) 3 rotations about the cube edges with angle $\pi$:
\[
\alpha_{4}=\pi\left(1,0,0\right),\quad\alpha_{5}=\pi\left(0,1,0\right), \quad
\alpha_{6}=\pi\left(0,0,1\right).
\]
\item ($8C_2$ -- order 2) 8 rotations about the face diagonals with angle $\pi$:
\begin{align*}
\alpha_{10} & =\pi\left(1,1,0\right),\quad\alpha_{11}=\pi\left(1,-1,0\right),\quad\alpha_{12}=\pi\left(1,0,1\right)\\
\alpha_{13} & =\pi\left(1,0,-1\right),\quad\alpha_{14}=\pi\left(0,1,1\right),\quad\alpha_{15}=\pi\left(0,1,-1\right).
\end{align*}
\item ($6C_3$ -- order 3) 6 rotations about the body diagonals with angle $\frac{2\pi}{3}n$,
$n\in\left\{ 1,2\right\} $:
\begin{align*}
\alpha_{4n+12} & =\frac{2\pi}{3}n\left(1,1,1\right),\quad\alpha_{4n+13}=\frac{2\pi}{3}n\left(1,1,-1\right),\\
\alpha_{4n+14} & =\frac{2\pi}{3}n\left(1,-1,1\right),\quad\alpha_{4n+15}=\frac{2\pi}{3}n\left(-1,1,1\right).
\end{align*}
\item ($E$ -- order 1) the identity,
\[
\alpha_{0}=\left(0,0,0\right).
\]
\end{enumerate}
The remaining $24$ elements and further $5$ conjugacy classes of Oh are constructed multiplying the vectors above by $-1$.

Since there are $5$ conjugacy classes in the group $O$, there are $5$ inequivalent \glspl*{irrep}. The character table of the  group $O$ is shown on Table \ref{TabCharacter}.

\begin{table}[b]

\caption{Character table of the group $\mathrm{O}\cong\mathcal{S}_4$}
\begin{centering}
\label{TabCharacter}
\begin{ruledtabular}
\begin{tabular}{cccccc}
 & $E$ & $8C_3$ & $6C_2$ & $6C_4$ & $3C_2$\\
\hline 
$A_{1g}$ & 1 & 1 & 1 & 1 & 1 \\
$A_{2g}$ & 1 & 1 & -1 & -1 & 1 \\
$E_g$ & 2 & -1 & 0 & 0 & 2 \\
$T_{1g}$ & 3 & 0 & -1 & 1 & -1 \\
$T_{2g}$ & 3 & 0 & 1 & -1 & -1 \\
\end{tabular}
\end{ruledtabular}
\par\end{centering}
\end{table}

The inverse matrix $T^{-1}$ of Eq. \eqref{getannotfromT} takes the form
\[
T^{-1}=\frac{1}{24}\left(\begin{array}{ccccc}
1 & 8 & 6 & 6 & 3\\
1 & 8 & -6 & -6 & 3\\
2 & -8 & 0 & 0 & 6\\
3 & 0 & -6 & 6 & -3\\
3 & 0 & 6 & -6 & -3
\end{array}\right).
\]

\subsection{Weights in individual irreducible representation copies} \label{IrrepCopies}

Let $G$ be a finite group with \glspl*{irrep} labeled
by $\alpha$. Let $D$ be a (reducible) representation of the group.
We can decompose $D$ into isotypic blocks $D_{\alpha}^{m_{\alpha}}$
as
\[
D=\bigoplus_{\alpha}D_{\alpha}^{m_{\alpha}}, 
\]
where $D_{\alpha}$ denotes an \gls*{irrep} and $m_{\alpha}$
is its multiplicity in $D$, that is
\[
D_{\alpha}^{m_{\alpha}}=D_{\alpha}^{\left(1\right)}\oplus D_{\alpha}^{\left(2\right)}\oplus\dots\oplus D_{\alpha}^{\left(m_{\alpha}\right)}.
\]

The decomposition of $D$ into isotypic blocks is unique. However,
the decomposition of an isotypic block into its component \glspl*{irrep} 
is only fixed up to $O(m_{\alpha})$ orthogonal transformations. To
see this, let us consider an arbitrary group element $g\in G$. The
representation of $g$ according to the \gls*{irrep} $D_{\alpha}$ is $D_{\alpha}\left(g\right)$.
If the \gls*{irrep} is $n_{\alpha}$ dimensional, then $D_{\alpha}\left(g\right)$
is an $n_{\alpha}\times n_{\alpha}$matrix. In other words, we can
introduce a set of $n_{\alpha}$ orthonormal vectors $e_{\alpha,i}$,
$i\in\left\{ 1,2,\dots n_{\alpha}\right\} ,$ so that
\[
D_{\alpha}\left(g\right)\equiv\sum_{ab}\left(e_{\alpha,a}\circ e_{\alpha,b}\right)\left[D_{\alpha}\left(g\right)\right]_{ab}.
\]
A vector $v_{\alpha}$ is said to transform according to the irreducible
representation $D_{\alpha}$, if
\[
v_{\alpha}=\sum_{i=1}^{\dim D_{\alpha}}e_{\alpha,i}v_{\alpha,i}.
\]
The transformation of $v$ by the group element $g$ is then written as
\bea
D_{\alpha}\left(g\right)v&=&\sum_{ab=1}^{\dim D_{\alpha}}\left(e_{\alpha,a}\circ e_{\alpha,b}\right)\left[D_{\alpha}\left(g\right)\right]_{ab}\sum_{i=1}^{\dim D_{\alpha}}e_{\alpha,i}v_{i}
=\sum_{a,i=1}^{\dim D_{\alpha}}e_{\alpha,a}\left[D_{\alpha}\left(g\right)\right]_{ai}v_{i}.
\eea
Likewise, the representation of $g$ in $D$ is written as
\[
D\left(g\right)=\sum_{\alpha}\sum_{n=1}^{m_{\alpha}}\sum_{ab=1}^{\dim D_{\alpha}}\left(e_{\alpha,a}^{\left(n\right)}\circ e_{\alpha,b}^{\left(n\right)}\right)\left[D_{\alpha}\left(g\right)\right]_{ab}.
\]

The core of the redundancy is that within each isotypic block, we
can introduce a new set of basis vectors,
\[
\tilde{e}_{\alpha,a}^{\left(k\right)}=\sum_{i=1}^{m_{\alpha}}c_{i}^{\left(k\right)}e_{\alpha,a}^{\left(i\right)};\quad\sum_{i=1}^{m_{\alpha}}\left|c_{i}^{\left(k\right)}\right|^{2}=1.
\]
The normalization condition of the coefficients $c_{i}^{\left(k\right)}$
ensures that the set of basis vectors $\left\{ \tilde{e}_{\alpha,a}^{\left(k\right)}\right\} $
is also orthonormal. Let us consider a generic vector 
\[
v_{m\alpha}=\sum_{n=1}^{m_{\alpha}}\sum_{i=1}^{\dim D_{\alpha}}e_{\alpha,i}^{\left(n\right)}v_{m\alpha,i}^{\left(n\right)}.
\]
The projection amplitude of the vector $v_{m\alpha}$ onto the \gls*{irrep}
$D_{\alpha}^{\left(k\right)}$ is
\bea
w_{k}\left(v_{m\alpha}\right)&=&\sum_{j=1}^{\dim D_{n}}e_{\alpha,j}^{\left(k\right)}\sum_{n=1}^{m_{\alpha}}\sum_{i=1}^{\dim D_{\alpha}}e_{\alpha,i}^{\left(n\right)}v_{m\alpha,i}^{\left(n\right)}
=\sum_{j=1}^{\dim D_{\alpha}}v_{m\alpha,j}^{\left(k\right)}.
\eea

What is the largest amplitude of $v_{m\alpha}$ in a single \gls*{irrep}?
To answer this, we need to take into account the redundancy of the
definition of \glspl*{irrep}. The amplitude on a transformed \gls*{irrep} $\tilde{D}_{\alpha}^{\left(k\right)}$takes
the form
\begin{equation}
\tilde{w}_{k}\left(v_{m\alpha}\right)=\sum_{j=1}^{\dim D_{n}}\tilde{e}_{\alpha,j}^{\left(k\right)}\sum_{n=1}^{m_{\alpha}}\sum_{i=1}^{\dim D_{\alpha}}e_{\alpha,i}^{\left(n\right)}v_{m\alpha,i}^{\left(n\right)}
=\sum_{l=1}^{m_{\alpha}}c_{l}^{\left(k\right)}\sum_{j=1}^{\dim D_{n}}v_{m\alpha,j}^{\left(l\right)}
=\sum_{l=1}^{m_{\alpha}}c_{l}^{\left(k\right)}w_{l}\left(v_{m\alpha}\right).
\end{equation}
We would like to calculate
\begin{align}
\max_{\left\Vert c\right\Vert ^{2}=1}\sum_{l=1}^{m_{\alpha}}c_{l}w_{l}\left(v_{m\alpha}\right)=\max_{\left\Vert c\right\Vert ^{2}=1}\sum_{j=1}^{\dim D_{\alpha}}\sum_{l=1}^{m_{\alpha}}v_{m_\alpha,j}^{\left(l\right)}c_{l}
=\left(\max_{\left\Vert c\right\Vert ^{2}=1}\left(\sum_{j=1}^{\dim D_{\alpha}}\sum_{l=1}^{m_{\alpha}}v_{m\alpha,j}^{\left(l\right)}c_{l}\right)^{2}\right)^{\frac{1}{2}}.
\end{align}
Introducing $V_{lj}=v_{m\alpha,j}^{\left(l\right)}$, we then obtain
\begin{align}
\max_{\left\Vert c\right\Vert ^{2}=1}\sum_{l=1}^{m_{\alpha}}c_{l}w_{l}\left(v_{m\alpha}\right)
=\left[\max_{\left\Vert c\right\Vert ^{2}=1}\sum_{l,p=1}^{m_{\alpha}}c_{l}\left(\sum_{i,j=1}^{\dim D_{\alpha}}V^{\phantom\dagger}_{lj}V_{ip}^{\dagger}\right)c_{p}\right]^{\frac{1}{2}}
=\left[\lambda_{\max}\left(\sum_{i,j=1}^{\dim D_{\alpha}}V^{\phantom\dagger}_{lj}V_{ip}^{\dagger}\right)\right]^\frac{1}{2}.
\end{align}
This result is independent of the initial basis chosen and well-defined.

It can be useful to know the maximal weight of a single \gls*{irrep} copy in an eigenvector. For example, this can provide useful information on the shape of the electron density. If the (multiparticle) eigenvector is dominated by a single \gls*{irrep}, the electron density will be reminiscent of the geometry of a single-particle wavefunction.

\section{Further details of the fitting procedure and results}
In this section, we provide several additional plots that characterize the performance and details of our fitting procedure. 
\subsection{\NiCl}
\begin{figure}[h]
\begin{centering}
\includegraphics[width=8cm]{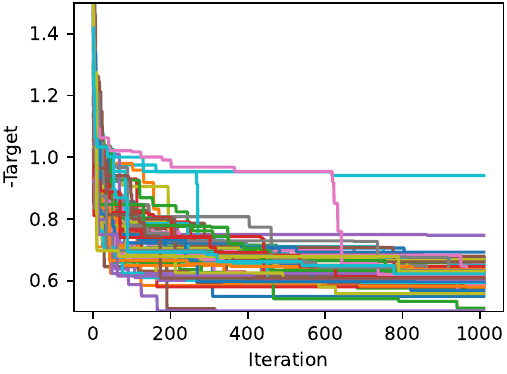}
\par\end{centering}
\caption{\NiCl: Decrease of the sum normalized $L_1$ distance function for $60$ runs
of $1000$ iterations\label{FigNiCl2_decrease}}
\end{figure}
\begin{figure}[h]
\begin{centering}
\includegraphics[width=0.4\columnwidth]{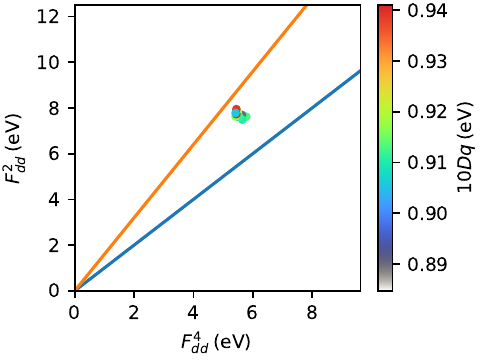}\\
\includegraphics[width=0.4\columnwidth]{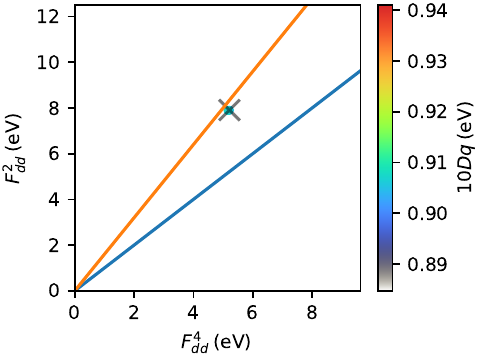}
\par\end{centering}
\caption{Top: distribution of $11$ \gls*{GPR} evaluations with $d\protect\leq1.2\chi^2_{\mathrm{min},\mathrm{GPR}}$ for \NiCl.
Bottom: results of subsequent greedy optimization starting from the
$11$ best \gls*{GPR} points. The chosen point is denoted by a gray cross ($\times$).\label{FigNiCl2BestPts}}
\end{figure}

\begin{figure*}[h]
\centering
\includegraphics[scale=0.8]{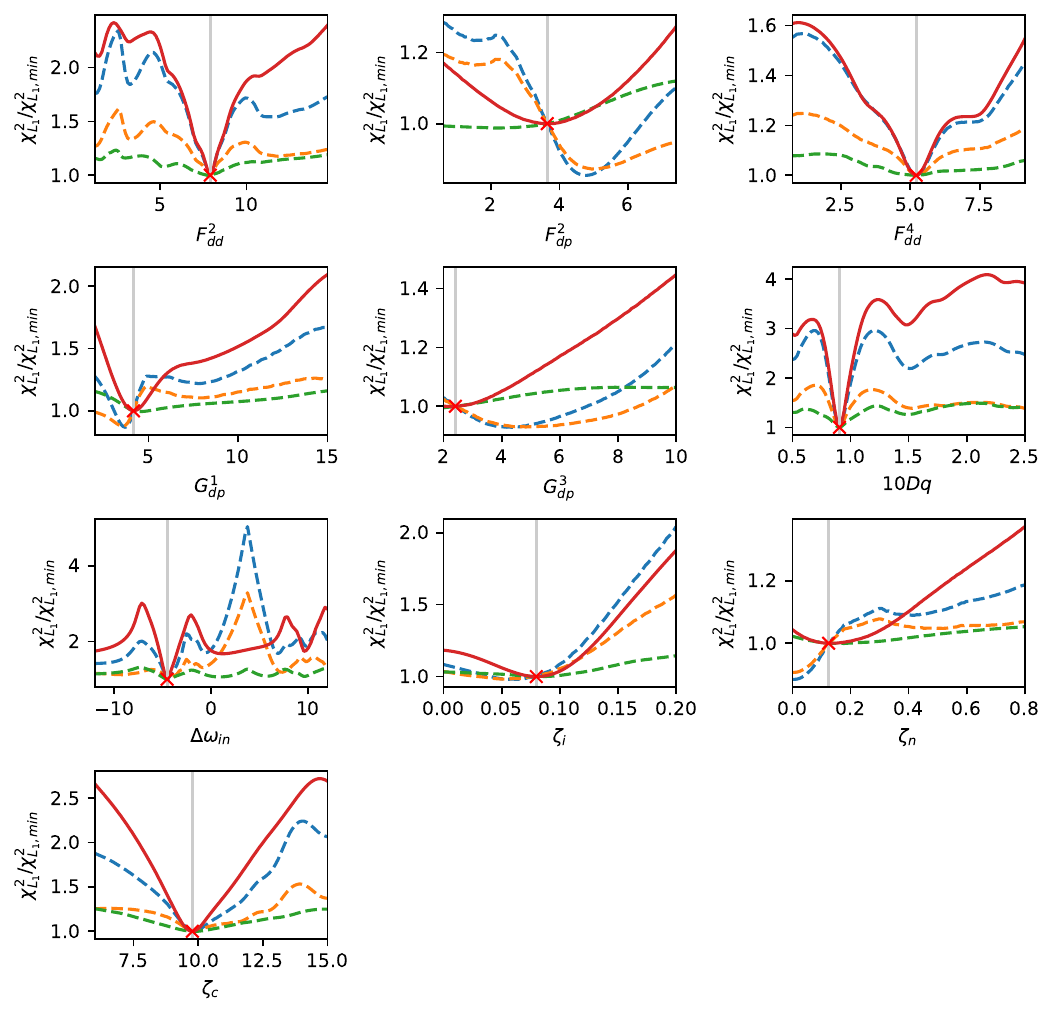}
\caption{\NiCl: The behavior of various distance measures around the fine-tuned minimum. Solid red: L1 sum normalized, dashed blue: L1 maximum normalized,
dashed orange: L2 sum normalized, dashed green: magnitude of gradient, maximum normalized\label{NiCl2Distances}
The fit is highly sensitive to initial Slater and crystal field parameters but weakly depend on intermediate state parameters, especially spin-orbit couplings.}
\end{figure*}

\clearpage
\subsection{\Hematite}
\begin{figure}[h]
\begin{centering}
\includegraphics[width=0.5\columnwidth]{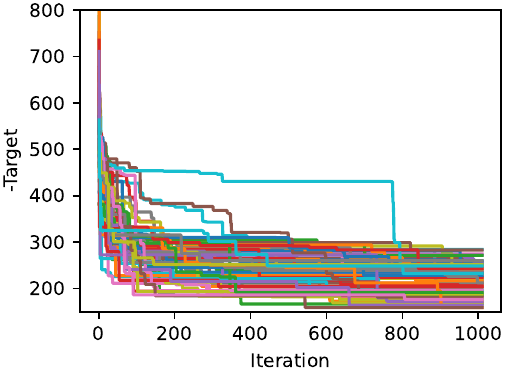}\\
\includegraphics[width=0.5\columnwidth]{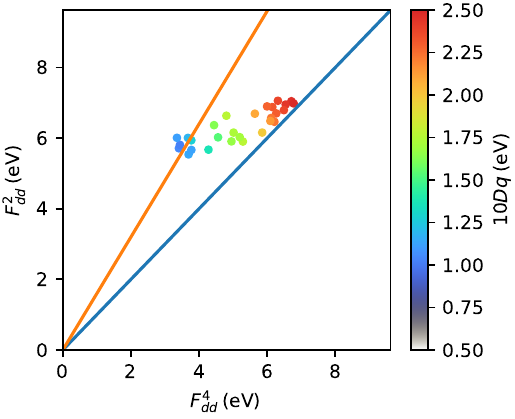}
\par\end{centering}
\caption{Top: Decrease of the maximum normalized L1 distance function for \Hematite over 
$60$ runs of $1000$ iterations. Bottom: distribution of the best $28$ GPR evaluations serving as starting point of the greedy refinement.\label{FigFe2O3_decrease}}
\end{figure}
\begin{figure}[h]
\centering
\includegraphics[scale=0.8]{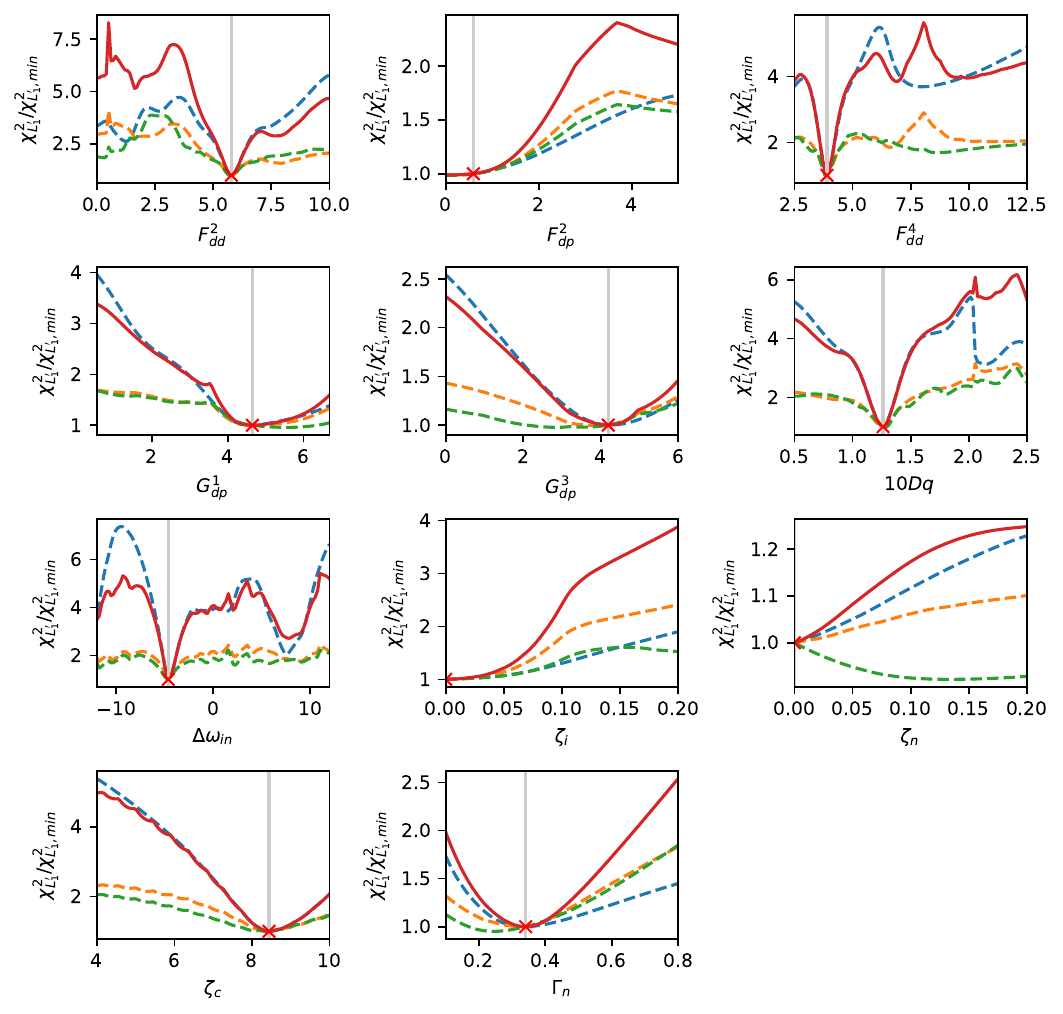}
\caption{\Hematite: The behavior of various distance measures around the fine-tuned minimum. Solid red: L1 \textit{maximum} normalized, dashed blue: L1 sum normalized, dashed orange: L2 sum normalized, dashed green: magnitude of gradient, maximum normalized. The fit is highly sensitive to initial Slater and crystal field parameters but has a weaker dependency on intermediate state parameters, especially the spin-orbit coupling $\zeta_n$.\label{Fe2O3Distances}}
\end{figure}

\clearpage
\subsection{\Osmate}

\begin{figure}[h]
\begin{centering}
\includegraphics[width=8cm]{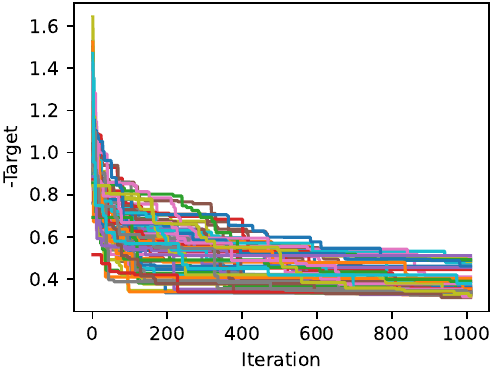}
\par\end{centering}
\caption{\Osmate: Decrease of the sum normalized L1 distance function for $60$ runs
of $1000$ iterations\label{Ca3LiOsO6_decrease}}
\end{figure}
\begin{figure}[h]
\begin{centering}
\includegraphics[width=0.4\columnwidth]{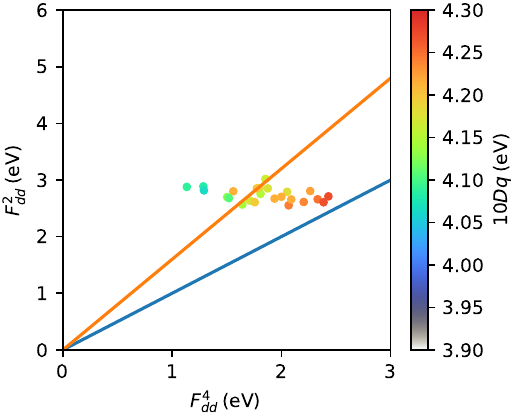}
\includegraphics[width=0.4\columnwidth]{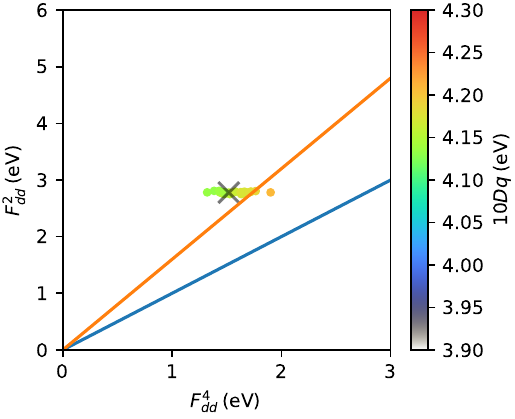}
\par\end{centering}
\caption{Left: distribution of $24$ \gls*{GPR} evaluations with $d\protect\leq1.06\chi^2_{\text{min},\text{GPR}}$ for \Osmate.
Right: results of subsequent greedy optimization starting from the
$24$ best \gls*{GPR} points. The final accepted point is denoted by a gray cross ($\times$). \label{FigCa3LiOsO6BestPts}}
\end{figure}

\begin{figure}[h!]
\centering
\includegraphics[scale=0.8]{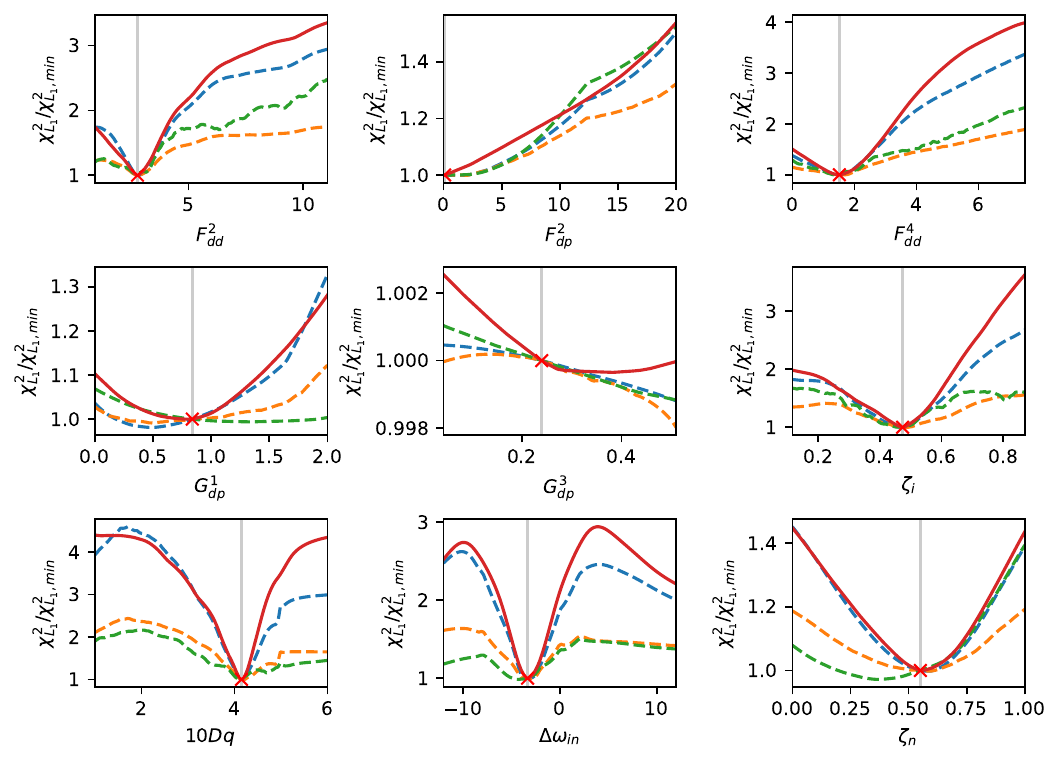}
\caption{\Osmate: The behavior of various distance measures around the fine-tuned minimum.
Solid red: L1 sum normalized, dashed blue: L1 maximum normalized,
dashed orange: L2 sum normalized, dashed green: magnitude of gradient,
maximum normalized. The fit is highly sensitive to initial Slater and crystal field parameters but has a weaker dependency on intermediate state parameters.\label{Ca3LiOsO6Distances}}
\end{figure}

\FloatBarrier
\bibliography{refs}